\documentclass[11pt,notitlepage,oneside,a4paper]{report}
\usepackage[latin1]{inputenc}
\usepackage[nottoc]{tocbibind}
\usepackage{slashed}
\usepackage{amssymb}
\usepackage[english]{babel}
\usepackage[font=small,labelfont=bf]{caption}
\usepackage{feynmp}
\usepackage{graphicx}
\usepackage{cite}
\usepackage{mcite}

\newcommand{\half}{\frac{1}{2}}
\newcommand{\dd}{\mathrm{d}}
\newcommand{\Tr}{\mathrm{Tr}\,}
\newcommand{\piT}{\pi T}  

\newcommand{\gE}{g_\mathrm{E}}
\newcommand{\mE}{m_\mathrm{E}}
\newcommand{\mD}{m_\mathrm{D}}
\newcommand{\pE}{p_\mathrm{E}}
\newcommand{\pM}{p_\mathrm{M}}
\newcommand{\pMy}{p_{\mathrm{M}1}}
\newcommand{\pMk}{p_{\mathrm{M}2}}
\newcommand{\MSbar}{\overline{\mathrm{MS}}}
\newcommand{\order}[1]{\mathcal{O}(#1)}

\newcommand{\NRQCD}{$\textrm{NRQCD}_3$}
\newcommand{\xbot}{\mathbf{x}_\bot}
\newcommand{\Ei}{\mathrm{Ei}}
\newcommand{\ds}{\displaystyle}

\newcommand{\dF}{d_\mathrm{F}}
\newcommand{\dA}{d_\mathrm{A}}
\newcommand{\CF}{C_\mathrm{F}}
\newcommand{\CA}{C_\mathrm{A}}
\newcommand{\TF}{T_\mathrm{F}}

\begin{document}

\thispagestyle{empty}

\begin{flushright}
HU-P-D144
\end{flushright}

\vfill

\begin{center}

{\Large \bf Applications of dimensional reduction\\
	to electroweak and QCD matter }
\vspace{1cm}

Mikko Vepsäläinen\footnote{E-mail: Mikko.T.Vepsalainen@helsinki.fi}
\vspace{0.5cm}

\textit{ Theoretical Physics Division, Department of Physical Sciences and \\
	Helsinki Institute of Physics\\
	P.O. Box 64, FIN-00014 University of Helsinki,
	Finland}
\end{center}

\vspace{1cm}

\begin{abstract}
This paper is a slightly modified version of the introductory part of a doctoral dissertation also containing the articles \mbox{hep-ph/0311268}, \mbox{hep-ph/0510375}, \mbox{hep-ph/0512177} and \mbox{hep-ph/0701250}. The thesis discusses effective field theory methods, in particular dimensional reduction, in the context of finite temperature field theory. We first briefly review the formalism of thermal field theory and show how dimensional reduction emerges as the high-temperature limit for static quantities. Then we apply dimensional reduction to two distinct problems, the pressure of electroweak theory and the screening masses of mesonic operators in hot QCD, and point out the similarities. We summarize the results and discuss their validity, while leaving all details to original research articles.
\end{abstract}

\vfill

\newpage

\pagenumbering{roman}

\tableofcontents

\nocite{Laine:2003bd,Gynther:2005dj,Gynther:2005av,Vepsalainen:2007ke}

\begin{fmffile}{diagrams}

\chapter{Introduction}
\setcounter{page}{1}
\pagenumbering{arabic}

The physics of interactions between elementary particles is described to an amazing accuracy by the standard model of particle physics. It ties three of the four fundamental interactions, namely the electromagnetic, weak and strong interactions, together under the conceptual framework of relativistic quantum field theory. Scattering processes and bound states involving few particles are well described by the model, although many open questions, mostly related to strongly interacting states, remain. In the energy region currently accessible to experiments we have therefore full reason to believe that this theory is correct.

When matter is heated high above everyday temperatures, its neutral constituents are torn apart into an interacting plasma of elementary particles. At temperatures of the same order or higher than the particle masses this necessitates combining quantum statistical mechanics with relativistic field theory. The interactions between individual particles are still governed by the standard model interactions, but the effects of hot medium change their long-distance behavior and give rise to many-particle collective modes.

Experimentally such extreme conditions are accessible in relativistic heavy ion collisions currently produced at the Relativistic Heavy Ion Collider (RHIC) in Brookhaven, and, starting this year, also at Large Hadron Collider (LHC) at CERN. In these experiments two heavy nuclei collide against each other, forming a finite volume of extremely hot matter. The matter described by the theory of strong interactions, quantum chromodynamics (QCD), goes through a phase transition to a deconfined phase of color-charged particles forming a quark-gluon plasma, which then rapidly cools as it expands. This kind of temperatures were also present in the very early universe, whose expansion is sensitive to the equation of state of both QCD and electroweak matter.

The formalism for finite temperature quantum field theory arises naturally from the path integral quantization of field theories. The time coordinate is extended to complex values to account for varying the fields over statistical ensemble, and the functional integral is over all field configurations periodic or antiperiodic in the imaginary time. When temperature is larger than any other scale in the process, the excitations in the imaginary time can be integrated out and the physics of static quantities is described by a three-dimensional effective theory. This is known as dimensional reduction \cite{Ginsparg:1980ef,*Appelquist:1981vg}. The effective theory can be systematically derived, and it exhibits the same infrared behavior as the full theory. At finite temperatures the main advantage in using a dimensionally reduced effective theory in perturbative computations is the ability to systematically treat the various infrared divergences, as well as the resummations needed to cure them, in a simpler setting.

Dimensional reduction has been successfully applied over the years to compute many bosonic quantities both perturbatively and in combination with lattice simulations. In the QCD sector, the three-dimensional formulation known as EQCD has made it possible to perturbatively compute the pressure up to the last perturbative order $g^6 \ln\,g$ \cite{Shuryak:1977ut,*Chin:1978gj,Kapusta:1979fh,Toimela:1982hv,Arnold:1994ps,*Arnold:1995eb,Zhai:1995ac,Braaten:1996jr,Kajantie:2002wa,Kajantie:2003ax}, and the result has also been extended to nonzero chemical potentials \cite{Vuorinen:2003fs}. Lattice implementations of EQCD have been used to compute the static correlation lengths of various gluonic operators \cite{Reisz:1992er,*Karkkainen:1992jh,*Karkkainen:1993wu,Kajantie:1997tt, Laine:1998nq,*Laine:1999hh, Hart:1999dj,Hart:2000ha,Cucchieri:2001tw}. There are also recent developments in formulating an effective theory preserving the spontaneously broken $Z(3)$ symmetry of the deconfined phase \cite{Vuorinen:2006nz}, which is explicitly broken in EQCD \cite{Kajantie:1998yc}. Besides QCD, the electroweak symmetry breaking has also been solved in detail using lattice simulations in dimensionally reduced effective theory \cite{Kajantie:1995dw,Kajantie:1995kf,Kajantie:1996mn, Kajantie:1996qd,Karsch:1996yh,Gurtler:1997hr}, motivated by the possibility of a first order electroweak phase transition being the origin of the observed baryon asymmetry in the universe.

There are only few applications of dimensional reduction to fermionic observables, because the fermion fields are integrated out from the three-dimensional effective theory. This simplifies the computation of bosonic quantities tremendously, but the accessible fermionic observables are then limited to those that can be inferred from vacuum or bosonic ones, such as quark number susceptibilities $\chi_{ij}=\partial^2p/\partial\mu_i\partial\mu_j$ \cite{Vuorinen:2002ue}. Systematic application of dimensional reduction to fermionic operators was developed in \cite{Huang:1996tz}, inspired by the progress in heavy quark effective theories.

The use of dimensional reduction is restricted to time-independent quantities. It should be mentioned here that for real-time computations there exists another scheme of resumming the light particle self-energy corrections to regulate some of the infrared divergences, namely the hard thermal loop (HTL) approximation \cite{Braaten:1989mz,*Braaten:1990az,Frenkel:1989br}. Both schemes succesfully resum the one-loop infrared divergences, but in general the HTL Green's functions are more complicated, since they carry the full analytic structure of the original theory. It is also very hard to systematically improve the HTL approximation beyond the leading order.

In this thesis we study two applications of dimensional reduction to the standard model, the perturbative evaluation of the electroweak pressure and the next-to-leading order correction to screening masses of mesonic operators. The thesis is organized as follows. In chapter \ref{chap:basics} we first review the formalism of thermal quantum field theory, and then discuss dimensional reduction in the context of general effective theories in section \ref{sec:dimensional_reduction}. In chapter \ref{chap:pressure} we go through the computation of the electroweak theory pressure, with special attention given to the behavior near the phase transition. We combine the result with the previously known QCD pressure in section \ref{sec:pressure_numerics} and study the convergence of the series and the deviation from the ideal gas for physical values of parameters. Results for a simpler, weakly coupled SU(2) + Higgs theory are also shown for comparison.

In chapter \ref{chap:correlators} we review our work on meson correlators. After a short motivation using linear response theory, we compute the leading order correlators at zero and finite density. Then we proceed to derive a dimensionally reduced effective theory for the lowest fermionic modes and solve the $\order{g^2}$ corrections to screening masses. Finally, we compare with recent lattice determinations of the masses and discuss the differences. Chapter \ref{chap:lopetus} contains our conclusions.

\chapter{Thermal field theory}
\label{chap:basics}

In this chapter we will first review how the thermodynamical treatment of quantum field theory can be formulated in terms of Euclidean path integrals. We then proceed to discuss dimensional reduction, which is the underlying effective theory method used in all the research papers included in this thesis.

\section{Basic thermodynamics of quantum fields}

The statistical properties of relativistic quantum field theory are most naturally described using the grand canonical ensemble. Since particles can be spontaneously created and annihilated, the microcanonical or canonical ensembles with fixed particle numbers cannot be built, but instead one would have to use the conserved quantities like electric charge. To avoid this kind of complicated constraints on field configurations, it is generally easier to fix the mean values of energy and conserved commuting number operators using the Lagrange multipliers $\beta=1/T$ and $\mu_i$, respectively. This is the grand canonical ensemble.

The thermodynamical properties of the system are given by the partition function and its derivatives. In quantum mechanics the partition function is defined as the trace of the density matrix $\rho$,
\begin{equation}
 Z(T,V,\mu_i) \equiv \Tr \rho =  \Tr e^{-\beta(H-\mu_i N_i)},
\label{eq:Z_gen_def}
\end{equation}
where $H$ and $N_i$ are the Hamiltonian and conserved number operators, respectively. The thermal average of an operator is then defined as
\begin{equation}
 \langle A \rangle = \frac{1}{Z}\Tr \rho A\, ,
\end{equation}
and the usual thermodynamic quantities like pressure, entropy, energy and particle numbers are given by the partial derivatives
\begin{eqnarray}
 p = T\frac{\partial \ln Z}{\partial V}, && S = \frac{\partial T\ln Z}{\partial T} \nonumber \\
 N_i = T\frac{\partial \ln Z}{\partial \mu_i}, && E = -pV +TS +\mu_i N_i.
\label{eq:thermo_quantities}
\end{eqnarray}

In quantum mechanics the evaluation of the trace in Eq.~(\ref{eq:Z_gen_def}) is simple, one just takes any complete orthonormal basis $\{|n\rangle\}$, preferably eigenstates of $H-\mu N$ if these are known, and sums over $\langle n|\rho|n\rangle$. The same procedure can in principle be applied to field theory, where the sum over basis vectors is replaced by a functional integral in the space of field configurations. 

Field theories are usually defined in the Lagrangian formalism, and finding the Hamiltonian function required for computation of the partition function in Eq.~(\ref{eq:Z_gen_def}) can be quite involved, in particular in the context of gauge theories. One has to fix the gauge and then carefully separate the canonical variables from auxiliary ones depending on the chosen gauge \cite{Weinberg:qft2}. In addition to the usual canonical equations of motion, the fields are constrained by the gauge condition and the field equation for the auxiliary field, which can be interpreted as the Gauss' law.

Once the Hamiltonian has been found, we can insert a complete set of eigenstates $|\phi(\mathbf{x});t\rangle$ of the field operator $\hat{\phi}(\mathbf{x})$ in the Heisenberg picture to compute the partition function. This gives
\begin{equation}
 Z(T,V,\mu_i) = \int [\dd \phi]\, \langle \phi(\mathbf{x});t|e^{-\beta(H-\mu_i N_i)}|\phi(\mathbf{x});t\rangle,
\label{eq:Z_as_trace}
\end{equation}
where the integration is over all canonical variables. From the time-dependence of the field operator it follows that
\begin{equation}
 \hat{\phi}(\mathbf{x},t) = e^{iHt}\hat{\phi}(\mathbf{x},0)e^{-iHt} \quad \Rightarrow \quad
	|\phi(\mathbf{x});t\rangle = e^{iHt}|\phi(\mathbf{x});0\rangle.
\end{equation}
Eq.~(\ref{eq:Z_as_trace}) can then be viewed as the transition amplitude for the field to return to the same state after an imaginary time $-i\beta$, when the time-development is given by the Hamiltonian $H-\mu_i N_i$,
\begin{equation}
 Z(T,V,\mu_i) = \int [\dd \phi]\, \langle \phi(\mathbf{x});t-i\beta|\phi(\mathbf{x});t\rangle.
\label{eq:Z_as_amplitude}
\end{equation}
Dividing the time interval into infinitesimally small pieces and inserting at every point a complete set of position and momentum eigenstates this can be cast into a path integral form (for details see e.g.\cite{Kapusta:finite,Weinberg:qft1})
\begin{equation}
 Z(T,V,\mu_i) = \int \! \mathcal{D}\phi \mathcal{D}\pi \exp\left[ i\int_t^{t-i\beta}\!\dd t' \int \!\dd^3 x \,
	\dot{\phi}(\mathbf{x},t')\pi(\mathbf{x},t') -\mathcal{H}(\phi,\pi) +\mu_i\mathcal{N}_i(\phi,\pi) \right],
\label{eq:Z_as_H_path}
\end{equation}
where $\mathcal{H}$ and $\mathcal{N}$ are the Hamiltonian and number densities, respectively, and $\dot{\phi} \equiv \partial_t \phi$. When $\mathcal{H}-\mu_i\mathcal{N}_i$ is at most quadratic in canonical momenta, the momentum integration can be done. In gauge theory it is useful to first reintroduce the Gauss' law by treating the temporal gauge field component $A_0^a$ as an independent variable, which, when integrated over, would be replaced by the stationary value satisfying Gauss' law.

Performing the momentum integrations, we get back to the Lagrangian formulation
\begin{equation}
 Z(T,V,\mu_i) = \int\!\mathcal{D}\Phi \exp\left[ i\int_t^{t-i\beta}\!\dd t' \int\!\dd^3 x\, \mathcal{L}'(\Phi,\dot{\Phi}) \right],
\label{eq:Z_as_L_path}
\end{equation}
where the integration is now over both canonical and auxiliary fields. The Lagrangian $\mathcal{L}'$ usually differs from the one we started with. In particular, the momenta in Eq.~(\ref{eq:Z_as_H_path}) must be replaced with the values solved from
\begin{equation}
 \dot{\phi}(\mathbf{x},t) = \frac{\delta}{\delta \pi(\mathbf{x},t)}\left( H[\phi,\pi] -\mu_i N_i[\phi,\pi]\right),
\end{equation}
so that in the end we have
\begin{equation}
 \mathcal{L}' = \pi(\phi,\dot{\phi})\dot{\phi} -\mathcal{H}(\phi,\pi(\phi,\dot{\phi}))
	+\mu_i\mathcal{N}_i(\phi,\pi(\phi,\dot{\phi})).
\label{eq:L_prime}
\end{equation}
Moreover, in a gauge theory one usually includes an additional gauge fixing term into the Lagrangian using Grassmannian ghost fields in order to have less constraints on the integration variables.

As can be seen in Eq.~(\ref{eq:Z_as_amplitude}), the partition function is computed as an integral over amplitudes with the same field configuration at both end points, $\phi(t-i\beta,\mathbf{x})= \phi(t,\mathbf{x})$. For fermionic variables it follows from the anticommutation properties of Grassmann variables that the trace has to be computed with antiperiodic condition $\psi(t-i\beta,\mathbf{x})= -\psi(t,\mathbf{x})$ instead. Both boundary conditions can be verified by inspecting the two-point function, taking into account the correct time ordering of the fields \cite{Kapusta:finite}.

When extending the time coordinate to complex values, the integration path is no longer unique. It can be chosen to fit the problem in question, with some minor restrictions. The time arguments of the operators whose thermal averages we are computing should obviously lie on the integration path. Also, the imaginary part of $t$ should be nonincreasing in order to have a well-defined propagator. There are two conventional choices for the path, leading to two different ways of computing at finite temperatures.

First, one can choose to include the whole real axis by first integrating from $-t_0$ to $t_0$, then down to $t_0-i\sigma$, with $0 \leq \sigma \leq \beta$, back to $-t_0-i\sigma$ and finally down to $-t_0-i\beta$,
in the end letting $t_0\to\infty$ (see e.g.~\cite{Lebellac:thermal}). This approach leads to the so-called real-time formalism, which has the advantage that one can directly compute real-time quantities without having to analytically continue the final results to Minkowski space. However, in this formalism the number of degrees of freedom is doubled, with unphysical fields living on the lower horizontal part of the integration path and mixing with the physical ones. This in turn requires the propagators to be extended to $2\times 2$ matrices, leading to complicated perturbation theory. We will not use the real-time formalism in this thesis.

A simpler choice is to integrate down the vertical line $t(\tau)=t_0-i\tau$, $\tau=0..\beta$, which leads to the so-called imaginary time formalism. The choice of $t_0$ does not affect the results, so one can choose $t_0=0$ and replace the time coordinate in Eq.~(\ref{eq:Z_as_L_path}) by $\tau=it$:
\begin{equation}
 Z(T,V,\mu_i) = \int_\mathrm{per.} \!\mathcal{D}\Phi \exp\left[ \int_0^\beta \!\dd \tau \int\!\dd^3 x\, 
	\mathcal{L}_E'(\Phi,\dot{\Phi}) \right].
\end{equation}
The functional integral is over periodic or antiperiodic fields as described above, and the Euclidean Lagrangian $\mathcal{L}_E'$ is the same as in Eq.~(\ref{eq:L_prime}), rotated to Euclidean space with the replacements
\begin{equation}
 \begin{array}{lll} \displaystyle
	t=-i\tau\, , & \gamma_0^{E}=\gamma^0_{M}\, , & A_0^{E} = iA^0_{M}\, ,  \\
	\partial_t = i\partial_\tau\, , & \gamma_i^{E}=-i\gamma^i_{M}\, , & A_i^{E} = A^i_{M}\, , \rule{0pt}{3ex}
 \end{array}
\end{equation}
where `E' and `M' stand for Euclidean and Minkowski space quantities, respectively. In the following we will always work in Euclidean space unless otherwise mentioned, and drop the `E' superscripts. In the above equation, $A_\mu$ represents any four-vector, in particular the gauge field components. There is no doubling of degrees of freedom in this formalism, and for static quantities, such as the free energy or screening masses, it is usually simpler to compute in imaginary time. Other results
have to be analytically continued to real time arguments, and while in principle this can be done with some mild regularity assumptions \cite{Cuniberti:2001hm}, in practice some additional model assumptions are required to carry out the continuation.

Because the fields are required to be periodic, the imaginary time direction can be viewed as a closed circle with circumference $\beta=1/T$. The momentum component in a compact dimension is quantized, so the fields can be decomposed in the momentum space as Fourier series
\begin{eqnarray}
 \phi(\tau,\mathbf{x}) = T\sum_{n=-\infty}^\infty \phi_n(\mathbf{x})e^{i\omega_n\tau}\, , \qquad
\omega_n= \left\{ \begin{array}{ll} 2n\pi T & \mathrm{(bosons)} \\ (2n+1)\pi T \quad & \mathrm{(fermions)} \end{array} \right. ,
\label{eq:matsubara_modes}
\end{eqnarray}
where $\omega_n$ are referred to as Matsubara frequencies \cite{Matsubara:1955ws}. From the gauge transformation rule for the gauge field components
\begin{equation}
 A_\mu \to \Omega A_\mu \Omega^{-1} -\frac{i}{g}(\partial_\mu \Omega)\Omega^{-1}, \qquad \Omega(x) = \exp[ig T^a \alpha^a(x) ]
\label{eq:mittamuunnos}
\end{equation}
it is easy to see that the gauge transformation functions $\alpha^a$ have to be periodic as well, so the ghost fields will have bosonic Matsubara frequencies despite of their anticommuting nature.

\subsection{Renormalization}

The thermal environment changes the boundary conditions and the propagators from their zero-temperature forms. Fortunately, this does not introduce any new ultraviolet divergences, but the usual renormalization procedure remains unchanged and the counterterms have precisely the same values as at $T=0$ (depending on the scheme). Intuitively this is easy to understand, since only the excitations with wavelengths $\gtrsim \beta$ can see the periodicity of the time direction, while the renormalization is only concerned with divergences related to the short distance behavior of Green's functions. The divergence structure is then precisely the same as in the zero-temperature theory and one can choose a $T$-independent renormalization scheme such as the $\MSbar$ scheme.

To see this in some more detail, we note that the free propagator at finite temperature can be viewed as an explicitly periodic combination of zero-temperature Euclidean propagators \cite{Lebellac:thermal},
\begin{equation}
 S_F(\tau,\mathbf{x};T) = \sum_{n=-\infty}^\infty S_F(\tau +n\beta,\mathbf{x};T=0),\qquad 0 \leq \tau < \beta\, .
\end{equation}
The zero-temperature ultraviolet divergences requiring renormalization arise from the short-distance singularities at $x^2=0$. The only term in the above sum where we can have $x^2=(\tau+n\beta)^2+\mathbf{x}^2=0$ is the $n=0$ term, which does not depend on temperature. The divergences of the thermal propagator are therefore correctly removed by the $T=0$ counterterms. At higher order diagrams these divergences are multiplied by $T$-dependent finite parts of the diagram, so the general proof of renormalizability and $T$-independence of counterterms is somewhat more involved, but it follows from a similar decomposition of propagator into a singular $T=0$ part and an analytic $T$-dependent part \cite{Collins:renormalization}.

As the parameters of the theory are renormalized, they also run with the scale according to the renormalization group equations. The actual equations are again the same as in $T=0$ theory, but the choice of renormalization point is complicated by the appearance of new scales $\pi T$ and $\mu$ in addition to the external scales present in the problem, as well as the the scales $gT$, $g^2 T$ generated dynamically by interactions. If these scales are very different, removing the large logarithms by a suitable choice of scale may prove difficult, and a careful analysis of the scale hierarchy is required to construct a good perturbative expansion.

While the ultraviolet divergences are unaffected by the finite temperature, at the infrared end the situation is very different. The finite extent of the temporal direction causes the field components with wavelengths $\gg 1/T$ to see the space effectively as three-dimensional, and this gives rise to many new infrared divergences. These will be treated in more detail in the following section.

\section{Dimensional reduction}
\label{sec:dimensional_reduction}

In this section we will review the rationale for dimensional reduction in the more general context of low-energy effective field theories. We will also discuss the finite-temperature infrared divergences and the resummations needed to get rid of them.

\subsection{Effective Lagrangians in general}
\label{sec:efftheory}

One of the fundamental properties of physics is that phenomena at some specific distance scale can be effectively described by a theory which does not depend on the physics at much shorter scales. This is fortunate, for otherwise we would not even be able to describe the trajectory of a thrown ball without knowledge of beyond the standard model physics. The same behavior, known as decoupling, is also present in quantum field theory, where it is by no means obvious that the heavy particles inevitably occurring as internal legs in Feynman diagrams can be neglected. The proof that the high-energy modes only contribute to long-distance phenomena by renormalization of the parameters and by corrections suppressed by inverse powers of the heavy masses is contained in the celebrated theorem of Appelquist and Carazzone \cite{Appelquist:1974tg}. From this point of view, every physical theory can be viewed as an effective theory, equivalent to the underlying more fundamental theory in some finite energy range.

Formally, if the underlying theory is known, the effective theory for light modes $\phi_l$ can be written as a path integral over the heavy modes $\Phi_h$,
\begin{equation}
  e^{iS_\mathrm{eff}[\phi_l]}  = \int \! \mathcal{D}\Phi_h\, \exp iS[\phi_l,\Phi_h],
\label{eq:gamma_eff_def}
\end{equation}
where the effective action $S_\mathrm{eff}[\phi_l]$ is in general a non-local functional of the light fields. Analytically the path integral can only be computed in the Gaussian approximation around some given field configuration $\bar{\Phi}_h$,
\begin{eqnarray}
 S[\phi_l,\Phi_h] &\approx & S[\phi_l,\bar{\Phi}_h] +\int \!\dd^d x \frac{\delta S}{\delta \Phi_h(x)} \Big|_{\Phi_h=\bar{\Phi}_h}
	\left(\Phi_h(x) -\bar{\Phi}_h(x)\right) \nonumber \\
 &&{}+\half \int \! \dd^d x\, \dd^d y \frac{\delta^2 S}{\delta \Phi_h(x)\delta \Phi_h(y)} \Big|_{\Phi_h=\bar{\Phi}_h} \!
\left(\Phi_h(x) -\bar{\Phi}_h(x)\right)\left(\Phi_h(y) -\bar{\Phi}_h(y)\right). \qquad \quad
\label{eq:gaussian_S}
\end{eqnarray}
Choosing $\bar{\Phi}_h$ to be a saddle point of the action, $\delta S[\phi_l,\Phi_h]/\delta \Phi_h=0$, the integration over $\Phi_h$ gives the effective action (for bosonic $\Phi_h$) as
\begin{equation}
 S_\mathrm{eff}[\phi_l] = S[\phi_l,\bar{\Phi}_h] 
	+\frac{i}{2} \Tr \ln \frac{\delta^2 S}{\delta \Phi_h(x)\delta \Phi_h(y)} \Big|_{\Phi_h=\bar{\Phi}_h}\, ,
\label{eq:eff_act_gen_result}
\end{equation}
where the last term depends on $\phi_l$ both directly through $S$ and through the saddle point condition which makes $\bar{\Phi}_h$ a functional of $\phi_l$. The Gaussian approximation corresponds to the one-loop level in heavy-loop expansion; if we want to go beyond that the path integral can no longer be computed analytically, but we have to resort to perturbation theory or some other approximation.

While the heavy fields can be integrated out as shown above, the resulting effective action is generally a complicated nonlocal functional of the light modes and cannot be cast in the form of an effective local Lagrangian density without some additional approximations. An often used method is the derivative expansion, where the non-local terms are expanded in the light field momenta $p$ over the heavy field mass $M$, leading to
\begin{equation}
 S_\mathrm{eff} = \int \! \dd^d x\, \mathcal{L}_\mathrm{eff} + \sum_n O_n \left(\frac{p}{M}\right)^n,
\label{eq:derivative_expansion}
\end{equation}
where $O_n$ represent operators suppressed by powers of the heavy mass. In terms of Feynman diagrams this means that the effective action is computed with only heavy fields on the internal lines, since the action is made local in the light fields.
The form of Eq.~(\ref{eq:derivative_expansion}) is precisely what should be expected based on the decoupling theorem: parameter renormalizations and heavy mass suppressed operators. There is a twist, however, since the light particle momenta need not be small when the non-local operator is embedded in a multi-loop graph and interacts with heavy fields, and the derivative expansion may then fail. For example, in the large-mass expansion at zero temperature \cite{Collins:renormalization,Lepage:1989hf} it is well known that one needs to take into account also the diagrams with light internal lines in order to get the correct low-energy effective Lagrangian. This will also be the case in the dimensionally reduced effective theory at high temperatures, as we will show later on.

\begin{figure}
\begin{center}
\begin{fmfgraph*}(100,50)
 \fmfbottom{l,r}
 \fmf{dashes}{l,v,r}
 \fmf{plain}{v,v}
 \fmfv{label=(a),l.a=-90}{v}
\end{fmfgraph*}
\begin{fmfgraph*}(100,50)
 \fmfleft{sw,nw}
 \fmfright{se,ne}
 \fmf{dashes}{sw,v1,nw}
 \fmf{dashes}{se,v2,ne}
 \fmf{plain,right,label=(b),l.d=13}{v1,v2}
 \fmf{plain,right}{v2,v1}
\end{fmfgraph*}
\begin{fmfgraph*}(100,50)
 \fmfleft{l}
 \fmfright{r}
 \fmf{dashes}{l,v1,v2,r}
 \fmf{plain,right,tension=0,label=(c),l.d=10}{v1,v2}
 \fmf{plain,right,tension=0}{v2,v1}
\end{fmfgraph*}
\end{center}
\caption{Diagrams in the effective action for a theory with two scalars. Solid lines represent the heavy field, dashed lines the light one.}
\label{fig:scalar_diagrams}
\end{figure}
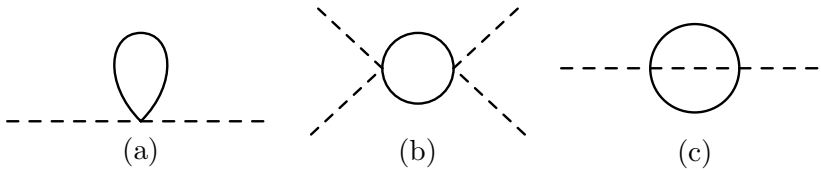

As an illuminating example, consider a theory with two scalar fields \cite{Dobado:effective},
\begin{equation}
 \mathcal{L} = \half \partial_\mu \phi \partial^\mu\phi -\half m^2 \phi^2 -V(\phi)
	+ \half \partial_\mu \Phi \partial^\mu\Phi -\half M^2 \Phi^2 +\half \lambda \phi^2\Phi^2,
\end{equation}
in the limit $m \ll M$. This is similar to the situation at finite temperature where $\phi$ can be thought as the static ($n=0$) Matsubara mode, while the heavy field mass is of the order $2\pi T$. In this model the dependence on the heavy field is quadratic, so we can exactly integrate out $\Phi$, giving
\begin{equation}
 S_\mathrm{eff} = S[\phi] + \frac{i}{2}\Tr \ln( -\partial^2 -M^2 +\lambda \phi^2)
	= S[\phi] - \frac{i}{2}\sum_{k=1}^\infty \frac{\lambda^k}{k} \Tr[ (\partial^2 +M^2)^{-1} \phi^2]^k\, ,
\end{equation}
where in the last step we have dropped a $\phi$-independent term and expanded in the small coupling $\lambda$. The first term in the sum ($k=1$) is represented by Fig.~\ref{fig:scalar_diagrams}(a) and contributes by a local term to the mass renormalization,
\begin{eqnarray}
 -\frac{i\lambda}{2} \Tr (\partial^2 +M^2)^{-1} \phi^2 &=&
	\frac{i\lambda}{2} \int \!\dd^d x\, \phi^2(x) \int_q \frac{1}{q^2 -M^2 +i\epsilon} \nonumber \\
 &=& -\frac{\lambda M^2}{2(4\pi)^2}\left( \frac{1}{\epsilon} +1 -\ln\frac{M^2}{\mu^2}\right)\int \!\dd^d x\, \phi^2(x)  ,
\end{eqnarray}
where we have used dimensional regularization to control the ultraviolet divergence in the momentum integration, with the conventions
\begin{equation}
 \int_q \equiv \left( \frac{e^\gamma \mu^2}{4\pi}\right)^\epsilon \int \! \frac{\dd^d q}{(2\pi)^d}, \qquad d=4-2\epsilon\, .
\label{eq:dimreg_int_def}
\end{equation}
Here $\mu$ is the (arbitrary) dimensional regularization scale, modified to include the constants typical of the $\MSbar$ scheme.

The $\lambda^2$-term, however, already shows where the derivative expansion causes problems. A straightforward computation of the diagram in Fig.~\ref{fig:scalar_diagrams}(b) gives
\begin{eqnarray}
 -\frac{i\lambda^2}{4} \Tr [(\partial^2 +M^2)^{-1} \phi^2]^2 &=& \frac{\lambda^2}{4(4\pi)^2}\int \! \dd^d x\, \dd^d y \int_k
	\phi^2(x) \phi^2(y) e^{-ik\cdot(x-y)} \times \nonumber \\
 &&\times \left( \frac{1}{\epsilon} -\ln\frac{M^2}{\mu^2} -\int_0^1 \!\dd t\, \ln\left[ 1-t(1-t)\frac{k^2}{M^2}\right]\right).
\end{eqnarray}
The first two $k$-independent terms contribute to the renormalization of the 4-point vertex. The remaining logarithm is a non-local operator connecting two $\phi^2$ products at different points. For small $k^2$ the integrand can be expanded in $k^2/M^2$, leading to a series of local four-point derivative couplings of the form $\phi (\partial^2/M^2)^n \phi$. However, when this operator is part of a larger diagram there is no guarantee that $k^2$ is small.

For example, the diagram in Fig.~\ref{fig:scalar_diagrams}(c) with one light and two heavy internal lines is not produced by the effective theory expanded this way. All loop momenta can be large, and therefore the expansion in $k^2/M^2$ is not reliable. Computing this diagram is rather nontrivial \cite{Ramond:1989yd}, but one can show that if the ultraviolet divergences are removed in the $\MSbar$ scheme, the diagram does not vanish in the limit $M \to \infty$. To have an explicit decoupling where all graphs containing heavy internal lines are suppressed one should use a renormalization scheme where the counterterm is the negative of the graph expanded in the light masses and momenta \cite{Collins:renormalization}. At finite temperatures this may be difficult because of the additional infrared divergences. Moreover, we would prefer to use the $\MSbar$ scheme where the counterterms are already known to high order and have a simple structure.

Because of the difficulties in integrating out the heavy fields as described above, at higher orders it is usually safer to construct the effective Lagrangian explicitly by matching the Green's functions. The decoupling theorem states that in a renormalizable theory the parameters in the effective theory can be chosen in such way that the Green's functions of light fields differ from those computed in the full theory by terms suppressed by powers of the heavy mass,
\begin{eqnarray}
 \lefteqn{ G_N(p_1,\ldots,p_N;g,G,m,M,\mu) = \langle 0|T\phi(p_1)\ldots\phi(p_N)|0\rangle_\mathrm{full} } \nonumber \\
 &=& z^{-N/2} G_N^*(p_1,\ldots,p_N;g^*,m^*,\mu)
	\left[ 1 +\mathcal{O}\left( 1/M^a \right) \right] \nonumber \\
 &=& z^{-N/2} \langle 0|T\phi^*(p_1)\ldots\phi^*(p_N)|0\rangle_\mathrm{eff}\left[ 1+\mathcal{O}\left( 1/M^a \right) \right],
\end{eqnarray}
where $M$ and $G$ are the masses and couplings in terms involving heavy fields, while those for terms with only light fields are labeled $m,g$. The corresponding effective theory parameters are $m^*,g^*$ and $\phi^*=z^{1/2}\phi$. We can use this information directly and write down the most general light mode Lagrangian which respects the symmetries of the original theory, and then compute a number of $N$-point functions (usually $N=2,3,4$ is enough) in both theories at some conveniently chosen external momenta to fix the parameters. We will see more detailed examples of this procedure in the following section.

\subsection{Three-dimensional effective theory at high temperature}
\label{subsec:dimred}

Field theories at finite temperature contain many new mass scales in addition to those given by the parameters of the zero-temperature Lagrangian. Besides the temperature itself there are dynamically generated scales related to collective modes and screening phenomena, and the particle masses are modified by thermal effects as well. Renormalizing the theory in the minimal subtraction scheme gives rise to logarithms of the type $\ln(m^2/\mu^2)$, where $m$ can be any of the different scales in the theory. In particular, large scales do not decouple but instead give contributions that grow logarithmically with the scale. This seems to make perturbation theory useless in theories with vastly different mass scales, since we cannot choose a renormalization scale that simultaneously makes all the logarithms small. As a result, terms in the perturbative expansion contain powers of large logarithms in addition to small coupling and need not decrease at higher orders.

To be more specific, in gauge theory the electric and magnetic screening scales are of order $gT$ and $g^2 T$, respectively, and thus there is a clear hierarchy of scales in the small coupling region where we would want to use perturbation theory. The solution is, as discussed above, either to use a more complicated renormalization scheme or to formulate an effective theory and continue using the $\MSbar$ scheme \cite{Weinberg:1980wa}. As it turns out, it is simpler to carry out the computations using the effective theory. We will mostly concentrate on gauge theories in what follows, in particular on QCD and electroweak theory.

In the imaginary time formalism we can write the four-dimensional theory in terms of the Matsubara modes of Eq.~(\ref{eq:matsubara_modes}). For generic bosonic and fermionic fields the free part of the action (without any chemical potentials, although they could easily be included) is
\begin{eqnarray}
 S_0 &=& \int_0^\beta \! \dd\tau \int \! \dd^3 x\,  \phi^\dagger[ -\partial^2 + m_b^2] \phi
	+\bar{\psi}(\slashed{\partial} +m_f)\psi \nonumber \\
 &=& T\!\!\sum_{n=-\infty}^\infty \int\!\dd^3 x\,  \phi_n^\dagger\left( -\partial_i^2 +[(2\pi nT)^2+m_b^2] \right) \phi_n
	+\bar{\psi}_n[ i(2n+1)\pi T \gamma_0 +\gamma_i\partial_i +m_f]\psi_n\, , \nonumber \\[-2ex] 
\label{eq:free_3d_action}
\end{eqnarray}
which can be viewed as a three-dimensional Euclidean theory of an infinite set of fields with masses $M_n^2 = \omega_n^2 + m^2$. If the temperature is much higher than the particle masses, we can use the arguments of the previous section and try to formulate an effective theory for the light modes with $M_n \ll T$, or the bosonic zero-modes since they are the only modes with $\omega_n=0$. This theory loses all dependence on the (imaginary) time coordinate, so we have effectively reduced the number of dimensions to three. From the point of view of modes with wavelengths much larger than $1/T$ the finite temporal direction of length $\beta$ has shrunk to a point.

While the dimensionally reduced theory cannot give any information about the time dependence of the theory, for static Green's functions the effective theory gives correct results up to corrections of order $m^2/(\pi T)^2$, where $m$ is any of the light masses. Note in particular that at high enough temperatures the highest unintegrated mass scales are the dynamically generated scales $\sim gT$, so the corrections to the effective theory are comparable with the higher orders of perturbation theory and both have to be taken into account to get a consistent perturbative expansion. To gain control over which operators to include, power counting rules have to be established for given momentum region. At higher orders it will be necessary to include nonrenormalizable operators into the effective theory, especially if one wishes to have a theory that produces \emph{all} static Green's functions to given order. In many cases, like when computing the free energy, it is sufficient to use only the couplings present already in the original theory, in which case the effective theory is super-renormalizable because of the lower dimensionality.

The main advantage in using an effective theory at high temperatures is in the infrared physics. In general, if the theory contains massless bosonic fields one expects more severe infrared singularities when going to finite temperature, since the Bose--Einstein factor in real-time propagators behaves as
\begin{equation}
 n_B(E) = \frac{1}{e^{\beta E}-1} = \frac{1}{e^{\beta k}-1} \to \frac{1}{\beta k} \quad \mathrm{as} \ k\to 0\, . 
\end{equation}
This can be also understood in the imaginary time formalism, where the zero Matsubara mode behaves like a massless particle in three dimensions, and lower dimensionality generally makes the infrared behavior worse. It is well known that in Yang--Mills theories perturbation theory at finite temperatures suffers from many infrared problems, becoming finally completely non-perturbative at $\order{g^6}$ \cite{Gross:1981br,Linde:1980ts}. These problems are related to massless particles, in particular to the gauge fields, whose screening by medium effects is not correctly reproduced by the na\"\i ve perturbation theory. By definition, the dimensionally reduced theory has the same infrared limit as the original theory, while being computationally simpler. The leading order contribution coming from scales of order $T$ can be included in the parameters of the effective theory via the matching procedure, which is infrared safe, and the infrared peculiarities can then be studied in a simpler setting. In particular, the dimensionally reduced effective theory does not contain any fermionic fields, which makes it easier to study non-perturbatively using lattice simulations.

The electric screening effects can be included by reorganizing the perturbative expansion. Computing the one-loop self-energy of a  zero-mode gauge field component $A_\mu$, we find that in the limit of vanishing momentum it behaves as
\begin{equation}
 \Pi_{\mu \nu}(\omega_n=0,\mathbf{k} \to 0) \propto g^2 T^2 \delta_{\mu 0}\delta_{\nu 0}\, .
\end{equation}
The temporal component develops a thermal mass of order $gT$, while the other components remain massless. In the soft limit where $k \lesssim gT$ it is not consistent to treat this self-energy as perturbation, but it should be included in the propagator instead. This means that we should sum all diagrams with an arbitrary number of self-energy insertions on the temporal gluon line to get consistent $\mathcal{O}(g^2)$ results, which is often referred to as resummation. In four dimensions one has to be careful not count any diagram twice because of this summation; usually this is done by adding and subtracting a term containing the self-energy in the Lagrangian,
\begin{equation}
 \mathcal{L} = \mathcal{L}_0 + \mathcal{L}_I = (\mathcal{L}_0 + \delta\mathcal{L}) + (\mathcal{L}_I -\delta \mathcal{L})
\end{equation}
and treating the subtracted term as an interaction. In the dimensionally reduced theory the resummation is simpler, since the thermal mass for $A_0$ comes out naturally from the matching procedure. Moreover, there is no risk of double counting diagrams, since the thermal mass is only created by $n \neq 0$ and fermionic modes (the mass can be computed in the $\mathbf{k}=0$ limit, and the dimensionless graphs vanish in dimensional regularization), which are not present in the effective theory. Note that the electric mass does not break the remaining gauge invariance, since when restricting to bosonic zero modes only we are also forced to only consider $\tau$-independent gauge transformations. The transformation rule in Eq.~(\ref{eq:mittamuunnos}) then boils down to
\begin{equation}
 A_0^a = 2\,\Tr A_0 T^a \to 2\, \Tr \Omega A_0 \Omega^{-1}T^a = \exp[ig\alpha^c\tau_{ab}^c] A_0^b\, ,
\end{equation}
so in the three-dimensional theory $A_0$ becomes a massive scalar transforming in the adjoint representation of the gauge group. The remaining gauge invariance in three dimensions prevents the spatial gauge field components from developing a mass term.

In the magnetic sector there are infinitely many diagrams that all contribute at order $g^6$, and, unlike for the electric mass, they appear with so different and complex topologies that they cannot be resummed in a simple way to tame the infrared singularities. In fact, there is no gauge-invariant magnetic mass term that could be included in the Lagrangian for perturbatively computing beyond $\mathcal{O}(g^6)$, but instead the magnetic screening has to be treated non-perturbatively. In the very low momentum region the fields with thermal masses $\sim gT$ can be integrated out as well, leaving a three-dimensional pure gauge theory with coupling $\tilde{g}_3^2 = g^2 T$, which is the only dimensionful parameter in the Lagrangian. In this theory there is no small dimensionless parameter to do perturbation theory with, but the infrared dynamics of nonabelian gauge theory is inherently nonperturbative.

To see how the matching of parameters in the dimensionally reduced theory goes in practice, we will take a closer look at the mass parameters, following to some extent \cite{Kajantie:1995dw,Braaten:1996jr}. The masses can be found by comparing the static two-point functions computed in both theories. For simplicity, we will use a scalar particle with a small zero-temperature mass $m \lesssim gT$ as an example and work to order $g^4$, which is sufficient for many computations, in particular for determining the free energy to order $g^5$ as in \cite{Gynther:2005dj,Gynther:2005av}.

In the full theory the inverse propagator can be written as
\begin{equation}
  k^2 + m^2 + \Pi(k^2) = k^2 + m^2 + \overline{\Pi}(k^2) +\Pi_0(k^2),
\label{eq:m_match_full}
\end{equation}
where $\overline{\Pi}(k^2)$ includes the diagrams with at least one heavy internal line, while $\Pi_0(k^2)$ is the contribution of $n=0$ modes only. In the effective theory the same function reads
\begin{equation}
	k^2 + m_3^2 +\Pi_3(k^2).
\label{eq:m_match_eff}
\end{equation}

The contribution coming from the non-static modes, $\overline{\Pi}(k^2)$, is of order $g^2T^2$, and the matching has to carried out in the region where the effective theory is valid, $k \lesssim gT$. Since integration over massive modes is infrared safe, the renormalized self-energy $\overline{\Pi}(k^2)$ has no infrared divergence and can be expanded in $k^2/T^2$,
\begin{equation}
	\overline{\Pi}(k^2)=\overline{\Pi}(0)+k^2\frac{\dd}{\dd k^2}\overline{\Pi}(0)+\mathcal{O}\left(g^2\frac{k^4}{T^2}\right),
\end{equation}
where the terms left out are of order $g^6 T^2$. Further expanding each term in loop expansion with coupling $g$,
\begin{equation}
 \overline{\Pi}(k) = \sum_{n=1}^\infty \overline{\Pi}^{(n)}(k),\qquad 
	\mathrm{where}\quad \overline{\Pi}^{(n)}(k) \sim \order{g^{2n}},
\end{equation}
the inverse propagator in Eq.~(\ref{eq:m_match_full}) reads, including terms up to $\order{g^4}$,
\begin{equation}
 k^2\left(1+\frac{\dd}{\dd k^2}\overline{\Pi}^{(1)}(0)\right) +m^2+\overline{\Pi}^{(1)}(0)+\overline{\Pi}^{(2)}(0) +\Pi_0(k^2).
\label{eq:prop_full_expand}
\end{equation}

The massive modes correspond to poles in the propagator, or the zeros of the inverse propagator, so we set the expressions in Eqs.~(\ref{eq:m_match_eff}),(\ref{eq:prop_full_expand}) equal to zero and solve for $k^2$. Equating the pole locations in both theories, we find the matching condition
\begin{equation}
	m_3^2+\Pi_3(k^2) = \left(1-\frac{\dd}{\dd k^2}\overline{\Pi}^{(1)}(0)\right)
	\left[ m^2+\overline{\Pi}^{(1)}(0)+\overline{\Pi}^{(2)}(0) +\Pi_0(k^2) \right].
\end{equation}
By construction, the infrared behavior contained in the soft self-energies $\Pi_0$ and $\Pi_3$ is the same in both theories, so this relation is infrared safe. The difference is of order $g^5$,
\begin{equation}
 \Pi_3(k^2) = \Pi_0(k^2)\left[ 1+ \order{k^2/T^2}\right], \qquad \Pi_3(k^2) \sim g_3^2 m_3 \approx g^3 T^2\, ,
\end{equation}
so, working at order $g^4$, we can drop all terms containing $\Pi_0,\Pi_3$ from the matching condition. We are then left with an equation for the three-dimensional mass parameter
\begin{equation}
	m_3^2 = m^2 +\overline{\Pi}^{(1)}(0)+\overline{\Pi}^{(2)}(0) -\left(m^2 +\overline{\Pi}^{(1)}(0)\right) 
	\frac{\dd}{\dd k^2}\overline{\Pi}^{(1)}(0).
\label{eq:m_matching_result}
\end{equation}
As a by-product we also found the field normalization factor to order $g^2$, since from looking at the coefficients of $k^2$ in both propagators we can write
\begin{equation}
 \phi_{3d}^2 = \frac{1}{T}\left[ 1+\frac{\dd}{\dd k^2}\overline{\Pi}^{(1)}(0) \right]\phi_{4d}^2\, .
\end{equation}
The factor $1/T$ here stems from the overall factor $T$ in Eq.~(\ref{eq:free_3d_action}), which is conventionally absorbed into the fields and couplings of the 3d theory.

It should be noted that Eq.~(\ref{eq:m_matching_result}) only contains contributions from the heavy scale $T$, whereas the infrared sensitive parts $\Pi_0$ and $\Pi_3$ drop out. The mass parameter $m_3$ regulates the infrared behavior of the dimensionally reduced theory, but it is a completely perturbative quantity and should not be confused with the actual screening lengths that are sensitive to infrared physics. In particular, the thermal mass of the adjoint scalar $A_0$ in the dimensionally reduced theory agrees with the electric screening mass $m_\mathrm{el}$ only at order $g^2$, beyond which $m_\mathrm{el}$ becomes sensitive to the magnetic screening \cite{Arnold:1995bh}, while $m_3$ on its part is given to $\order{g^4}$ by the completely perturbative expression in Eq.~(\ref{eq:m_matching_result}).

Apart from the gauge fields, the only other elementary boson in the standard model is the Higgs field, which has a negative mass parameter $-\nu^2$ in the phase of unbroken $SU(2)\times U(1)$ symmetry. Near the electroweak phase transition the Higgs field mass is a special case in the power counting, since the $T=0$ mass parameter and the thermal corrections almost cancel each other, giving
\[
	m_3^2 \sim -\nu^2 + g^2 T^2 \sim g^3 T^2
\]
or smaller, depending on how close to the phase transition we choose to work. To have a better separation of scales, it is necessary to integrate out the fields with masses $\sim gT$ when computing close to the electroweak phase transition, as we did in \cite{Gynther:2005av}. This leads to a theory containing only the Higgs field and spatial gluons. The thermal mass $m_3^2(T)$ is the leading term in the Higgs field effective potential, which drives the phase transition.

The above matching computation gives another example of how the expansion in loops and momenta can be identified when the the correct momentum region is known. At high temperatures, the mass parameters can be estimated as $gT$ and the momenta at most of the same magnitude, in the region where dimensional reduction is valid. The required level of matching is determined by the problem in question and the accuracy goal one wants to reach. For example, for computing the free energy to order $g^5$ we needed the couplings only at tree-level, but the mass parameters to two-loop ($g^4$) order.

A more general analysis given in \cite{Kajantie:1995dw} states that in order to have a theory which gives the same light field Green's functions as the full theory up to corrections of order $\order{g^4}$, we need to match the parameters at least to this order. To be more precise, the coupling constants are required to one-loop level $g_3^2 = T(g^2+g^4)$ and adjoint scalar (temporal gauge field component) masses to two-loop accuracy $\mE^2 = T^2(g^2+g^4)$. If the theory contains a light scalar field such as the Higgs field, its thermal mass should be computed to three-loop level $m_3^2=-\nu^2+T^2(g^2+g^4+g^6)$, since the first terms cancel each other, and the mass is of order $g^4 T^2$ close to the phase transition. The same analysis shows that beyond $\order{g^4}$ it is necessary to include non-renormalizable 6-dimensional operators into the effective theory.

Apart from the simple power counting, the importance of the higher order operators inevitably resulting from the reduction step is difficult to estimate. In \cite{Jakovac:1994xg,Kajantie:1995dw} it is noted that in both abelian and SU(2) Higgs models these operators are further suppressed by small numerical coefficients in addition to powers of the coupling constant, and thus give only very small contributions. The operators following from the second reduction step, where the scales $\sim gT$ are integrated out to give a pure gauge theory, can be consistently treated as perturbations with respect to the tree-level Lagrangian, as discussed in \cite{Braaten:1994na}.

For matching purposes we still need to compute some Green's functions in the full theory, but using the effective theory this only has to be done once, after which the computations can be carried out in the simpler effective theory. For both QCD \cite{Arnold:1994ps,*Arnold:1995eb,Braaten:1996jr} and electroweak theory \cite{Kajantie:1995dw} the matching has been carried out explicitly to order $g^4$, and for a generic theory containing scalars, fermions and gauge fields the rules given in \cite{Kajantie:1995dw} can be used to find the parameters of the effective theory. The QCD coupling has even been matched to two-loop [$g_3^2 = T(g^2+g^4+g^6)$] level in \cite{Laine:2005ai}.

While the effective theory approach saves us from computing multiple complicated sum-integrals, at finite temperatures the main advantages of dimensional reduction lie in the easy way to organize the resummations and separating the contributions of different scales. Eventually non-perturbative methods such as lattice simulations are needed to handle the infrared limit, but the dimensional reduction methods allow us to work out the parameters with completely perturbative methods, and then apply the computationally intensive methods to the simpler three-dimensional theory. Lattice simulations in the dimensionally reduced theory are easier because there is one spatial dimension less, no fermions and the shortest scales $\lesssim 1/T$ have been integrated out.

\chapter{Pressure of the standard model}
\label{chap:pressure}

At high temperatures the local $SU(2)_L \times U(1)_Y$ gauge symmetry of electroweak theory is restored. The phase transition is driven by the Higgs field, whose effective potential is modified by thermal corrections in such way that the vacuum expectation value of the field vanishes when the temperature is raised. Because of the possibility of the phase transition being strongly first order and contributing to the baryon number asymmetry, the effective potential has been extensively studied both by 1-loop \cite{Anderson:1991zb,Carrington:1991hz,Dine:1992wr} and 2-loop \cite{Arnold:1992rz,Farakos:1994kx,Fodor:1994bs} perturbative calculations and by dimensional reduction \cite{Kajantie:1995dw} combined with lattice simulations \cite{Kajantie:1995kf,Kajantie:1996mn,Kajantie:1996qd,Karsch:1996yh,Gurtler:1997hr}. In those works it was shown that in the standard model the electroweak phase transition is a crossover for realistic Higgs masses.

Apart from the effective potential computations, the thermodynamics of electroweak theory has not been studied in detail. In \cite{Gynther:2005dj,Gynther:2005av} we computed the most fundamental thermodynamic quantity, the free energy, for electroweak theory at high temperatures. This computation is very similar to the evaluation of the free energy in QCD, with the main differences coming from the presence of a light scalar field driving the phase transition and the multitude of scales and couplings leading to a very complicated general structure. Together with the QCD result and the few terms mixing the strong and electroweak couplings, this computation gives us the free energy of the full standard model.

Partial derivatives of the free energy give the basic thermodynamical quantities as in Eq.~(\ref{eq:thermo_quantities}). It should be noted here that we are computing in the grand canonical ensemble, whose partition function gives the grand potential $\Omega = -T \ln Z$, but at zero chemical potentials this can be identified with the free energy $F = \Omega + \mu_i N_i$. In the thermodynamical limit $V\to\infty$ the free energy density equals the pressure, $F = -pV$, so for simplicity we will we talking about pressure from now on.

The energy density and pressure are particularly interesting, since they control the expansion of the universe at its very early stages. Temperatures higher than the electroweak crossover cannot be reached experimentally, but they were present in the early universe. The relic densities of particles decoupling from the ordinary matter are sensitive to the evolution of the universe, which in turn is governed by the equation of state. Recent measurements of the cosmic microwave background suggest a sizeable amount of cold dark matter, which could be explained by weakly interacting massive particles (WIMPs) (see \cite{Bertone:2004pz} for a review). Given a theory describing WIMPs, we need to know the evolution of the universe at the time of their decoupling as well as at later times to make predictions of the present situation. In \cite{Hindmarsh:2005ix} it is estimated that a 10\% change in the equation of state leads to 1\% difference in relic densities, which is visible in future microwave observations.

\section{Perturbative evaluation of the pressure}
\label{sec:sm_pressure}

The electroweak sector of the standard model is given by the Euclidean Lagrangian
\begin{eqnarray}
 \mathcal{L} &=& \frac{1}{4}G_{\mu\nu}^a G_{\mu\nu}^a + \frac{1}{4}F_{\mu\nu} F_{\mu\nu} +D_\mu\Phi^\dagger D_\mu\Phi 
	-\nu^2\Phi^\dagger\Phi + \lambda(\Phi^\dagger\Phi)^2 +\bar{l}_L \slashed{D}l_L +\bar{e}_R\slashed{D}e_R \nonumber \\
 &&{} +\bar{q}_L\slashed{D}q_L +\bar{u}_R\slashed{D}u_R +\bar{d}_R\slashed{D}d_R + ig_Y\left(\bar{q}_L\tau^2\Phi^* t_R 
	-\bar{t}_R(\Phi^*)^\dagger\tau^2q_L\right),
\label{eq:sm_lagrangian}
\end{eqnarray}
where $G_{\mu\nu}^a = \partial_\mu A_\nu^a -\partial_\nu A_\mu^a +g\epsilon^{abc}A_\mu^b A_\nu^c$ and $F_{\mu\nu} =\partial_\mu B_\nu -\partial_\nu B_\mu$ are the field strengths of the weak and hypercharge interactions, $\Phi$ is the Higgs field and the covariant derivatives act on the chiral fermion fields and the Higgs field as usual (for details, see Eq.~(2.3) of \cite{Gynther:2005dj}). We only include the Yukawa coupling for the top quark, since for other particles the Yukawa couplings (which are proportional to particle masses in the broken symmetry phase) are orders of magnitude smaller.

When the Euclidean action is given, the pressure can be computed as the logarithm of the partition function,
\begin{equation}
 p(T) = \lim_{V\to\infty} \frac{T}{V} \ln \int \!\mathcal{D}A\mathcal{D}\psi\mathcal{D}\bar{\psi}\mathcal{D}\Phi 
	\exp\left[ -\int_0^\beta \!\dd \tau \int \! \dd^3 x\, \mathcal{L}(A,\bar{\psi},\psi,\Phi)\right],
\label{eq:pressure_integral}
\end{equation}
where the path integral is over all fields in the Lagrangian. As described in the previous chapter, a straightforward perturbative evaluation of the path integral Eq.~(\ref{eq:pressure_integral}) fails because of infrared divergences. The solution is to resum a class of diagrams by means of an effective theory, using dimensional reduction.

In the first level of dimensional reduction all non-static modes, in particular all fermions, are integrated out. This leads to an effective theory $S_\mathrm{E}$, whose parameters are matched by perturbative computations in the full theory with no resummations,
\begin{equation}
 p(T) \equiv \pE(T) +\lim_{V\to\infty} \frac{T}{V}\ln \int \!\mathcal{D}A_k \mathcal{D}A_0 \mathcal{D}\Phi
	\exp\left(-S_\mathrm{E}\right).
\label{eq:sheavyred}
\end{equation}
Note in particular the appearance of parameter $\pE(T)$, which is the contribution of the non-static modes, or scales $\sim \pi T$, to the pressure. This parameter can be also viewed as the matching coefficient of the unit operator by looking at the (unnormalized) expectation value of unit operator in both theories,
\begin{equation}
 \langle 1 \rangle_\mathrm{full} = \Tr 1\cdot\rho_\mathrm{full} = Z_\mathrm{full} = e^{-F/T}
	= e^{-F_\mathrm{E}/T +\ln Z_\mathrm{E}} = e^{-F_\mathrm{E}/T} \langle 1 \rangle_\mathrm{E}\, ,
\end{equation}
where $F=-pV$. Since the matching is infrared safe, all parameters of $S_\mathrm{E}$ and also $\pE$ are series in $g^2$. In addition to curing some of the infrared problems, this approach makes full use of the scale hierarchy $T\gg gT \gg g^2 T$ by separating the contribution from each scale into successive effective theories, whose contributions enter at different levels of perturbation theory. For example, it is easy to see that the dimensionally reduced theory $S_\mathrm{E}$ in Eq.~(\ref{eq:sheavyred}) starts to contribute at level $T\mE^3 \sim g^3 T^4$.

The effective theory $S_\mathrm{E}$ still contains two different scales $gT$ and $g^2 T$, the latter of which is related to non-perturbative magnetic screening effects. If one wishes to go further using perturbation theory, it is useful to integrate out the electric scales $gT$ as well, giving
\begin{equation}
 p(T) \equiv \pE(T) +\pM(T) +\frac{T}{V}\ln \int \!\mathcal{D}A_k \exp\left(-S_\mathrm{M}\right),
\label{eq:heavyred}
\end{equation}
where $S_\mathrm{M}$ only contains the spatial gauge fields. Close to the phase transition this step is more complicated because the scalar mass is very light, and deserves a separate discussion in section \ref{sec:pt_integration}. The only dimensional parameter in $S_\mathrm{M}$ is the gauge coupling $\tilde{g}_3^2 \approx g^2 T$, so this theory begins to contribute at order $T\tilde{g}_3^6 \sim g^6 T^4$, this term being completely non-perturbative. We have only kept terms of order $g^5$ in our calculation of the pressure, so this non-perturbative contribution can be dropped. The purpose of this second reduction step is that the computation of $\pM$ in the first effective theory $S_\mathrm{E}$ can be considered as a matching computation, without having to worry about the resummations needed for spatial gauge fields.

The electroweak theory contains many dimensionless coupling constants, and we need to establish a power counting between them in order to determine which terms to include in the perturbative expansion. We have decided to use the weak gauge coupling as reference, and, denoting the strong and hypercharge couplings by $g_s$ and $g'$, respectively, make the simple choice
\begin{equation}
 \lambda \sim g'^2 \sim g_s^2 \sim g_Y^2 \sim g^2, \qquad \nu^2 \lesssim g^2 T^2\, ,
\end{equation}
which corresponds to three-loop expansion in all couplings. Numerically this is not the best choice, since the strong and Yukawa couplings are large compared to electroweak couplings, and we underestimate their importance. However, trying to incorporate higher orders of $g_Y$ would require four-loop sum-integrals, which we do not know how to perform. The strong coupling is even harder, since the $g_s^6$ order suffers from the same infrared problems as any nonabelian gauge theory. In \cite{Kajantie:1995dw} the rule $g'^2 \sim g^3$ is used, but there is no danger and practically no extra work in keeping terms of order $g'^5$ as well. It should be also kept in mind that the one-loop renormalization group running of the couplings is such that $g'(T)$ grows with temperature, whereas $g(T)$ decreases.

After all the preparations are done, it remains to actually compute the pressure. We start by evaluating $\pE$ to three-loop order. At the 4d full theory level the Higgs mass parameter $\nu^2$ is treated as a perturbation, so we expand the propagators in $\nu^2$. This is possible since the matching procedure is infrared safe. The resulting massless sum-integrals can be evaluated using the methods developed in \cite{Arnold:1994ps,*Arnold:1995eb} and can be conveniently read from the Appendix A of \cite{Braaten:1996jr}. The largest work lies in writing down all the required diagrams with correct symmetry and group theory factors, reducing them to integrals given in \cite{Braaten:1996jr} and summing everything together. Note that the diagrams with only static modes do not have to explicitly subtracted, since they vanish in the dimensional regularization due to lack of dimensionful parameter.

Schematically, the generic form of $\pE$ is
\begin{eqnarray}
 \pE(T) &=& T^4\left[\alpha_{E1} +\sum_i g_i^2\alpha_{Ei} +\frac{1}{(4\pi)^2}\sum_{ij} g_i^2 g_j^2\alpha_{Eij} \right] \nonumber \\
 &&{}+\nu^2 T^2 \left[ \alpha_{E\nu} + \frac{1}{(4\pi)^2} \sum_i g_i^2 \alpha_{Ei\nu} \right]
	+\frac{\nu^4}{(4\pi)^2}\alpha_{E\nu\nu} + T^4\cdot{\cal O}(g^6),
\label{eq:4dgenpres}
\end{eqnarray}
where the summation is over $g_i^2=g^2,g'^2,g_s^2,g_Y^2,\lambda$ and the values of all nonzero coefficients (all combinations except $\alpha_{E\lambda s},\alpha_{Es\nu}$, and $\alpha_{Es},\alpha_{Ess}$ since we exclude the pure QCD terms) can be found in the Appendix A of \cite{Gynther:2005dj}. To keep track of different contributions they are given in terms of the group theory constants, which for SU(2) read $\TF=1/2$, $\CF=3/4$, $\dF=2$, $\CA=2$ and $\dA=3$.

We have normalized the pressure so that (the real part of) the pressure at the symmetric phase vanishes at zero temperature, $p(T=0)=0$, in order to exclude the large vacuum energy contribution and the related divergences. This normalization is already taken into account in Eq.~(\ref{eq:4dgenpres}), where we have subtracted a term proportional to $\nu^4$ computed at zero temperature. The $T=0$ computation differs from the high-temperature expansion in Eq.~(\ref{eq:4dgenpres}), and the difference is contained in the remaining coefficient $\alpha_{E\nu \nu}$.

The renormalization of the 4d parameters does not remove all divergences, but $1/\epsilon$ terms can be found in most of the coefficients of $\order{g^4}$ terms, corresponding to infrared divergences that cancel against similar terms in $\pM$. Having stated above that the matching computation is infrared safe, we should elaborate on the nature of these divergences and their cancellation a bit more. Dimensional regularization simultaneously handles both the infrared and ultraviolet limit, and it is not easy to tell the divergences apart.

The electroweak theory is known to be renormalizable. The computation of $\pE$ cannot thus contain any ultraviolet divergences, since they are removed by the counterterms. However, there are diagrams that are both ultraviolet and infrared divergent and vanish in dimensional regularization. We can see how they behave through the following simple example. Consider the logarithmically divergent integral
\begin{equation}
 \int_k \frac{1}{k^4} \equiv \left(\frac{e^\gamma \mu^2}{4\pi}\right)^\epsilon \int \!\frac{\dd^{4-2\epsilon} k}{(2\pi)^{4-2\epsilon}} \frac{1}{k^4},
\end{equation}
which vanishes in dimensional regularization. This can be written as a sum of two integrals, one divergent at the ultraviolet and the other at the infrared momenta,
\begin{eqnarray}
 \lefteqn{ 
 \int_k \frac{1}{k^2(k^2+m^2)} +\frac{m^2}{k^4(k^2+m^2)}
 = \int_k \int_0^1 \dd x \frac{1}{[k^2+xm^2]^2} +\frac{2m^2(1-x)}{[k^2+xm^2]^3} } \qquad && \nonumber \\
 &=& \frac{1}{16\pi^2}\left(\frac{e^\gamma \mu^2}{m^2}\right)^\epsilon \int_0^1\dd x\, \Gamma(\epsilon) x^{-\epsilon}
	+\Gamma(1+\epsilon)(1-x)x^{-1-\epsilon} \nonumber \\
 &=& \frac{1}{16\pi^2}\left( \frac{1}{\epsilon_\mathrm{UV}} - \frac{1}{\epsilon_\mathrm{IR}} \right).
\end{eqnarray}
The renormalization counterterms remove the ultraviolet divergence $1/\epsilon_\mathrm{UV}$ here, leaving the infrared divergent part $-1/\epsilon_\mathrm{IR}$. The vanishing diagram therefore contributes with an infrared divergence when renormalized.

The scaleless diagrams at finite temperature are precisely those with only static modes, and no summations over the Matsubara frequencies. The dimensionally reduced effective theory contains the same diagrams, but with self-energy corrections resummed to give masses on some propagators. These masses do not change the ultraviolet behavior of the diagram, but regularize the infrared limit, so the divergence structure is just $1/\epsilon_\mathrm{UV}$, which precisely cancels against $-1/\epsilon_\mathrm{IR}$ from the full theory computation.

The contribution of the scales $gT$ can be calculated from the path integral in Eq.~(\ref{eq:sheavyred}) once the dimensionally reduced theory is known. Before matching, we need to consider the most general renormalizable (in 3d) Lagrangian respecting the symmetries of the full theory,
\begin{eqnarray}
S_\mathrm{E} &=& \int\!\dd^3x\, \frac{1}{4}G_{ij}^a G_{ij}^a +\frac{1}{4}F_{ij}F_{ij} +(D_i\Phi)^\dagger(D_i\Phi)
 	+m_3^2\Phi^\dagger\Phi +\lambda_3(\Phi^\dagger\Phi)^2  \nonumber \\
 && {}+\half(D_i A_0^a)^2 +\half\mD^2 A_0^a A_0^a +\frac{1}{4}\lambda_A (A_0^a A_0^a)^2 +\half(\partial_i B_0)^2
	+\half\mD'^2 B_0 B_0 +\frac{1}{4}\lambda_B B_0^4 \nonumber \\
 && {}+h_3\Phi^\dagger\Phi A_0^a A_0^a +h_3'\Phi^\dagger\Phi B_0 B_0
	-\half g_3 g_3' B_0\Phi^\dagger A_0^a \tau^a\Phi \, ,
\label{eq:3daction}
\end{eqnarray}
where we have included masses and quartic self-interactions for the scalar fields $A_0,B_0$, the former temporal components of the gauge field. The field $A_0$ transforms in the adjoint representation of SU(2), wheres $B_0$ does not interact with the gauge fields due to the abelian nature of U(1).

In order to relate the pressure $\pM$ to the full theory we need to know the parameters of the effective theory $S_\mathrm{E}$ in terms of the full theory couplings. For the couplings the leading order results are sufficient, since the two-loop diagrams in the effective theory are already of order $g^4$. The matching then boils down to absorbing the factor of $T^{1/2}$ to couplings to give them the correct dimensions,
\begin{equation}
\begin{array}{rclrcl}
\displaystyle
	g_3^2 &=& g^2 T\,, & g_3'^2 &=& g'^2 T\,, \\
	\lambda_3 &=& \lambda T\,, & \lambda_{A,B} &=& \mathcal{O}(g^4), \\
	h_3 &=& \frac{1}{4}g^2 T\,, & h_3' &=& \frac{1}{4}g'^2 T\, .
\end{array}
\label{eq:couplingmatch}
\end{equation}
The quartic couplings for the adjoint scalars are of higher order than we need in our computation.

The matching of the mass parameters is more complicated. Since the leading order (one-loop) diagrams are of the order $Tm^3 \sim g^3 T^4$, we need order $g^4$ terms in the expressions for the masses to get the pressure up to $\order{g^5}$. Moreover, the two-loop diagrams contain ultraviolet divergences, so we need also the $\order{\epsilon}$ terms for the masses when using dimensional regularization. The mass parameters in electroweak theory have already been computed in \cite{Kajantie:1995dw} apart from the $g^2\epsilon$ terms, which we have evaluated in \cite{Gynther:2005dj}. The masses are found by matching the two-point functions at vanishing external momentum as in Eq.~(\ref{eq:m_matching_result}).

The general form of the adjoint scalar masses is
\begin{eqnarray}
 \mD^2 &=& T^2 \left[ g^2 \left(\beta_{E1} +\beta_{E2}\epsilon +\mathcal{O}(\epsilon^2)\right) 
	+\frac{g^4}{(4\pi)^2}\left(\beta_{E3}+\mathcal{O}(\epsilon)\right) +\mathcal{O}(g^6) \right. \nonumber \\
 && \left. +\frac{g^2}{(4\pi)^2}\left(\beta_{E\lambda}\lambda +\beta_{Es}g_s^2 +\beta_{EY}g_Y^2
	+\beta_{E'}g'^2 +\beta_{E\nu}\frac{-\nu^2}{T^2} \right) \right],
\label{eq:matchmd}
\end{eqnarray}
and similarly for $\mD'$. The coefficients $\beta_{Ex}$ can be found in the Appendix B.1 of \cite{Gynther:2005dj}. It should be noted that there are no divergences in these coefficients, but the renormalization of the 4d theory is enough to make the adjoint scalar masses finite. In the 3d theory these parameters are renormalization group invariant to this order, and only start running at order $g^6$ with terms proportional to $\lambda_A^2$ and $g_3^2\lambda_A$ \cite{Farakos:1994kx}.

The fundamental scalar mass has $1/\epsilon$ divergences that are not removed by the renormalization of the full theory. These are again related to the infrared limit of the static modes and are removed by the counterterms in the effective theory. Looking from the dimensionally reduced theory, the matching procedure produces the bare mass which we can either split into the renormalized mass and counterterms, or continue using the mass parameter with $1/\epsilon$ terms included as we did in \cite{Gynther:2005dj}. Either way, the divergences will cancel in the final result for the pressure, and $m_3$ itself is not a physical parameter we would be interested to study in detail.

Using the $\MSbar$ scheme in the effective theory to renormalize the Higgs mass, we get the finite result
\begin{eqnarray}
\lefteqn{ m_3^2(\Lambda) = -\nu^2 +T^2\left( \frac{1}{4}\CF g^2 +\frac{1}{16}g'^2 +\frac{1}{6}(\dF+1) \lambda +\frac{1}{12}N_c g_Y^2
	\right) } \nonumber \\
&&{}+\epsilon\, T^2\left( g^2 \beta_{A2} +g'^2\beta_{B2} +\lambda\beta_{\lambda 2} +g_Y^2\beta_{Y2} \right)
	+\frac{-\nu^2}{(4\pi)^2} \left( g^2\beta_{\nu A} +g'^2\beta_{\nu B} +\lambda\beta_{\nu\lambda} + g_Y^2\beta_{\nu Y} \right)
	\nonumber \\
&&{}+T^2\left[ \frac{g^4}{(4\pi)^2}\beta_{AA} +\frac{g'^4}{(4\pi)^2}\beta_{BB} +\frac{g^2 g'^2}{(4\pi)^2}\beta_{AB}
	+\frac{\lambda g^2}{(4\pi)^2}\beta_{A\lambda} +\frac{\lambda g'^2}{(4\pi)^2}\beta_{B\lambda}
	+\frac{\lambda^2}{(4\pi)^2}\beta_{\lambda\lambda} \right. \nonumber \\
	&&\left. {}+\frac{g^2 g_Y^2}{(4\pi)^2}\beta_{AY} +\frac{g'^2 g_Y^2}{(4\pi)^2}\beta_{BY} 
	+\frac{g_s^2 g_Y^2}{(4\pi)^2}\beta_{sY} +\frac{\lambda g_Y^2}{(4\pi)^2}\beta_{\lambda Y}
	+\frac{g_Y^4}{(4\pi)^2}\beta_{YY} \right],
\label{eq:m3matched}
\end{eqnarray}
which depends on the $\MSbar$ renormalization scale $\Lambda$ replacing $\mu$ in Eq.~(\ref{eq:dimreg_int_def}). The parameters are linear combinations of $\zeta'(-1)$, $\gamma_E$ and $\ln(\Lambda/4\pi T)$, and they are given explicitly in the Appendix B.2 of \cite{Gynther:2005dj}. The mass counterterm can be read from the matching computations,
\begin{equation}
	\delta m_3^2 = \frac{T^2}{(4\pi)^2\epsilon}\left( -\frac{81}{64}g^4 +\frac{7}{64}g'^4 +\frac{15}{32}g^2 g'^2
	-\frac{9}{4}\lambda g^2 -\frac{3}{4}\lambda g'^2 +3\lambda^2 \right).
\label{eq:m3divergence}
\end{equation}
This coincides with the counterterm computed directly from the effective theory $S_\mathrm{E}$,
\begin{equation}
	\frac{1}{(4\pi)^2\epsilon}\left( -\frac{39}{64}g_3^4 +\frac{5}{64}g_3'^2 +\frac{15}{32}g_3^2 g_3'^2
	-\frac{9}{4}\lambda_3 g_3^2 -\frac{3}{4}\lambda_3 g_3'^2 +3\lambda_3^2 +\frac{3}{2}h_3^2 -3h_3 g_3^2 +2h_3'^2 \right),
\label{eq:m3divergence3}
\end{equation}
when the relations (\ref{eq:couplingmatch}) between couplings are taken into account. Since the 3d theory is super-renormalizable and has only a finite number of divergent graphs, the counterterm in Eq.~(\ref{eq:m3divergence3}) is actually an exact result, without any higher order corrections \cite{Farakos:1994kx}.

Since we are computing only vacuum diagrams in the effective theory, matching the fields is not required, apart from what was included in the mass parameter computations. All the required parameters are then known, and the pressure $\pM$ can be computed by evaluating all one-particle irreducible vacuum diagrams up to three-loop level in terms of these parameters. Apart from a gauge boson loop with both Higgs and $A_0$ self-energy corrections, all the required 3d integrals are computed in \cite{Braaten:1996jr}.
Because of the massive propagators the general structure of the result is much more complicated than for $\pE$:
\begin{eqnarray}
\lefteqn{ \frac{\pM(T)}{T} = \frac{1}{4\pi} \dF \left(m_3^2+\delta m_3^2 \right)^{3/2} \left[ \frac{2}{3}
	+\epsilon\left(\frac{16}{9}+\frac{4}{3}\ln\frac{\mu_3}{2m_3}\right) \right]
	+\frac{1}{4\pi}\left( \frac{1}{3}\dA \mD^3 +\frac{1}{3} \mD'^3 \right) } \nonumber \\
&& {}+\frac{1}{(4\pi)^2} \bigg[-\dF (\dF+1) \lambda_3 m_3^2 -\dF\dA h_3 m_3 \mD -\dF h_3' m_3 \mD'  \nonumber \\
&& {}-\left. \left( \CF g_3^2 +\frac{1}{4}g_3'^2\right)\dF m_3^2 \left( \frac{1}{2\epsilon} 
	+\frac{3}{2} +2\ln \frac{\mu_3}{2m_3}\right) -\CA\dA g_3^2\mD^2 \left( \frac{1}{4\epsilon} +\frac{3}{4}
	+\ln \frac{\mu_3}{2\mD}\right) \right] \nonumber \\
&& {}+\frac{1}{(4\pi)^3}\Big[ g_3^4 m_3 B_{AAf} +g_3'^4 m_3 B_{BBf} +g_3^2 g_3'^2 m_3 B_{ABf} +g_3^4 \mD B_{AAa}
	+g_3^2\lambda_3 m_3 B_{A\lambda f} \nonumber \\
&& {}+g_3'^2 \lambda_3 m_3 B_{B\lambda f} +\lambda_3^2 m_3 B_{\lambda \lambda f} +h_3^2 m_3 B_{hhf} +h_3^2 \mD B_{hha}
	+h_3'^2 m_3 B'_{hhf} +h_3'^2 \mD' B'_{hhb}\nonumber \\
&& {}+g_3^2 g_3'^2 m_3 2b(m_3) +g_3^2 g_3'^2\mD b(\mD) +g_3^2 g_3'^2 \mD' b(\mD')+\frac{\dF}{4m_3}(\dA h_3\mD +h_3'\mD')^2
	\nonumber \\
&& {}+\dF^2 m_3^2 \left( \frac{\dA h_3^2}{2\mD} +\frac{h_3'^2}{2\mD'}\right) +g_3^4\CA\CF\dF\frac{1}{3}
	\left( \frac{m_3^2}{\mD}\ln\frac{\mD+m_3}{m_3} +\frac{\mD^2}{m_3}\ln\frac{\mD+m_3}{\mD} \right) \nonumber \\
&& {}+\dF(\dF+1)\lambda_3(\dA h_3\mD +h_3'\mD') +g_3^2 h_3\mD B_{Aha} +g_3'^2 h_3'\mD' B'_{Bhb} +g_3^2 h_3'\mD' B'_{Ahb}
	\nonumber \\
&& {}+ g_3'^2 h_3\mD B_{Bha} +g_3^2 h_3 m_3 B_{Ahf} \bigg].
\label{eq:3dpres}
\end{eqnarray}
The coefficients $B_{xyz}$ and the coefficient function $b(x)$ are linear combinations of $1/\epsilon$, $\ln 2$, $\pi^2$ and $\ln(\mu_3/M)$, where $\mu_3$ is the 3d dimensional regularization scale and $M$ can be any combination of the different mass parameters $\mD,\mD',m_3$. The detailed expressions can be found in the Appendix C of \cite{Gynther:2005dj}.

The part of the pressure coming from the electric scales in Eq.~(\ref{eq:3dpres}) has many new features that are not present in the corresponding computation for QCD. In particular, the only dimensional parameters in dimensionally reduced QCD (known as EQCD) are $g_3$ and $\mD$, so the possible terms are, for dimensional reasons, $\mD^3$, $g_3^2\mD^2$ and $g_3^4\mD$, with divergent coefficients containing $\ln(\mu_3/\mD)$. This is in sharp contrast with the abundance of different terms in Eq.~(\ref{eq:3dpres}); not only are there many combinations of couplings and masses, but also completely new kinds of expressions like $m_i^2/m_j$ and $\ln(m_i/m_j)$.

All the coefficients $B_{xyz}$ and $b(x)$ of the $\order{g^5}$ terms have ultraviolet divergences, but they cancel against the $1/\epsilon$ terms in the mass counterterm $\delta m_3^2$. The renormalized mass in Eq.~(\ref{eq:m3matched}) depends on the renormalization scale $\Lambda$ through logarithms $\ln(\Lambda/4\pi T)$, which come with the divergences as usual. They cancel against the corresponding logarithms $\ln(\mu_3/M)$ in Eq.~(\ref{eq:3dpres}), leaving terms like $g^4 \mD\ln(\mD/T) \sim g^5 \ln\, g$. This kind of terms are not present in EQCD, where all the mass parameters are finite at $\order{g^4}$. If we choose $\mu_3 = \Lambda$, the scale dependence in $\pM$ vanishes completely at $\order{g^3}$ and $\order{g^5}$ when the running of the 4d couplings is taken into account.

The remaining $1/\epsilon$ terms shown explicitly at two-loop level in Eq.~(\ref{eq:3dpres}) cancel against the infrared divergences in $\pE$, Eq.~(\ref{eq:4dgenpres}). Also there the cancellation between terms coming from scales $\pi T$ and $gT$ results in large logarithms of order $g^4 \ln\, g$. These terms are also present in the QCD pressure, where they were originally derived from the requirement that the pressure should not depend on the scale at $\order{g^4}$ \cite{Toimela:1982hv}.

The presence of terms proportional to $\ln(g)$ shows that we cannot choose the scale in such way that the large logarithms would completely vanish. The use of an effective theory to separate the contributions from different scales is often advocated by the absence of large logarithms, but as we see, the infrared divergences mix the different scales, and logarithms of $\ln(gT/T)$ are left in the final result. A stronger argument for formulating the problem in terms of effective theories is the proper handling of resummations and isolating the non-perturbative infrared behavior into a simpler theory, as discussed in section~\ref{subsec:dimred}.

\section{Pressure near the phase transition}
\label{sec:pt_integration}

One of the most interesting properties of the electroweak theory is the crossover phase transition. When the temperature is lowered, the effective potential of the Higgs field develops a new minimum at some finite value $\langle \Phi \rangle \neq 0$. This phase transition gives masses to all quarks and leptons (except neutrinos) as well as to those gauge bosons mediating the weak interactions that correspond to the broken part of the eletroweak $\mathrm{SU}(2)_L \times \mathrm{U}(1)_Y$ symmetry.

In the deconfinement phase transition of QCD the strong coupling grows very large and perturbation theory cannot be used to study the phase transition. This is not the case in the electroweak symmetry breaking, since the confinement radius of weak interactions is tremendously large ($g$ has a Landau pole at $1/\Lambda_\mathrm{EW} \approx 10^6$~m) compared to the relevant distance scales at the electroweak transition temperatures ($1/T_c \sim 10^{-18}$~m for $T_c \sim 200$~GeV). We can then try to use perturbation theory in studying the electroweak phenomena close to the phase transition, in particular to extend our previous computation of the pressure down to transition temperatures. Besides the pressure playing a central role in describing thermodynamics near the transition, this also allows us to test the validity of dimensional reduction at the phase transition, something that cannot be done in QCD.

Note that perturbation theory is not able to describe the transition itself correctly with physical Higgs masses. Both two-loop effective potential calculations \cite{Farakos:1994kx}, although limited to $m_H \lesssim m_W$, and the $\epsilon$-expansion analysis extrapolated to $\epsilon=1$ \cite{Arnold:1993bq} suggest the presence of a first order phase transition for large Higgs masses, while in reality the transition is of crossover type for $m_H \gtrsim 72$~GeV \cite{Kajantie:1996mn,Karsch:1996yh,Gurtler:1997hr}. In the unbroken phase the Higgs field has a non-vanishing vacuum expectation value, which we would have to incorporate in our computations to build the perturbative expansion around the true, physical vacuum. Neverthless, because the coupling constant stays small in the transition we are able to perturbatively compute physical quantities while approching the transition from above, as long as we stay in the symmetric vacuum.

The pressure computed in Eqs.~(\ref{eq:4dgenpres}),(\ref{eq:3dpres}) cannot be directly continued down to the phase transition, because it becomes singular as we approach the critical temperature (which exists in perturbation theory). The reason for this is that the thermally corrected Higgs mass $m_3^2$ in Eq.~(\ref{eq:m3matched}) becomes very small at the phase transition, and finally turns negative a little below the transition. At leading order it is this negative mass parameter that makes the symmetric phase unstable and drives the phase transition. The singular effects of small thermal Higgs mass can be seen in $\pM$, Eq.~(\ref{eq:3dpres}), which contains terms like $\mD^2/m_3$ and $\mD \ln(\mu_3/m_3)$.

\begin{figure}
\centering
\includegraphics[width=0.92\textwidth]{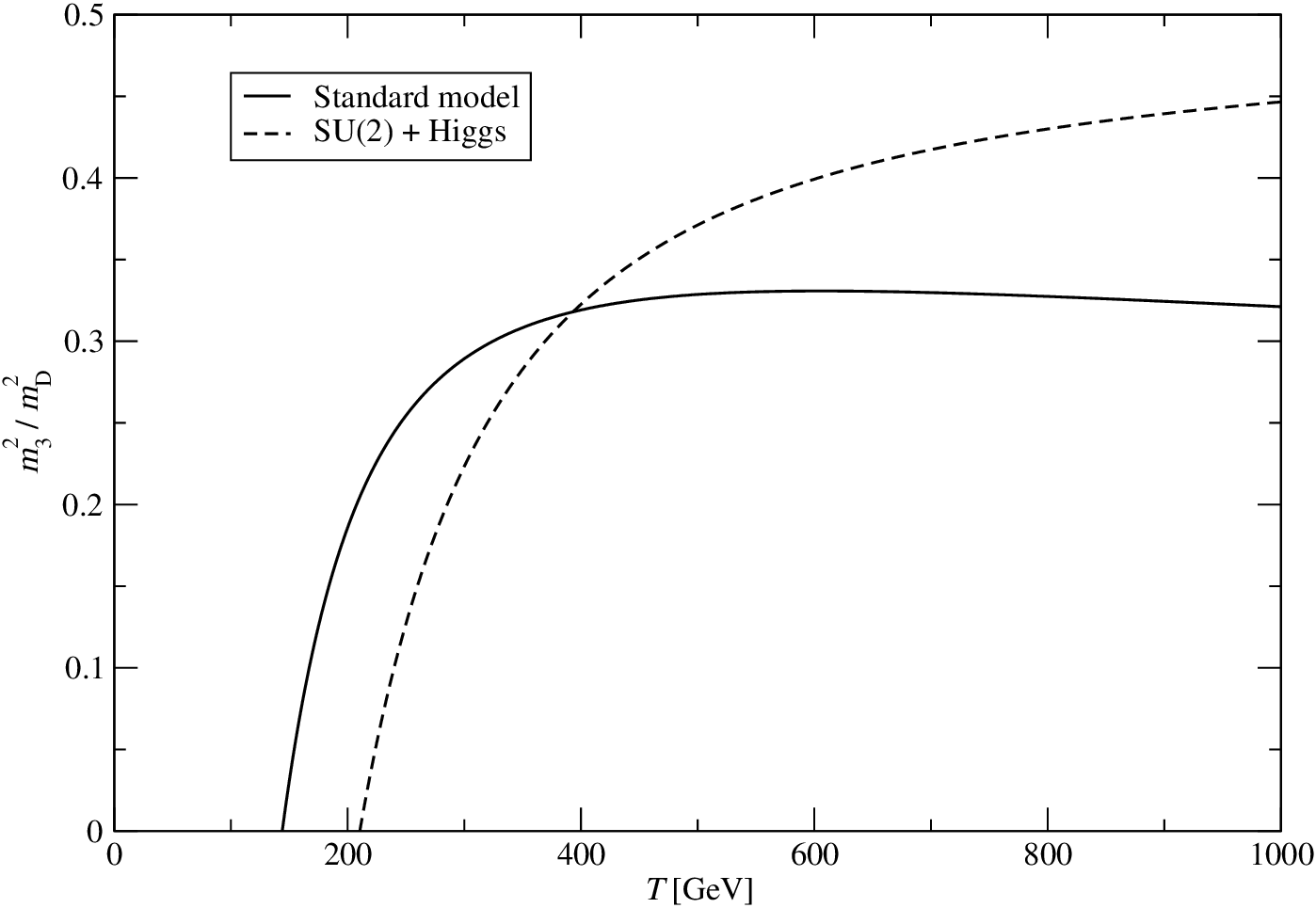}
\caption{The ratio of the fundamental scalar thermal mass to the SU(2) adjoint scalar mass. The regularization scale is chosen as $\Lambda=2\pi T$.}
\label{fig:massratio}
\end{figure}

The origin of this problem is in our power counting, which assumed $m_3 \sim \mD \sim gT$. Close to $T_c$ this assumption fails and we run into the same infrared problems as in the original theory, since the resummation of non-static modes no longer gives a finite mass to the Higgs field. In Fig.~\ref{fig:massratio} we have plotted the ratio of the renormalized Higgs mass to the SU(2) adjoint scalar mass in both the full standard model (including strong interactions) and the SU(2) + Higgs theory which we have used as a weakly coupled toy model. The numerical values of the parameters are given in section~\ref{sec:pressure_numerics}, and the renormalization scale for $m_3$ is chosen as $\Lambda=2\pi T$. As the figure shows, the mass ratio drops steeply as the temperature approaches the phase transition, and at $T-T_c \lesssim 50$~GeV the assertion $m_3 \sim \mD$ clearly fails.

The solution is to resum yet another class of diagrams, the adjoint scalar self-energy corrections on the fundamental scalar line. We then have one more level of dimensional reduction, and instead of Eq.~(\ref{eq:heavyred}) the pressure is given by
\begin{equation}
 p(T) \equiv \pE(T) +\pMy(T) +\pMk(T) +\frac{T}{V}\ln \int \!\mathcal{D}A_k \mathcal{D}\Phi \exp\left(-S_\mathrm{M}\right),
\label{eq:lightred}
\end{equation}
where $\pE$ is the same as before, Eq.~(\ref{eq:4dgenpres}), and $\pMy$ is computed from the effective theory $S_\mathrm{E}$ in Eq.~(\ref{eq:3daction}), treating now the light scalar mass $m_3$ as a perturbation and expanding the integrals in $m_3^2/\mD^2$. The contribution from scales $m_3^2 \lesssim g^3 T^2$ is contained in the contribution $\pMk$, which is computed from an effective theory containing only the fundamental scalar field and the spatial components of the gauge bosons,
\begin{equation}
S_{\mathrm{E}2} = \int\!\dd^3 x\, \frac{1}{4} G^a_{ij}G^a_{ij} + \frac{1}{4}F_{ij}F_{ij} + (D_i\Phi)^\dagger(D_i\Phi)
                      + \widetilde{m}_3^2\Phi^\dagger\Phi + \tilde{\lambda}_3(\Phi^\dagger\Phi)^2 \, ,
\end{equation}
where the gauge couplings and the scalar self-coupling do not get any matching corrections at this level, $(\tilde{g}_3^2,\;\tilde{g}_3'^2,\;\tilde{\lambda}_3) = (g_3^2,\;g_3'^2,\;\lambda_3)$. The mass parameter, however, now also resums the adjoint scalar loops, which we have to compute up to one-loop level ($g^3$ in our power counting),
\begin{eqnarray}
\widetilde{m}_3^2 &=& m_3^2 - \frac{1}{4\pi}\left(\dA h_3 \mD + \frac{1}{4}g_3'^2 \mD'\right) \nonumber \\
 && {}-\frac{1}{2\pi}\left[ \dA h_3 \mD\left(1 +\ln\frac{\mu_3}{2\mD} \right)
	+\frac{1}{4}g_3'^2 \mD' \left(1+\ln\frac{\mu_3}{2\mD'} \right) \right] \epsilon + \mathcal{O}(g^4).
\label{eq:m3mass}
\end{eqnarray}
Apart from the $\order{\epsilon}$ terms, this expression has been previously computed in \cite{Kajantie:1995dw}. The correction is of the same order $g^3$ as the leading term $m_3^2$, showing that resummation is necessary to get consistent results.

For simplicity, we have neglected the order $g^4$ corrections to the scalar mass. This can be justified by power counting arguments, since for $m_3^2 \sim g^3 T^2$ the two-loop corrections would contribute parametrically at order $m_3 g_3^2 \sim g^{11/2} T^3$, which is strictly speaking higher than $\order{g^5}$ we are considering here. Dropping terms suppressed by $\sqrt{g}$ may not be numerically justified, but we expect their effect to be small, in particular because this only concerns four out of more than a hundred degrees of freedom in the standard model. The practical reason is that we want to avoid computing all three-loop diagrams in the effective theory, as well as the renormalization and scale dependence of the Higgs mass at two-loop level.

Evaluating all the three-loop vacuum diagrams of theory $S_\mathrm{E1}$ and two-loop diagrams of $S_\mathrm{E2}$ (there are only three of them), and setting the number of fermion families to $n_\mathrm{F}=3$, we get
\begin{eqnarray}
\frac{\pMk}{T} &=& \frac{\dF}{6\pi}\widetilde{m}_3^3 -\frac{\widetilde{m}_3^2}{(4\pi)^2}\left[\dF(\dF+1)\tilde{\lambda}_3
	+\frac{1}{2}\dF\left( \CF\tilde{g}_3^2 +\frac{1}{4}\tilde{g}_3'^2\right) \left( \frac{1}{\epsilon} +3
	+4\ln\frac{\tilde{\mu}_3}{2\widetilde{m}_3}\right)\right],\hspace{1cm}
\label{eq:m2result}
\end{eqnarray}
\begin{eqnarray}
 \frac{\pMy}{T} &=&  \frac{1}{4\pi}\left( \frac{1}{3}\dA \mD^3 +\frac{1}{3} \mD'^3 \right)
	+\frac{1}{(4\pi)^2} \CA\dA g_3^2\mD^2 \left( -\frac{1}{4\epsilon} -\frac{3}{4} - \ln \frac{\mu_3}{2\mD}\right) \nonumber \\
 &&{}-\frac{1}{(4\pi)^3\epsilon}\left[ \frac{1}{4}\CA\CF g_3^4\mD +\dA h_3^2\mD +h_3'^2\mD'
	+\frac{1}{4}\CF g_3^2 g_3'^2(\mD+\mD') \right]\frac{\dF}{2}  \nonumber \\ 
 &&{}+\frac{1}{(4\pi)^3}\left\{ g_3^4\mD \left[ \CA^2\dA\left( -\frac{89}{24} +\frac{11}{6}\ln 2 -\frac{\pi^2}{6} \right) 
	+\CA\CF\dF\left( -\half -\frac{3}{4}\ln\frac{\mu_3}{2\mD} \right) \right] \right. \nonumber \\
 &&{}+g_3^2 g_3'^2\CF\dF \frac{1}{4}\left[ (\mD+\mD')\left( -4 -2\ln\frac{\mu_3}{\mD+\mD'} \right)
	-\mD\ln\frac{\mu_3}{2\mD} - \mD'\ln\frac{\mu_3}{2\mD'} \right]  \nonumber \\
 &&{}+\left. h_3^2\mD\dA\dF\left( -4-3\ln\frac{\mu_3}{2\mD}\right) +h_3'^2\mD'\dF\left( -4-3\ln\frac{\mu_3}{2\mD'}\right) \right\}.
\label{eq:m1result}
\end{eqnarray}
The $\order{g^4}$ two-loop divergences in $\pMy$ cancel against $\pE$ as before. In addition, there are $1/\epsilon$ terms left in $\pE$ whose coefficients by themselves are of order $g^2$ but combine to a term proportional to $m_3^2$, which cancels against the divergence in $\pMk$. Unlike in the high-temperature case, there are also $\order{g^5}$ divergences in $\pMy$, but these go away when the one-loop corrections in $\widetilde{m}_3^2$ multiplying the ``sunset'' diagram in $\pMk$ are included. The scale dependence cancels along with divergences, if we take into account the running of the couplings in the full theory, and set all regularization scales to be the same, $\Lambda = \mu_3 = \tilde{\mu}_3$. The phase transition can now be safely approached, since the result (\ref{eq:m2result}) for $\pMk$ is perfectly well-behaved as $\widetilde{m}_3^2$ goes to zero.

\section{Numerical results}
\label{sec:pressure_numerics}

To see how the multitude of terms we have computed affects the physical pressure, we have to supply some numbers for the parameters of the 4d Lagrangian, Eq.~(\ref{eq:sm_lagrangian}). Apart from the yet undiscovered Higgs particle mass, the standard model parameters have been measured to great precision in collider experiments, in particular LEP. The values of couplings can be determined from their tree-level relations to various mass parameters,
\begin{equation}
\begin{array}{rclrcl}
\displaystyle \nu^2(m_Z) & = & \displaystyle \half m_H^2\,, &
\displaystyle \lambda(m_Z) & = & \displaystyle \frac{1}{\sqrt{2}}G_\mu m_H^2\,, \\
\displaystyle g^2(m_Z) & = & \displaystyle 4\sqrt{2}G_\mu m_W^2\,, & \hspace{1cm}
\displaystyle g'^2(m_Z) & = & \displaystyle 4\sqrt{2}G_\mu\left(m_Z^2-m_W^2\right), \rule{0pt}{3ex} \\
\displaystyle g_Y^2(m_Z) & = & \displaystyle 2\sqrt{2}G_\mu m_t^2\,, &
\displaystyle \alpha_s(m_Z) & = & \displaystyle 0.1187\,, \rule{0pt}{3ex}
\end{array}
\label{eq:sm_parameter_values}
\end{equation}
where $m_W=80.40$~GeV, $m_Z=91.19$~GeV and $m_t=174$~GeV are the masses of the W and Z bosons and the top quark, respectively, and $G_\mu=1.664\cdot 10^{-5}\;\mathrm{GeV}^{-2}$ is the Fermi coupling constant \cite{Eidelman:2004wy}. The cited values are what we have used in \cite{Gynther:2005dj,Gynther:2005av}, and they remain practically unchanged in the more recent Review of Particle Physics \cite{Yao:2006px}, with only the strong coupling being slightly smaller, $\alpha_s(m_Z) = 0.1176$. Searches for the Higgs particle give its mass a lower limit $m_H \gtrsim 114$~GeV but leave it otherwise unknown. We have used the value $m_H=130$~GeV in all our analysis. The pressure of the full standard model is very insensitive to the Higgs mass when we are not close to the phase transition, although $T_c$ itself depends on $m_H$. In \cite{Gynther:2005dj} we have shown that increasing $m_H$ to 200~GeV causes a relative change of $10^{-3}$ in the pressure.

In addition to the standard model, we have also studied the simpler SU(2) + Higgs theory, for which the corresponding results can be found from those computed above by setting $g'^2=g_s^2=g_Y^2=n_\mathrm{F}=0$. Besides simpler analytic expressions, this model has some advantages over the standard model when we want to study the behavior of dimensional reduction near the critical temperature. The Higgs field drives the phase transition, but it only represents four of the 106.75 effective degrees of freedom in the standard model, so its effects on the pressure are hard to see. In SU(2) + Higgs model the corresponding number is only 10. Also, this toy model is weakly coupled, with $\lambda\approx 0.20$ and $g^2\approx 0.61$ given by Eq.~(\ref{eq:sm_parameter_values}), so the perturbative expansion converges better than in the presence of large couplings $g_s$ and $g_Y$. More details on computations in this model can be found in \cite{Gynther:thesis}. Thermodynamics of SU(2) + fundamental Higgs theory have also been studied on lattice \cite{Csikor:1996sp}.

\begin{figure}
\centering
\includegraphics[width=0.92\textwidth]{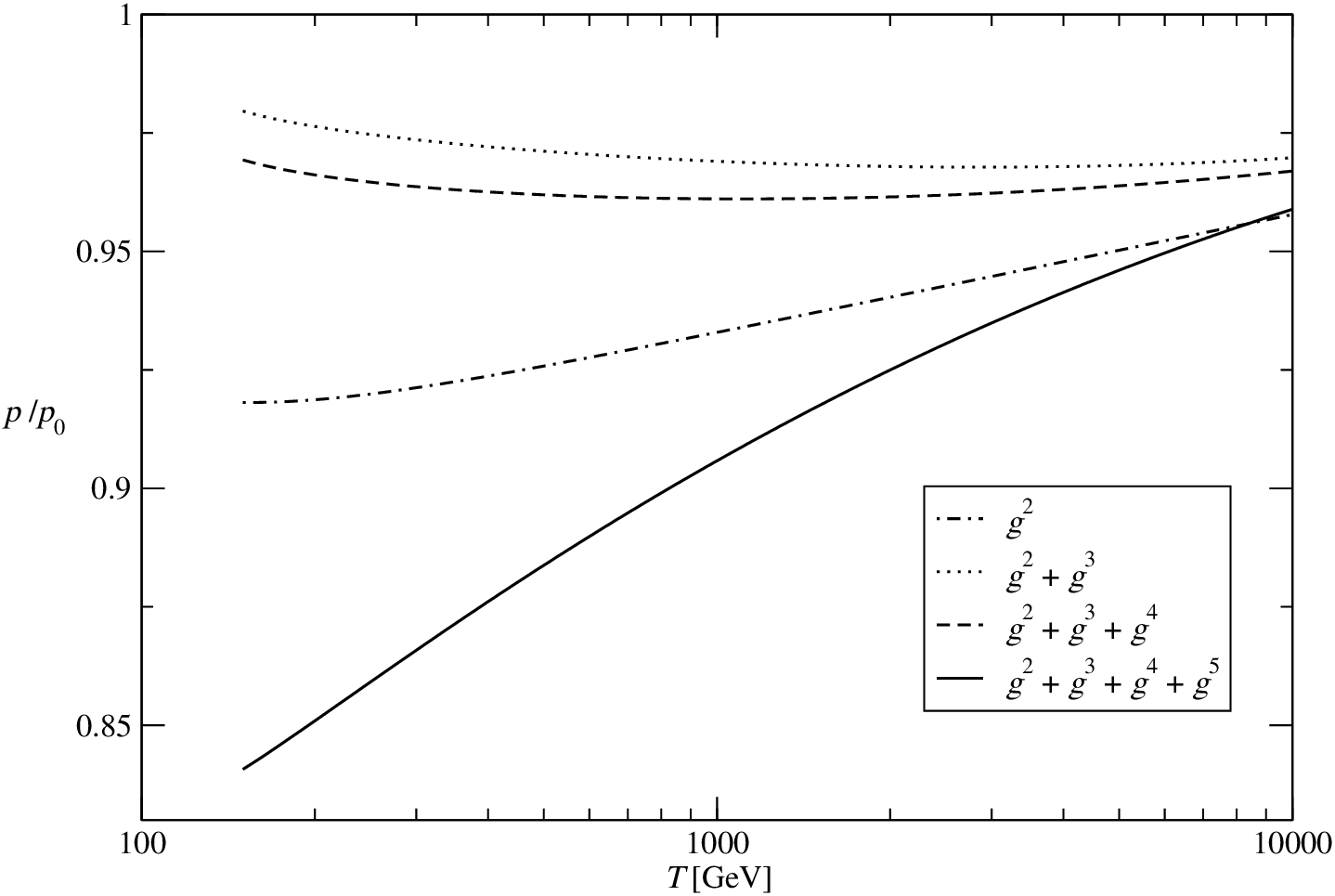}
\caption{Pressure of the standard model at different orders of perturbation theory.}
\label{fig:p_sm_orders}
\end{figure}

In Fig.~\ref{fig:p_sm_orders} we have plotted the pressure of the full standard model up to $\order{g^5}$, combining the results in Eqs.~(\ref{eq:4dgenpres}),(\ref{eq:3dpres}) and the pure QCD pressure taken from \cite{Braaten:1996jr,Arnold:1994ps,*Arnold:1995eb,Zhai:1995ac,Kajantie:2002wa} (with one-loop quark diagrams subtracted to avoid double counting),
\begin{equation}
 p(T) = \pE(T) + \pM(T) + p_\mathrm{QCD}(T) +T^4 \cdot \order{g^6}.
\end{equation}
The radiative electroweak corrections also need to be taken into account in the two-loop electric gluon mass $m_\mathrm{3E}^2$ which we insert in $p_\mathrm{QCD}(T)$ above. They have been computed in \cite{Gynther:2005dj}, and used in the leading order term $2m_\mathrm{3E}^3/3\pi$.

Although the physical pressure is independent of the renormalization scale, the perturbative expansion depends on the scale through the renormalization of parameters at orders higher than those included in the computation, and we need to fix the scale to define the couplings. As the remaining scale dependence is cancelled by higher order terms, the magnitude of the unknown corrections can be estimated by varying the scale. We have chosen to use  $\Lambda=2\pi T$, having shown that the scale dependence is indeed weak.

The pressure in all our plots is normalized to the Stefan--Boltzmann result of non-interacting gas of relativistic particles,
\begin{equation}
 p_0 = \frac{\pi^2 T^4}{90}\left( 2 + 2\dA + 2(N_c^2-1) +2\dF +2\,\frac{7}{8}n_\mathrm{F}[ \dF +1 +N_c(\dF+2)] \right)
	= \frac{\pi^2 T^4}{90}  \left\{ \begin{array}{l} 106.75 \\ 10 \end{array} \right.
\label{eq:sm_dofs}
\end{equation}
for standard model and SU(2) + Higgs, respectively. The number multiplying $T^4$ is actually $\alpha_{E1}$ of Eq.~(\ref{eq:4dgenpres}) + the gluon contribution.

Fig.~\ref{fig:p_sm_orders} shows that the perturbative expansion does not converge very well at moderate temperatures, but instead the $\order{g^5}$ correction is even larger than any of the preceeding terms. This behavior is known in QCD, and the strong coupling constant is still large at electroweak temperatures, $g_s^2(m_Z) \approx 1.48$. In the full standard model the strongly interacting degrees of freedom sum up to 79, or 74\% of the number in Eq.~(\ref{eq:sm_dofs}), so the terms with gluon exchange clearly dominate the higher order corrections, together with the few terms containing Yukawa interactions. The pressure lies 5-10\% below the ideal gas result, and begins to converge very slowly at temperatures in TeV range.

\begin{figure}
\centering
\includegraphics[width=0.92\textwidth]{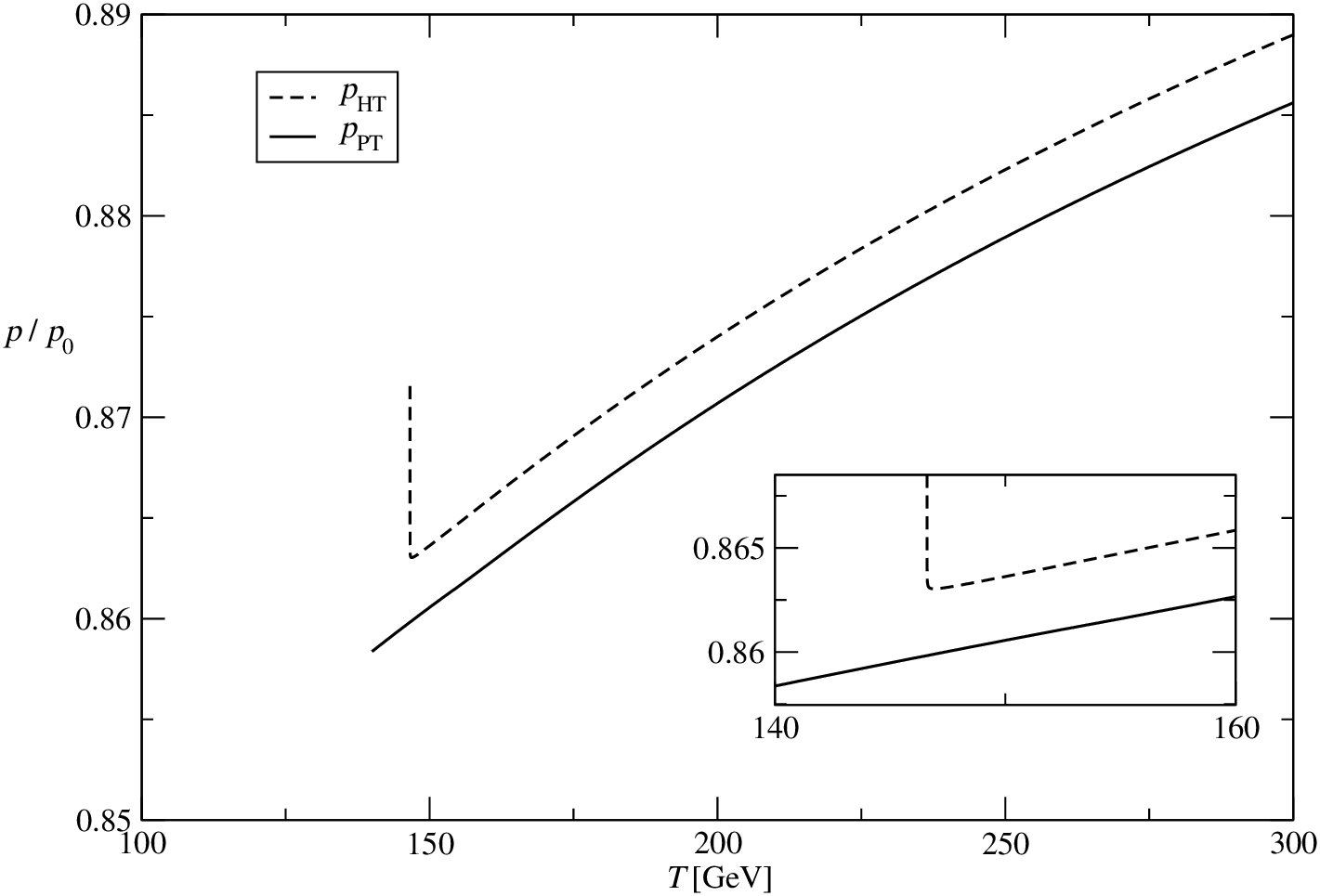}
\caption{The pressure of the standard model near the phase transition. Here $p_\mathrm{HT}$ stands for the high-temperature calculation, $p_\mathrm{PT}$ is resummed for the light scalar mass.}
\label{fig:overlay}
\end{figure}

Close to the phase transition we have to use the resummed result of Eq.~(\ref{eq:lightred}) for the pressure. In Fig.~\ref{fig:overlay} we have plotted both the high-temperature result $p_\mathrm{HT}(T)$ and light scalar mass resummed result $p_\mathrm{PT}(T)$. As the figure shows, the high-temperature result is very sharply peaked at $T_c$, whereas the corrected computation goes through the transition smoothly. Of course, below $T_c$ the system goes to the nonsymmetric ground state whose pressure is larger than the symmetric phase pressure plotted in Fig.~\ref{fig:overlay}, since a thermal system always tries to minimize the free energy, or maximize the pressure. When $\widetilde{m}_3^2$ becomes negative slightly below $T_c$, the symmetric phase pressure develops an imaginary part, which can be interpreted as the decay rate of the unstable symmetric phase \cite{Weinberg:1987vp}. In this region we have plotted the real part of the pressure in our figures.

The two curves in Fig.~\ref{fig:overlay} differ by a term roughly proportional to $T^4$, but the difference is only about 0.3\%. This is because in $p_\mathrm{PT}$ we have resummed another class of diagrams that are not present in $p_\mathrm{HT}$. We have also left out all three-loop diagrams in $S_\mathrm{E2}$, which would contribute at order $\widetilde{m}_3 g^4$, and the $\order{g^4}$ corrections to $m_3^2$. In particular the terms with strong and Yukawa couplings might be important even at this order.

\begin{figure}
\centering
\includegraphics*[width=0.92\textwidth]{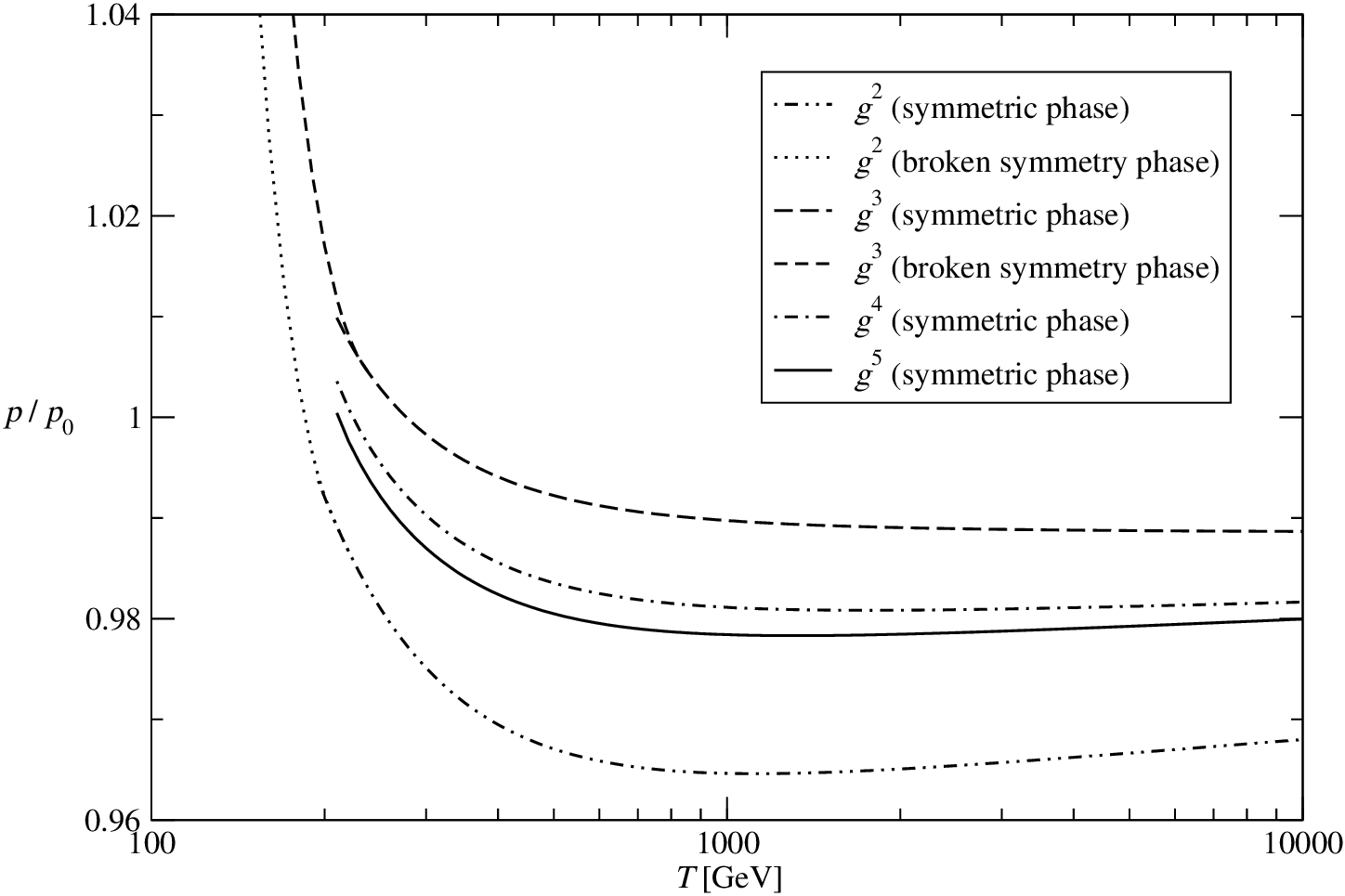}
\caption{The pressure of SU(2) + Higgs theory.}
\label{fig:pressure_su2h}
\end{figure}

\begin{figure}
\centering
\includegraphics*[width=0.92\textwidth]{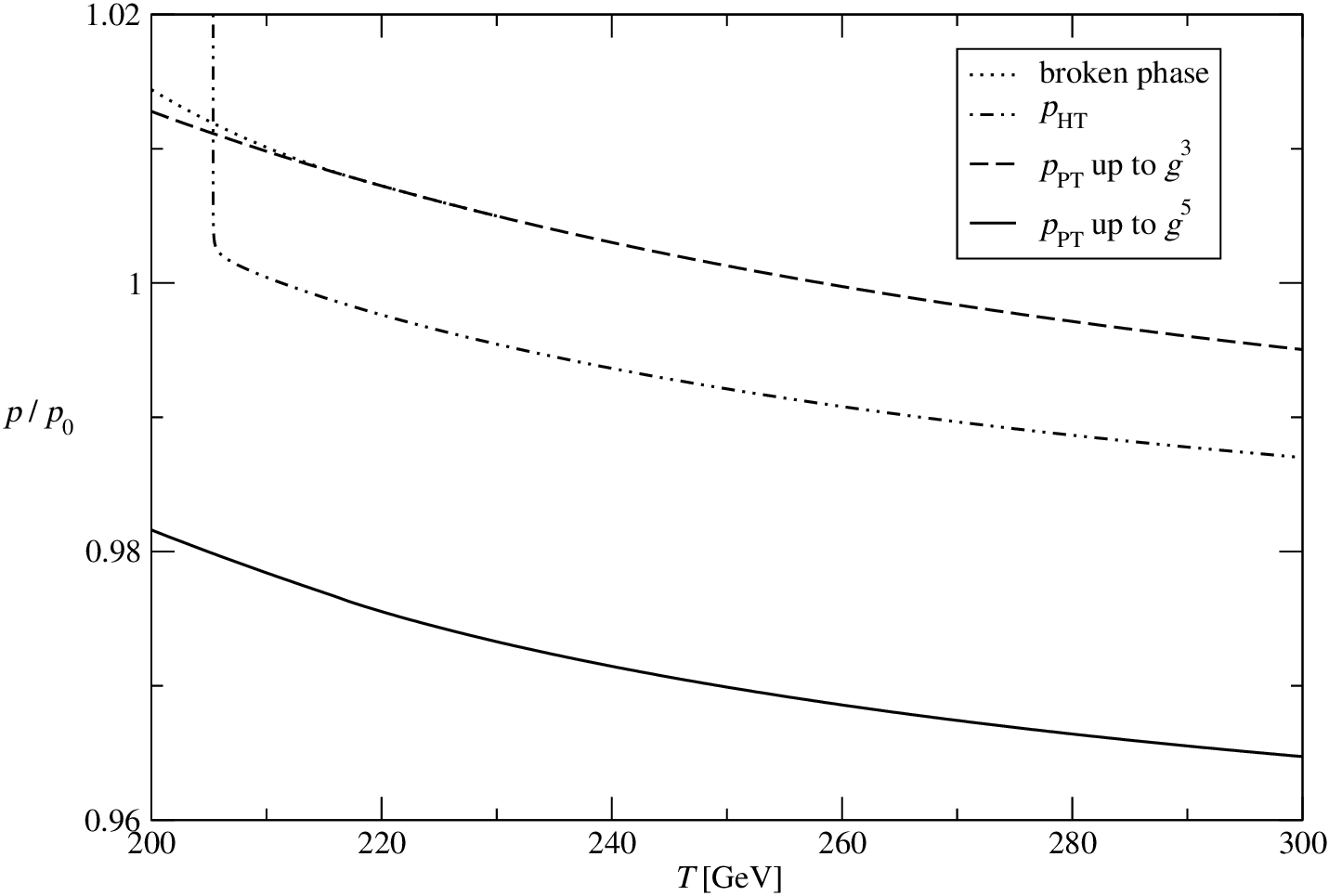}
\caption{The pressure of SU(2) + Higgs theory near the phase transition. Here $p_\mathrm{HT}$ stands for the high-temperature calculation, while $p_\mathrm{PT}$ has Higgs mass corrections resummed. Also shown is the $\order{g^3}$ pressure in the broken symmetry phase.}
\label{fig:pressure_su2h_pt}
\end{figure}

Corresponding plots for the simpler SU(2) + fundamental Higgs theory are shown in Fig.~\ref{fig:pressure_su2h} and Fig.~\ref{fig:pressure_su2h_pt}, with parameters taken from Eq.~(\ref{eq:sm_parameter_values}) using $m_W=80$~GeV and $m_H=130$~GeV. In both figures we have also plotted an approximation of the broken phase pressure below $T_c$ to indicate the phase transition at $T_c\approx 220$~GeV. It can be derived from the two-loop (order $g^3$) computations of the effective potential \cite{Arnold:1992rz} and the pressure in the symmetric phase by using $p_\mathrm{BP}(T,\phi)=p_\mathrm{SP}(T) - V_\mathrm{eff}(T,\phi)$. Fig.~\ref{fig:pressure_su2h} should not be trusted near the transition, but away from $T_c$ it shows that in the weakly coupled model the perturbative expansion converges nicely and settles to a level which is about 2\% below the ideal gas pressure. Note that in the one-loop renormalization $\lambda(2\pi T)$ grows with $T$, so there is no reason to expect free theory behavior even at very high temperatures.

Close to the phase transition Fig.~\ref{fig:pressure_su2h_pt} shows a similar peak in $p_\mathrm{HT}$ as in the standard model, while $p_\mathrm{PT}$ does not see the transition at all until $\widetilde{m}_3^2$ becomes negative at about 15~GeV below $T_c$. The constant (times $T^4$) difference between the two pressures is larger than in the standard model, about $2\%$. The couplings are all small here, so we expect that the three-loop diagrams in in $S_\mathrm{E2}$ and the $\order{g^4}$ corrections to the fundamental scalar mass are not important. However, in this model larger fraction of the overall pressure comes from the Higgs sector, so the result is more sensitive to different resummations on the scalar propagator.

We have computed the pressure of the full standard model to order $g^5$ both near the phase transition and at high temperatures. In principle, it is possible to go one step further and compute the coefficient of the last perturbatively accessible term of order $g^6\ln\,g$. This computation has been carried out in QCD \cite{Kajantie:2002wa}, but because the QCD pressure dominates the standard model pressure and $g_s$ is large, we expect advances in understanding the QCD pressure to be more important than four-loop diagrams in electroweak theory. Another computable term would be the neglected three-loop diagrams and $\order{g^4}$ mass corrections in $S_\mathrm{E2}$ close to the phase transition, contributing parametrically at order $g^{5.5}$, but these affect only the Higgs sector and are probably too small to have any physical implications.

\chapter{Mesonic correlation lengths}
\label{chap:correlators}

Dimensional reduction is only useful in cases where the physical quantity of interest is time-independent. We can then average over any time coordinates, or, in terms of momentum space Green's functions, take the limit $p_0 \equiv \omega \to 0$ on all external momenta. This is trivially so in the case of thermodynamic potentials computed in the previous chapter, since they are computed from vacuum diagrams, having no external legs at all.

Another class of observables undergoing a dimensional reduction are various screening correlators, which describe the response of the system to a time-independent external perturbation. At low momenta they are usually dominated by simple imaginary poles in the momentum space, leading to spatial correlators that at large scales decay exponentially with the distance. The characteristic scale of this exponential fall-off is referred to as the screening length and its inverse the screening mass. A typical example would be the electric field of a point charge immersed in electromagnetic plasma, which at large distances is screened by the electric mass $m_\mathrm{el}^2 \approx e^2T^2/3$.  It should be noted that in general the masses of real-time bound states can be very different from the corresponding screening masses.

Static correlators of bosonic operators have been succesfully studied using dimensionally reduced effective theories. For example, various gluonic correlation lengths have been measured by implementing the three-dimensional theory of static gluons (EQCD) on lattice \cite{Reisz:1992er,*Karkkainen:1992jh,*Karkkainen:1993wu,Kajantie:1997tt, Laine:1998nq,*Laine:1999hh, Hart:1999dj,Hart:2000ha,Cucchieri:2001tw}. In these works the fermionic modes are integrated out as in sections \ref{sec:dimensional_reduction} and \ref{sec:sm_pressure}, leaving a theory of soft gluonic excitations around the perturbative vacuum. However, this is not the only option when deriving an effective theory, but we can as well expand around any other saddle point of the action, corresponding to a choice $\bar{\Phi}_h \neq 0$ in Eq.~(\ref{eq:gaussian_S}). In particular, not all fermionic modes need to be integrated out, but the effective theory can constructed around some specific fermionic state. This opens a possibility to study also fermionic correlators using dimensional reduction.

Of particular interest are operators consisting of a light quark-antiquark pair propagating in the hot medium. In \cite{DeTar:1985kx,*DeTar:1987ga} it was suggested that at scales comparable to the magnetic scale $1/g^2T$ the spectrum of quark-gluon plasma consists of color-singlet modes only, while the colored excitations are dynamically confined. The lowest lying excitations at these scales would then be the various glueball modes and the mesonic and baryonic states consisting of two and three quarks, respectively. In order to better understand the long-distance behavior of the plasma, the properties of these states have been measured in detail on lattice. Most of these studies have been devoted to Euclidean correlators, or the screening properties of these operators, due to the inherently Euclidean nature of lattice simulations. In particular the spectrum of the hadronic screening masses has been carefully measured, the first simulations dating back 20 years \cite{DeTar:1987ar,*DeTar:1987xb,Gottlieb:1987gz}.

While combining perturbative calculations with lattice simulations using dimensional reduction has been very useful when measuring the glueball spectrum, the hadronic screening masses have been measured using expensive 4d simulations. However, the need for analytical tools is even greater in the fermionic sector, where the lattice simulations have difficulties in treating the light dynamical quarks correctly. The situation is yet worse when we allow for quark chemical potentials, which make the fermion determinant complex and ruin the conventional importance sampling. On the other hand, operators built out of quark fields are usually less infrared sensitive, so perturbation theory should more applicable in computing their properties.

The first attempts to determine the screening masses of mesonic states at high temperatures using dimensional reduction were more of a qualitative nature, since they knowingly left out corrections of the same order as the leading term \cite{Hansson:1992kb,*Hansson:1994nb, Koch:1992nx,*Koch:1994zt}. In particular, the scale inside the logarithm in the two-dimensional Coulomb potential $\sim \ln r$ was not fully identified. A more systematic approach was developed by Huang and Lissia in \cite{Huang:1996tz}, where it was shown that the dimensionally reduced theory for fermionic modes can be formulated in terms of massive non-relativistic quarks in 2+1 dimensions. They also discussed the correct power counting of different operators, and computed one-loop corrections to the quark self-energy and the quark-gluon interaction vertex. However, although the effective theory was derived in order to calculate screening quantities, the authors did not proceed to compute any masses in that work.

Following \cite{Huang:1996tz}, we used similar methods in \cite{Laine:2003bd} to derive an effective three-dimensional theory for the lowest fermionic modes $\pm \pi T$, which dominate the mesonic correlator at large distances. This theory takes the form of non-relativistic quarks coupled to EQCD, or the spatial gluons and an adjoint scalar field. Because the fermionic sector of the theory is very similar in form and power counting to the effective theory for heavy quarks in four dimensions known as ``non-relativistic QCD'', we have named the reduced theory \NRQCD. Using this theory, we were able to compute the next-to-leading order correction to mesonic screening masses in perturbation theory, and this computation was extended to finite quark chemical potentials in \cite{Vepsalainen:2007ke}. In this chapter we will review these results.

\section{Linear response theory and screening phenomena}

The correlation functions usually computed in theoretical calculations are related to physically measurable quantities through linear response theory. Our short presentation here follows mostly \cite{Lebellac:thermal}, and is somewhat biased towards screening physics.

Consider perturbing the system in equilibrium with some external probe, described by an interaction Hamiltonian $V(t)$ which vanishes for $t<0$. In the Schr\"odinger picture the time-development of an unperturbed state is given by the time-independent Hamiltonian $H$, while the effect of $V(t)$ can be written in terms of a time-development operator $U(t)$,
\begin{eqnarray*}
 |\psi_S(t)\rangle &=& e^{-iHt}|\psi_S(0)\rangle \equiv e^{-iHt}|\psi_H\rangle, \\
 |\psi_S'(t) \rangle &=& e^{-iHt}U(t)|\psi_S(0)\rangle,
\end{eqnarray*}
where $U(t)$ satisfies
\begin{eqnarray}
 i\partial_t |\psi_S'(t) \rangle &=& H|\psi_S'(t) \rangle +e^{-iHt} i\partial_t U(t) |\psi_S(0)\rangle
	= (H+V)|\psi_S'(t) \rangle \nonumber \\
 & \Rightarrow & i\partial_t U(t) = e^{iHt} V(t) e^{-iHt} U(t), \qquad U(t)=1 \quad \mathrm{for} \quad t<0\, .
\label{eq:U_time_dev}
\end{eqnarray}
In the last equation $V_H(t)$, the potential in the unperturbed Heisenberg picture is seen to emerge. If $V$ is small, $U(t)$ can be solved recursively as a series in $V$ by integrating Eq.~(\ref{eq:U_time_dev}),
\begin{equation}
 U(t) = 1 -\int_0^t \! \dd t_1\, V_H(t_1) - \int_0^t \! \dd t_1 \int_0^{t_1} \! \dd t_2 \, V_H(t_1)V_H(t_2) +\order{V^3}.
\end{equation}
The change in the expectation value of an arbitrary operator $\hat{O}(t)$ in the Schr\"odinger picture is then
\begin{eqnarray}
 \delta \langle \hat{O}(t)\rangle
 & \equiv & \langle \psi'_S(t)|\hat{O}(t)|\psi_S'(t) \rangle - \langle \psi_S(t)|\hat{O}(t)|\psi_S(t) \rangle \nonumber \\
 &=& -i\int_0^\infty \! \dd t' \langle \psi_H| \theta(t-t')[\hat{O}_H(t),V_H(t')]|\psi_H \rangle,
\label{eq:delta_Ot}
\end{eqnarray}
where the operators and the state vectors are now all in the Heisenberg picture with the unperturbed Hamiltonian $H$. In particular, Eq.~(\ref{eq:delta_Ot}) applies to the eigenstates of $H$, so we can sum over all states in the ensemble with appropriate weights, and replace the expectation value in a specific state by thermal average.

Typically, the external interaction can be written as a time-dependent c-number source $v(t,\mathbf{x})$ coupled to the system through some current $J$ built of field operators,
\begin{equation}
 V(t) = \int \!\dd^3 x\,  J(\hat{\phi}(t,\mathbf{x})) v(t,\mathbf{x}),
\label{eq:current_source_coupling}
\end{equation}
and the response of the system is measured through the same current. Eq.~(\ref{eq:delta_Ot}) can then be written in terms of the retarded correlation function $D_R$,
\begin{eqnarray}
\delta \langle J(\hat{\phi}(t,\mathbf{x})) \rangle
&=& -i\int_0^\infty \! \dd t' \int \!\dd^3 x'\, v(t',\mathbf{x}') 
	\langle \theta(t-t')[J(\hat{\phi}_H(t,\mathbf{x})),J(\hat{\phi}_H(t',\mathbf{x}')) ]\rangle \nonumber \\
& \equiv & - \int \! \dd^4 x' D_R(x-x')v(x'),
\label{eq:delta_J_coord}
\end{eqnarray}
where the lower limit of the time integration can be extended to $-\infty$ because $v(t,\mathbf{x})=0$ for $t<0$. In this expression the response of the system is clearly separated into the retarded propagator, which is specific to the thermal system in question, convoluted with a factor $v(x)$ depending on the details of the perturbation.

The excitations of the system manifest themselves as large responses to an external perturbation. Going into the momentum space, the Fourier transform of Eq.~(\ref{eq:delta_J_coord}) reads
\begin{equation}
 \delta \langle J \rangle(\omega,\mathbf{k}) = -i D_R(\omega,\mathbf{k}) v(\omega,\mathbf{k}),
\end{equation}
so this is equivalent to having a pole in the propagator for some values of frequency $\omega$ and momentum $\mathbf{k}$. In general there are some real-time excitations, whose frequencies depend on the momentum through the dispersion relation $\omega=\omega(\mathbf{k})$. In addition, at the static limit $\omega \to 0$ the propagator may contain poles at imaginary momenta, corresponding to perturbations that decay exponentially with distance due to plasma screening. The former are difficult to study in imaginary time formalism because of the need to analytically continue the imaginary time results, but the time-independent screening correlators are well-adapted for Euclidean computations.

It should be noted that the correlators often have very different dependence on $\omega$ and $\mathbf{k}$, since the thermal environment breaks the Lorentz invariance. The screening states and the real-time excitations describe completely different physics, and the corresponding poles need not be related at finite temperature. One should also be careful when taking limits $\omega,\mathbf{k} \to 0$, since different orders of limits may give different results, and the correct procedure depends on the physical situation. A good comparison of screening and real-time quantities in the context of solvable 2+1-dimensional Gross--Neveu model can be found in \cite{Huang:1995yt}.

\section{Mesonic screening states at high temperatures}

The Euclidean Lagrangian governing the behavior of quarks and gluons at finite temperature is
\begin{equation}
 \mathcal{L}_\mathrm{E} = \frac{1}{4}F_{\mu\nu}^a F_{\mu\nu}^a + \bar{\psi}(\gamma_\mu D_\mu + M)\psi\,,
\end{equation}
where the covariant derivative is defined as $D_\mu \psi \equiv \partial_\mu \psi -ig A_\mu^a T^a \psi$ and the gluon field strength $F_{\mu\nu} = i/g[D_\mu,D_\nu]$, $T^a$ being the generators in the fundamental representation of SU($N_c$). We consider $N_F$ flavors of degenerate quarks, so the quark field $\psi$ is an $N_F$-component vector in flavor space, and the mass matrix $M$ is proportional to unit matrix, $M=m \cdot 1_{N_F}$. For simplicity we will set $m=0$ in most of what follows.

Finite quark densities are included through chemical potentials $\mu_\mathrm{f}$ multiplying the quark number density operators $\mathcal{N}_\mathrm{f} = \bar{\psi}_\mathrm{f} \gamma_0 \psi_\mathrm{f}$. This is precisely the same structure as in the momentum time-component, so the actual effect when doing computations in perturbation theory is a shift in $p_0$, which in absence of chemical potential woud be one of the fermionic Matsubara frequencies, $i[(2n+1)\pi T] \to i[(2n+1)\pi T -i\mu_\mathrm{f}]$. In pure QCD all quark numbers are conserved separately, so the chemical potential for each flavor can be chosen independently. Weak interactions, on the other hand, mix different quark flavors, mostly inside SU(2) doublets but also between families, because the weak interaction eigenstates differ from the mass eigenstates (for a brief review on this mixing, see \cite{Yao:2006px}). Because of this, only certain combinations of the baryon number and the different lepton numbers are conserved in the full standard model. We assume that the time scales of chemical equilibration through weak interactions are much larger than the characteristic time scales of QCD processes we are studying, even when discussing static correlators, and continue to use independent chemical potentials for ech flavor. In the numerical studies we will use two distinct cases, isoscalar ($\mu_\mathrm{u}=\mu_\mathrm{d}\equiv \mu_S$) and isovector ($\mu_\mathrm{u}=-\mu_\mathrm{d} \equiv \mu_V$) chemical potentials for illustration, but the analytical results are applicable to general $\mu_\mathrm{f}$.

The quark fields $\bar{\psi}$,$\psi$ can be used to define mesonic operators of different spin and flavor structures. These operators can be thought of either as the currents coupling to external perturbations as in Eq.~(\ref{eq:current_source_coupling}) or as interpolating operators for physical particle states, although in the latter case it should be remembered that the real-time mesonic bound states do not survive at very high temperatures, and that their properties may be very different from the corresponding screening states. We denote
\begin{equation}
 O^a = \bar{\psi} F^a \Gamma \psi,
\label{eq:general_bilinear}
\end{equation}
where $\Gamma$ is one of $\{ 1,\gamma_5,\gamma_\mu,\gamma_\mu \gamma_5 \}$ for scalar, pseudoscalar, vector and axial vector objects $O^a=S^a, P^a, V_\mu^a, A_\mu^a$, respectively. The flavor strucure is written in terms of the identity matrix $F^s$ and the traceless matrices $F^a$, which satisfy
\begin{equation}
 F^s \equiv 1_{N_F}, \quad \Tr[F^a F^b] = \half \delta^{ab}, \quad a,b=1,\ldots,N_F^2-1\, .
\end{equation}

We wish to compute the static correlators of the above operators. The particular correlators we are interested in are defined as
\begin{equation}
 C_\mathbf{q}[ O^a, O^b] \equiv \int_0^{1/T}\! \dd \tau \int \! \dd^3 x \, e^{i\mathbf{q\cdot x}}
\langle O^a(\tau,\mathbf{x}) O^b(0,\mathbf{0})\rangle.
\label{eq:corr_def_q}
\end{equation}
The quantity in this equation seems different from the retarded propagator defined in Eq.~(\ref{eq:delta_J_coord}). The number of relevant propagator-like functions at finite temperature is large because of the possibilities of having either real or imaginary time coordinates as well as different orderings of the operators, but they are all related to the spectral function $\rho(\omega,\mathbf{q})$, which in turn is given by the analytic continuation of the Euclidean correlator in Eq.~(\ref{eq:corr_def_q}). All these relations can be found in standard textbooks (see e.g.~\cite{Lebellac:thermal}), and a good summary is given in \cite{Laine:2006ns}. The relevant relation for our computations is
\begin{equation}
 D_R(q_0=0,\mathbf{q}) = \int_{-\infty}^\infty \frac{\dd \omega}{\pi} \frac{\rho(\omega,\mathbf{q})}{\omega-i\epsilon}
 = \lim_{q_0 \to 0+} \int_{-\infty}^\infty \frac{\dd \omega}{\pi} \frac{\rho(\omega,\mathbf{q})}{\omega-iq_0}
 = \lim_{q_0 \to 0+} C_E(q_0,\mathbf{q}),
\end{equation}
where $C_E(q_0,\mathbf{q})$ is the Euclidean correlator in Eq.~(\ref{eq:corr_def_q}) with an arbitrary (Euclidean) momentum zero-component $q_0$.

We can use the rotational invariance of the system to choose the vector $\mathbf{x}$, the direction in which we are measuring the correlations, to point in the $x_3$ direction, and further average over the transverse $x_1x_2$-plane. The fundamental quantity we are studying here is then the $z$-dependent correlator
\begin{equation}
 C_z[O^a, O^b] = \int \! \dd^2 \xbot C_{(\xbot,z)} [O^a, O^b] = \int_0^{1/T}\!\dd \tau \int \! \dd^2 \xbot 
	\langle O^a(\tau,\xbot,z) O^b(0,\mathbf{0},0) \rangle.
\label{eq:pointplanedef}
\end{equation}

\begin{figure}
\begin{center}
\begin{fmfgraph*}(60,30)
 \fmfleft{l}
 \fmfright{r}
 \fmf{plain,right=0.5,label=(a)}{l,r}
 \fmf{plain,right=0.5}{r,l}
 \fmfv{d.sh=square,d.size=5,d.fill=empty}{l,r}
\end{fmfgraph*}
\hspace{1cm}
\begin{fmfgraph*}(60,30)
 \fmfleft{l}
 \fmfright{r}
 \fmf{plain,right=0.5,label=(b)}{l,r}
 \fmf{plain,right=0.5}{r,l}
 \fmfv{d.sh=square,d.size=5,d.fill=empty}{l,r}
 \fmffreeze
 \fmfipath{p[]}
 \fmfiset{p1}{vpath(__r,__l)}
 \fmfiset{p2}{vpath(__l,__r)}
 \fmfi{photon}{point length(p1)/2 of p1 -- point length(p2)/2 of p2}
\end{fmfgraph*}
\hspace{1cm}
\begin{fmfgraph*}(60,30)
 \fmfleft{l}
 \fmfright{r}
 \fmf{plain,right=0.5,label=(c)}{l,r}
 \fmf{plain,right=0.5}{r,l}
 \fmfv{d.sh=square,d.size=5,d.fill=empty}{l,r}
 \fmffreeze
 \fmfipath{p}
 \fmfipair{v[]}
 \fmfiset{p}{vpath(__r,__l)}
 \fmfiset{v1}{point length(p)/4 of p}
 \fmfiset{v2}{point 3length(p)/4 of p}
 \fmfi{photon}{v1{dir -135}..v2{dir135} }
\end{fmfgraph*}
\hspace{1cm}
\begin{fmfgraph*}(70,30)
 \fmfleft{l}
 \fmfright{r}
 \fmf{plain,right,tension=0.5}{l,v1}
 \fmf{plain,right,tension=0.5}{v1,l}
 \fmf{plain,right,tension=0.5}{r,v2}
 \fmf{plain,right,tension=0.5}{v2,r}
 \fmf{dbl_wiggly,tension=2,label=(d),l.d=21}{v1,v2}
 \fmfv{d.sh=square,d.size=5,d.fill=empty}{l,r}
\end{fmfgraph*}
\end{center}
\caption{Classes of diagrams contributing to the meson correlator. Diagrams of type (d) only apply to flavor singlets.}
\label{fig:meson_diagram_types}
\end{figure}
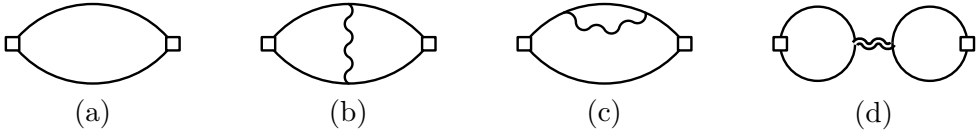

At very high temperature the QCD coupling is small due to asymptotic freedom, and the correlator $C_z$ can be computed using perturbation theory. The leading order result is given by the free theory diagram Fig.~\ref{fig:meson_diagram_types}(a) consisting of two noninteracting quarks propagating in the hot medium.
When the chemical potentials all vanish, this diagram can be computed in the momentum space, and apart from constants corresponding to terms $\sim \delta(\mathbf{x})$ in the coordinate space, the result is proportional to the function
\begin{equation}
 B_\mathrm{3d}(2\omega_n) \equiv \int\!\frac{\dd^3 p}{(2\pi)^3} 
	\frac{1}{[\omega_n^2+\mathbf{p}^2][\omega_n^2+(\mathbf{p}+\mathbf{q})^2]}
 = \frac{i}{8\pi q}\ln \frac{2\omega_n -iq}{2\omega_n +iq}\, ,
\end{equation}
summed over all fermionic momenta $\omega_n$ with coefficients that depend on the spin structure of the operator in question. At large distances, or low momenta, the behavior of the correlator is dominated by the lowest singularities of this expression, located at $q=\pm i2\pi T$. If we for a moment imagine rotating the three-dimensional theory of the lowest Matsubara modes to 2+1 dimensions and treating the direction of $\mathbf{q}$, or $x_3$, as a time coordinate, this singularity becomes a branch cut on the real $q_3$-axis, and can be understood as the threshold of producing two free quarks of mass $\pi T$. When the interactions are turned on, we expect this singularity to convert into a pole corresponding to a bound state in the 2+1-dimensional theory.

It should be noted that the temporal components of vector and axial vector correlators do not have the singularity associated with $B_\mathrm{3d}(2\omega_n)$, since carrying out the Dirac algebra gives these terms a prefactor $q^2+4\omega_n^2$, which regularizes the singularity at $q=\pm 2i\omega_n$. Computing in the configuration space, it is easy to see that these correlators are suppressed by powers of distance and decay even faster, since the contribution from the lowest singularity is removed. The longitudinal components $V_3$ and $A_3$ of those correlators vanish completely, apart from contact terms, because of the current conservation $\partial_\mu V_\mu=0$.

When the chemical potentials are turned on, even the free correlator becomes very hard to compute. This is due to the shift in the temporal momentum components, which causes the correlator to mix different Matsubara modes even after integrating over $\tau$-direction. If the two quarks have identical chemical potentials, which is the case for flavor singlets or isoscalar chemical potential, the correlator $C_z$ in Eq.~(\ref{eq:pointplanedef}) can be computed explicitly. For scalar operator the result is
\begin{eqnarray}
\label{eq:iso_vapaa_skalaari}
 C_z[S^a, S^b] &=& \delta^{ab} \frac{N_c T}{8\pi z \sinh 2\pi T z}\left( 2\pi T \coth 2\pi Tz \, \cos 2\mu z 
	+2\mu \sin 2\mu z +\frac{1}{z} \cos 2\mu z \right) \qquad \quad \\
 &=& \delta^{ab} \frac{N_c}{8\pi^2}\frac{1}{z^3}\left[ 1 -\left( \frac{7}{360} +\frac{1}{6}\frac{\mu^2}{\pi^2 T^2}
	+\frac{1}{12}\frac{\mu^4}{\pi^4 T^4}\right)(2\pi Tz)^4 +\mathcal{O}(z^5) \right] \nonumber \\
 &=& \delta^{ab} \frac{N_c T^2}{2}\frac{1}{z}e^{-2\pi Tz} \left[ \left(1+\frac{1}{2\pi Tz}\right)\cos 2\mu z 
	+\frac{\mu}{\pi T} \sin 2\mu z \right] +\mathcal{O}(e^{-4\pi Tz}), \nonumber
\end{eqnarray}
where we have also indicated the limiting behavior at small and large distances, respectively. For $\mu=0$ this result agrees with the previous zero density computation in \cite{Florkowski:1993bq}. The effect of a small chemical potential on the correlator is seen as oscillations with wavelength $l_\mu =\pi/\mu$ inside the zero density envelope, while for a large $\mu \gg \pi T$ the interaction with the particle bath is so strong that correlator oscillates wildly, averaging to zero.

The case of arbitrary chemical potentials is much harder because of the summations mixing different modes, but a good approximation can be found by only taking into account the lowest modes $\omega_n=\pm\pi T$, which dominate the correlator at large distances. The scalar correlator can then be written in terms of exponential integral function $\Ei(z)$, which we approximate at large $z$ to leading order in $1/z$, resulting in
\begin{eqnarray}
 C_z[S^a, S^b] &\approx & \sum_\mathrm{ij}F^a_\mathrm{ij}F^b_\mathrm{ji} \frac{N_c T^2}{2\pi}\left( 1-e^{-\Delta\mu_\mathrm{ij}/T} \right)
	\frac{1}{z}e^{-2\pi Tz}\left[ \left(\frac{1}{\Delta\mu_\mathrm{ij}} - \frac{\Delta\mu_\mathrm{ij}}{(2\pi T)^2 +
	\bar{\mu}_\mathrm{ij}^2}\right)\bar{\mu}_\mathrm{ij} \sin \bar{\mu}_\mathrm{ij}z \right. \nonumber \\
 &&{}\left. +\left(\frac{1}{\Delta\mu_\mathrm{ij}} + \frac{\Delta\mu_\mathrm{ij}}{(2\pi T)^2
	+\bar{\mu}_\mathrm{ij}^2}\right)2\pi T \cos \bar{\mu}_\mathrm{ij}z \right],
\label{eq:vapaayleinen_appr}
\end{eqnarray}
where $\Delta\mu_\mathrm{ij}\equiv \mu_\mathrm{i}-\mu_\mathrm{j}$ and $\bar{\mu}_\mathrm{ij}\equiv \mu_\mathrm{i}+\mu_\mathrm{j}$. Details of the computation and a more complete result for the general case can be found in the Appendix A of \cite{Vepsalainen:2007ke}. In this correlator the scale of oscillations is $\bar{\mu}_\mathrm{ij}$, a direct generalization of the isoscalar case. In particular, for isovector chemical potentials the oscillation vanishes (although it reappears at terms suppressed by $1/z$ as $\sin\Delta\mu_\mathrm{ij}z$, see \cite{Vepsalainen:2007ke}) and the correlator has the same functional form as in the $\mu=0$ case.

\begin{figure}
\centering
\includegraphics[width=0.92\textwidth]{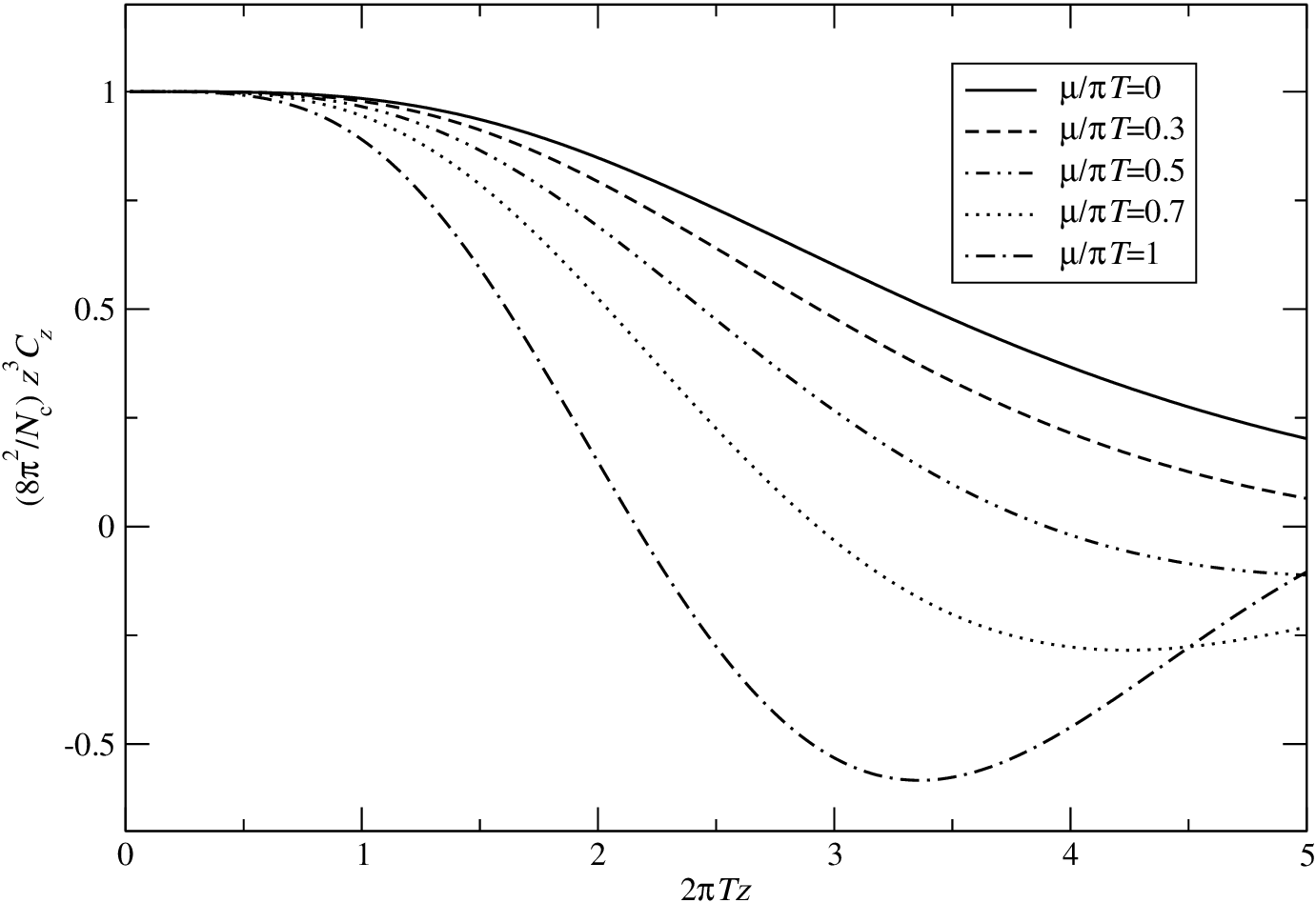}
\caption{The free scalar meson correlator for different values of isoscalar chemical potential.}
\label{fig:vapaa_scaled}
\end{figure}

Together all these results show that the free correlator falls off as $\exp(-2\pi Tz)$, regardless of chemical potentials. The finite density only shifts the singularity in $q_3$ by an imaginary part, which shows as oscillations in $C_z$. We define the screening mass as the coefficient of this exponential fall-off, or the real part of the pole location, because that is what determines the asymptotic behavior of the correlator. In Fig.~(\ref{fig:vapaa_scaled}) we have plotted the isoscalar correlator Eq.~(\ref{eq:iso_vapaa_skalaari}) for different values of chemical potential. The figure shows that when the wavelength of the oscillations $l_\mu =\pi/\mu$ is large, $\mu \lesssim 0.5\pi T$, the oscillatory behavior becomes apparent only at distances where the correlator is exponentially suppressed, and it is hard to discern the cosine term from an increased exponential fall-off. For larger $\mu$ this difference is obvious, and even more so when studying the asymptotic behavior of the correlator, so in the leading order of perturbation theory $m=2\pi T +\order{g^2}$ is the consistent definition also at finite density.

\section{Effective theory for the lowest fermionic modes}

Going beyond the leading order, the location of the singularity in the static meson correlator is modified by interactions with gluons. We would like to compute in perturbation theory the next-to-leading order corrections to the screening masses, which requires computing the diagrams of types Fig.~\ref{fig:meson_diagram_types}(b,c). Note that the flavor singlet correlations are also mediated by purely gluonic states, some of which have masses lower than $2\pi T$. These states are completely non-perturbative, so we only concentrate in this work on flavor non-singlets. The couplings of different quark operators to glueballs have been worked out in \cite{Laine:2003bd} and the masses of these gluonic operators for both zero and finite chemical potentials have been measured in \cite{Hart:2000ha}.

Computing the one-gluon diagrams is not enough, however, since one runs here into the same infrared problems as always when computing with light particles at finite temperature. In particular, for soft momenta (which implicitly requires bosonic zero mode $\omega_n=0$) the one-loop correction to the gluonic zero mode propagator is of the same order as the leading term and needs to be resummed into the propagator to get rid of infrared singularities arising from soft gluons. As discussed in previous chapters, at order $g^2$ this resummation gives the static temporal gluon component an electric mass $\mE\sim gT$, which we have to include in the soft gluon propagators of diagrams Fig.~\ref{fig:meson_diagram_types}(b,c).

Another class of diagrams requiring resummations consists of the graphs with soft gluon exchanges, as shown in Fig.~\ref{fig:meson_diagram_types}(b). The integration over $q_3$ when going to coordinate space gets the largest contribution from the poles at $q_3\approx \pm i\pi T$, and if the gluon momentum is small, the additional quark and gluon propagators in the diagram are also nearly on-shell, $1/\slashed{p} \sim \order{1/g^2 T}$, compensating for the factors of $g$ from the vertices. To get a consistent next-to-leading order correction we therefore have to sum over all diagrams with an arbitrary number of soft gluon exchanges. This is a common requirement for having a bound state in theory, since summing only a finite number of diagrams with free propagators cannot give rise to new singularities, and it applies as well to the screening states under discussion. We wish to avoid the exceedingly complicated formalism of relativistic bound states, and instead make use of the hierarchy between the Matsubara modes $\pi T$ and the momentum scales where the resummations become necessary, $p\sim gT$. This suggests using an effective theory to organize the resummations of the soft modes, while simply including the effects of the large momenta in the parameters.

The leading order computation shows that at large distances the screening correlator is dominated by the lowest Matsubara modes $\pm\pi T$, while the contribution coming from the other modes is exponentially suppressed. We concentrate on the lowest modes and integrate over all excitations of momenta $\sim \pi T$ around these modes, which gives an effective theory only containing the static ($\omega_n=0$) gluons and the fermionic modes with $\omega_n=\pm\pi T$. In particular, all the other fermionic states are off-shell by $2\piT \gg gT$, so this theory does not contain any creation or annihilation of quarks, but the only quark lines are those entering and leaving the diagrams as external legs. Because we are computing near the screening pole, these external quarks are almost on-shell, $p^2=0$, and the interactions with soft gluons do not change that very much. The relevant expansion parameter is then the ``off-shellness'' $p_0^2+\mathbf{p}_\bot^2 +p_3^2 \sim g^2 T^2$, which is of the same order as the momenta of the soft gluons. If we rotate the 3-dimensional action of the quark modes into the 2+1-dimensional Minkowski space, the Matsubara mode $\pm\piT$ can be viewed as a heavy quark mass, and the restriction to momenta much lower than this and the separation of quarks from antiquarks effectively makes the quarks non-relativistic in 2+1 dimensions. Note that there is nothing special in the lowest Matsubara modes, but we could equally well derive a similar theory for modes $\omega_n=3\piT$, for example. The reason we have singled out the lowest modes is because they dominate the screening correlator, which we aim to compute.

The difference between the purely gluonic EQCD and the effective theory considered here is the state around which the expansion takes place, and is dictated by the physical application we have in mind. EQCD is directed at computing either vacuum diagrams or gluonic correlators, which do not have any quarks on external legs. Any fermionic excitation is then off-shell at least by $2\piT$, and only contributes at the high-momentum integration. The mesonic operators, on the other hand, are built out of quark fields, which cannot be integrated out if we intend to compute with them. Still, given an external quark state, any additional quarks would again be very much off-shell, so expanding around that state in low momenta we can neglect all other fermionic excitations. To compare with QED, the relation between the two theories roughly corresponds to the Euler--Heisenberg effective Lagrangian for photon-photon interactions and the non-relativistic quantum mechanics used to compute hydrogen (or positronium) binding energies. In the former the electrons have been integrated out completely, whereas in the latter we can mostly ignore the contribution of the states with e.g.\ two electrons and a positron, not because the mass would be much higher ($3m_e$ vs.\ $m_e$), but because they are off-shell by $\sim 2m_e$, which is large compared to typical momenta $\sim \alpha m_e$.

The bosonic sector of the dimensionally reduced QCD is well-known \cite{Ginsparg:1980ef,*Appelquist:1981vg}, and in finite density it reads \cite{Hart:2000ha}
\begin{equation}
 \mathcal{L}_\mathrm{eff}^\mathrm{b} = \half \Tr F_{ij}^2 + \Tr [D_i,A_0]^2 + \mE^2 \Tr A_0^2 
	+\frac{ig^3}{3\pi^2}\sum_\mathrm{f} \mu_\mathrm{f} \Tr A_0^3
	+ \lambda^{(1)}_\mathrm{E}\left( \Tr A_0^2\right)^2 +\lambda^{(2)}_\mathrm{E}\Tr A_0^4\,,
\label{eq:su3_bosonic_L}
\end{equation}
where we have also included the cubic $A_0$ self-coupling proportional to quark chemical potentials. The parameters are found by matching gluonic 2- and 3-point functions. We only need the adjoint scalar mass to one-loop level and other couplings at tree level, so
\begin{equation}
 \mE^2 = g^2 T^2\left( \frac{N_c}{3} +\frac{N_F}{6} + \frac{1}{2\pi^2}\sum_\mathrm{f} \frac{\mu_\mathrm{f}^2}{T^2} \right),
 \qquad \gE^2 = g^2 T,  \qquad \lambda^{(1,2)} = \mathcal{O}(g^4 T). 
\end{equation}
The cubic and quartic $A_0$ self-interactions can be ignored, since they contribute to meson correlators only at order $g^6$ and higher. The bosonic theory in Eq.~(\ref{eq:su3_bosonic_L}) describes a 3-dimensional gauge theory with a massive adjoint scalar $A_0^a$, and has been extensively used to compute gluonic quantities at high temperatures, as discussed in the beginning of this chapter.

On the fermionic sector the tree-level Lagrangian is just a sum over the parts of the full theory Lagrangian containing modes $\omega_n=\pm\piT$. For a single mode $\omega_n$ this term reads
\begin{equation}
\mathcal{L}_\mathrm{q} = \bar{\psi}\left[ i\gamma_0 \omega_n +\gamma_0 \mu -ig\gamma_0 A_0 +\gamma_k D_k +\gamma_3 D_3\right]\psi,
\label{eq:fermion4d}
\end{equation}
where $k=1,2$ and $A_0$ is the gluonic zero mode. We have separated the $x_3$-direction from the other spatial components, anticipating the choice to measure correlations in that direction. Note that the interaction with static gluons does not mix different fermion Matsubara modes, so we have a separate term like Eq.~(\ref{eq:fermion4d}) for each mode we wish to compute with.

Using a non-standard representation for Dirac matrices (see \cite{Laine:2003bd,Vepsalainen:2007ke} for details) and decomposing the four-component spinor as
\begin{equation}
 \psi = \left( \begin{array}{c} \chi \\ \phi \end{array} \right) ,
\end{equation}
the Lagrangian can be written in a form where the fields $\chi$ and $\phi$ are light and heavy close to the pole $p_3=ip_0=i(\omega_n-i\mu)$, respectively, while the roles are reversed at the other pole $p_3=-ip_0$. Solving the equation of motion for the heavy component and expanding the resulting non-local operators in $1/p_0$, we get the non-relativistic Lagrangian
\begin{eqnarray}
 \mathcal{L}_\mathrm{q} &\approx & i\chi^\dagger\left[ p_0 -gA_0 +D_3 -\frac{1}{2p_0}\left( D_\bot^2 
	+\frac{g}{4i} [\sigma_i, \sigma_j] F_{ij} \right) \right] \chi \nonumber \\
 && {}+i\phi^\dagger\left[ p_0 -gA_0 -D_3 -\frac{1}{2p_0}\left( D_\bot^2 + \frac{g}{4i}[\sigma_i,\sigma_j]F_{ij}\right)
	\right] \phi +\mathcal{O}\left( \frac{1}{p_0^2} \right).
\label{eq:treelevelL}
\end{eqnarray}
All dependence on the chemical potential at this level is contained in the shift of the temporal momentum component, $p_0=\omega_n-i\mu$. The Lagrangian is easier to understand if we again imagine rotating to 2+1 dimensions and setting $z=it$. The zero-point energy is then given by $p_0$, while the other free terms combine to $-(i\partial_t+\nabla^2/2p_0)$, the standard nonrelativistic kinetic term with mass $p_0$. If we forget the relativistic origin of these terms, there is no reason for the zero-point energy and the mass parameter in the kinetic term to be the same, and as we will see shortly, the loop corrections will give these parameters different values. It should be noted that for $\omega_n>0$ the field $\phi$ has negative mass, and should be interpreted as the antiparticle of $\chi$.

Already at the tree-level the expansion in $1/p_0$ gives rise to an infinite number of terms, and beyond this we will have to take into account all possible terms allowed by symmetries. To limit the possibilities, a power counting has to be established. Requiring all the transverse momenta to be at most of the order of the electric mass, $\mathbf{p}_\bot \lesssim gT$, and that the terms in the action to be of order unity, we get
\begin{equation}
 \chi \sim 1/|\xbot| \sim gT \,, \qquad  A \sim (x_3/\xbot^2)^{1/2} \sim g^{1/2} T^{1/2}\, .
\end{equation}
The off-shellness $\Delta p_3 \equiv p_3\pm ip_0$ corresponds to the kinetic energy in the 2+1 dimensional theory, as can be verified from the poles in the quark propagators, Eqs.~(\ref{eq:chipropagator}),(\ref{eq:phipropagator}). For nearly on-shell quarks we can then estimate the derivative $\partial_3$ by
\begin{equation}
 \Delta p_3 \sim \mathbf{p}_\bot^2/p_0 \sim g^2 T  \quad \Rightarrow \quad  \partial_3 \sim g^2 T \qquad \textrm{acting on quarks}.
\end{equation}
Using this power counting, and keeping only terms required to give the screening mass to order $g^2$, the final form of the fermionic Lagrangian is
\begin{equation}
 \mathcal{L}_\mathrm{eff}^\mathrm{f} = i\chi^\dagger\left( M -\gE A_0 +D_3 -\frac{\nabla_\bot^2}{2p_0} \right)\chi 
	+i\phi^\dagger\left( M -\gE A_0 -D_3 -\frac{\nabla_\bot^2}{2p_0} \right)\phi\, ,
\label{eq:final_nrqcd}
\end{equation}
where only the zero-point energy, which we will henceforth denote by $M$, needs to be matched beyond tree-level. This matching is carried out by comparing the poles in the one-loop corrected quark propagator depicted in Fig.~\ref{fig:quark_se}. The details of the computation are given in the original papers, and the result is simply
\begin{figure}
\begin{center}
\begin{fmfgraph*}(80,30)
 \fmfset{arrow_len}{3mm}
 \fmfleft{l}
 \fmfright{r}
 \fmf{fermion,label=$p$}{l,v1}
 \fmf{fermion,label=$p-q$,tension=0.5}{v1,v2}
 \fmf{fermion}{v2,r}
 \fmf{photon,left,tension=0,label=$q$,label.side=right}{v1,v2}
\end{fmfgraph*}
\end{center}
\captionsetup{aboveskip=0pt}
\caption{One-loop correction to the quark self-energy.}
\label{fig:quark_se}
\end{figure}
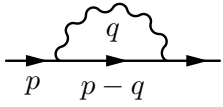
\begin{equation}
 M = p_0 +g^2\CF \frac{T^2}{8p_0}\left(1+\frac{\mu^2}{\pi^2 T^2}\right) =
	\omega_n -i\mu +g^2\CF \frac{T^2}{8(\omega_n-i\mu)}\left(1+\frac{\mu^2}{\pi^2 T^2}\right).
\label{eq:matched_M}
\end{equation}
It should be noted that for $\omega_n=\pm\piT$, which are the modes of interest here, the real part of $M$ does not depend on $\mu$.

The free quark propagators following from Eq.~(\ref{eq:final_nrqcd}) are
\begin{eqnarray}
 \langle \chi_u(p)\chi^*_v(q) \rangle &=& \delta_{uv}(2\pi)^3\delta(p-q) \frac{-i}{M+ip_3+\mathbf{p}_\bot^2/2p_0}
	\label{eq:chipropagator} \\
 \langle \phi_u(p)\phi^*_v(q) \rangle &=& \delta_{uv}(2\pi)^3\delta(p-q) \frac{-i}{M-ip_3+\mathbf{p}_\bot^2/2p_0}
	\label{eq:phipropagator}
\end{eqnarray}
or in the configuration space
\begin{eqnarray}
 \langle \chi_u(x)\chi^*_v(y) \rangle &=& -i\delta_{uv}\theta( \omega_n(x_3-y_3)) \frac{p_0}{2\pi(x_3-y_3) }
	e^{-M(x_3-y_3)-\frac{p_0(\xbot -\mathbf{y}_\bot)}{2(x_3-y_3)} } \\
 \langle \phi_u(x)\phi^*_v(y) \rangle &=& -i\delta_{uv}\theta( \omega_n(y_3-x_3)) \frac{p_0}{2\pi(y_3-x_3) }
	e^{-M(y_3-x_3)-\frac{p_0(\xbot -\mathbf{y}_\bot)}{2(y_3-x_3)} }.
\end{eqnarray}
In these equations it is obvious that for $\omega_n>0$ the field $\chi$ propagates forward and $\phi$ backward in $x_3$, the time coordinate of the 2+1-dimensional theory, confirming our interpretation of $\phi$ as the antiparticle of $\chi$. For negative modes these roles are reversed, with $\phi$ propagating forward in $x_3$. In \cite{Laine:2003bd} we expanded these propagators in $\mathbf{p}_\bot/p_0$ inside loop integrals to make sure that the transverse momenta are parametrically smaller than the heavy scale $\sim T$. However, as discussed in \cite{Luke:1999kz}, while this works well for (single) heavy quark effective theory, in NRQCD there are difficulties at two-loop level with this approach, and it is preferable to keep the kinetic terms summed into the propagators. For our modest purposes there is no real difference, but in \cite{Vepsalainen:2007ke} we chose not expand in transverse momenta. For consistency, this requires that we expand the gluon field in multipole expansion, which at this level boils down to disallowing transverse momentum transfer from gluons to quarks. The masses and the quark-antiquark potential turn out to not depend on the way we treat the propagators, whereas in order to compute the full correlators it is necessary to keep the kinetic terms resummed. As a check, we have shown that \NRQCD{} is able to reproduce the leading order scalar correlator in Eq.~(\ref{eq:iso_vapaa_skalaari}) exactly.

In terms of the new fields $\chi$ and $\phi$ the operators whose correlators we intend to compute can be written as
\begin{equation}
\begin{array}{rrcl}
 \ds S  :& \ds \quad \bar \psi \psi & = & \ds \chi^\dagger \phi + \phi^\dagger \chi \, , \\
 \ds P  :& \quad \bar \psi \gamma_5 \psi & = &  \chi^\dagger \sigma_3 \phi - \phi^\dagger \sigma_3 \chi \, , \\
 \ds V_0 :& \quad \bar \psi \gamma_0 \psi  & = &  \chi^\dagger \chi + \phi^\dagger \phi \,, \\
 \ds V_k :& \quad \bar \psi \gamma_k \psi & = & -\epsilon_{kl} (\chi^\dagger \sigma_l \phi -\phi^\dagger \sigma_l \chi), \\
 \ds V_3 :& \quad \bar \psi \gamma_3 \psi  & = &  i(\chi^\dagger \chi - \phi^\dagger \phi), \\
 \ds A_0 :& \quad \bar \psi \gamma_0 \gamma_5 \psi & = &  \phi^\dagger \sigma_3 \phi -\chi^\dagger \sigma_3 \chi \,, \\
 \ds A_k :& \quad \bar \psi \gamma_k \gamma_5 \psi  & = &  -i (\chi^\dagger \sigma_k \phi + \phi^\dagger \sigma_k \chi), \\
 \ds A_3 :& \quad \bar \psi \gamma_3 \gamma_5 \psi & = & -i (\chi^\dagger \sigma_3 \chi + \phi^\dagger \sigma_3 \phi).
\end{array}
\label{eq:opers_in_3d}
\end{equation}
It should be noted that the temporal and longitudinal components of vector and axial vector currents consist of terms like $\chi^\dagger\chi$ and $\phi^\dagger \phi$, so their correlators are proportional to $\theta(z)\theta(-z)$ and vanish at nonzero distances. For $V_3$ and $A_3$ this follows from current conservation, whereas the correlators for charges $V_0$ and $A_0$ are power-suppressed, as discussed in the previous section. Apart from those operators, the correlator in the effective theory is independent of the spin structure, up to a multiplicative constant, since $(\sigma_i)^2$. The flavor structure on the other hand is significant if we allow for finite chemical potentials.

\section{Solving the screening states}

Having derived the Lagrangian for \NRQCD, which is just the sum of the bosonic part in Eq.~(\ref{eq:su3_bosonic_L}) and the fermionic part in Eq.~(\ref{eq:final_nrqcd}), it remains to find the masses of mesonic operators. These correspond to bound states in the 2+1-dimensional theory, following from the summation of diagrams of type Fig.~\ref{fig:meson_diagram_types}(b,c) with an arbitrary number of soft gluon exchanges. In a nonrelativistic theory the resummation can be carried out by finding the static potential for a $\chi^*\phi$ pair, and then solving the resulting Schr\"odinger equation with this potential. Here ``static'' should be understood from the 2+1-dimensional point of view, which means that the potential will be valid for large $x_3$. All these computations can be performed using the effective theory just derived, since it is only the $\omega_n=0$ gluons with low momenta that need to be summed beyond the leading order.

The static potential is written as an expansion in $\gE^2r$,
\begin{equation}
 V(r) \sim \gE^2 \ln r + \gE^4 r + \mathcal{O}(\gE^6 r^2).
\end{equation}
Using either the power counting $1/r \sim \mathbf{p}_\bot \lesssim gT$ or the leading order Schr\"odinger equation we see that $\gE^2 r \sim g$. The leading Coulomb-type term $\gE^2 \ln r$ in the potential is then sufficient for computing $\order{g^2}$ corrections to screening masses, and can be evaluated perturbatively by computing all one-gluon diagrams in the effective theory. The leading logarithmic term already gives a confining potential, so there is no qualitative difference in dropping the linear term, which is parametrically of order $g^3$. 

The static potential of a $\phi^*\chi$ pair is computed by inserting a point-splitting in the correlator to give the quarks a small spatial separation in the transverse direction, and finding the Schr\"odinger-type equation satisfied by this correlator at $z\to\infty$ limit. The details of this computation can be found Appendix B of \cite{Vepsalainen:2007ke}, and the result is
\begin{equation}
 V(\mathbf{r}) = \frac{\gE^2 \CF}{2\pi}\left( \ln\frac{\mE r}{2} +\gamma_E -K_0(\mE r) \right),
\label{eq:potential}
\end{equation}
where $K_0$ is a modified Bessel function. The result is both ultraviolet and infrared finite once we have resummed the gluon self-energy corrections to an electric mass $\mE$, while letting $\mE\to 0$ we would again find the infrared divergences of the original theory. The screening masses at this order are not sensitive to the magnetic screening of spatial gluons, as this would manifest itself as divergences in the potential. The potential in Eq.~(\ref{eq:potential}) depends on the quark chemical potentials only through $\mE$, as the leading order is only sensitive to the propagation of gluons in the hot medium while the quarks simply act as static color charges.

The Schr\"odinger equation satisfied by the correlator was already found as an intermediate result when computing the potential, and it reads
\begin{equation}
 \left[ \pm(M_\mathrm{i} +M_\mathrm{j}) -\frac{1}{\pm 2\bar{p}_\mathrm{0ij}}\nabla^2_\mathbf{r}
	+ V(\mathbf{r}) \right]\Psi_0 = m_\mathrm{full}\Psi_0,
\label{eq:schrodinger}
\end{equation}
where we have separated the variables as
\begin{equation}
 C(\mathbf{r},z) = \Psi_0(\mathbf{r})e^{-m_\mathrm{full}z}\, ,
\end{equation}
and the $\pm$ signs apply for $\omega_n=\pm \piT$, respectively. The flavors of the quarks forming the meson are labelled with indices $i,j$, and in general the flavor symmetry between different mesons is broken by the different chemical potentials. The parameters $M$ and $p_0$ are generally complex when computing with finite chemical potentials, but for opposite modes they are related by $M_- = -M_+^*$ and $p_{0-}=-p_{0+}^*$, so the screening masses satisfy by $m_{\mathrm{full},-} = m_{\mathrm{full},+}^*$. Thus we only need to compute the masses for $\omega_n=\piT$, and in addition these relations guarantee that the full correlator, which is the sum over all separate Matsubara modes, behaves as
\begin{equation}
 C_z[O^a, O^b] \propto \sum_\mathrm{ij} F^a_\mathrm{ij}F^b_\mathrm{ji}\, 2\cos[\mathrm{Im}(m_\mathrm{full,ij})z
	-\alpha_\mathrm{ij}]\, \exp[-\mathrm{Re}(m_\mathrm{full,ij})z ],
\end{equation}
where $\alpha_\mathrm{ij}$ is the overall phase of the $\phi_\mathrm{i}^*\chi_\mathrm{j}$ correlator. This is of the same form as the leading order term, with the real part of the mass parameter giving an exponential decay while the imaginary part contributes to a cosine-like oscillation term. In particular, this correlator is real-valued even though the term coming from any single mode is in general complex.

The Schr\"odinger equation with potential Eq.~(\ref{eq:potential}) cannot be solved analytically, so we have to find the eigenvalues $m_\mathrm{full}$ using numerical computations. For this, we cast the equation into a dimensionless form, which depends on the values of the physical parameters only through the dimensionless combinations $\rho$ and $\hat{E}_0$ defined as
\begin{equation}
 \rho \equiv \frac{\gE^2\CF}{\pi \mE^2}\left( \frac{1}{\omega_n-i\mu_\mathrm{i}} + \frac{1}{\omega_n-i\mu_\mathrm{j}} \right)^{-1},
 \qquad \gE^2\frac{\CF}{2\pi}\hat{E}_0 \equiv m_\mathrm{full} -M_\mathrm{i} -M_\mathrm{j}\, .
\label{eq:dimless_parameters}
\end{equation}
One should note in $\rho$ the appearance of the reduced mass, typical of two-body problems. The physical screening mass is given by $\hat{E}_0$ which we solve numerically,
\begin{equation}
 \mathrm{Re}(m_\mathrm{full}) = \mathrm{Re}(M_\mathrm{i}+M_\mathrm{j}) +\gE^2\frac{\CF}{2\pi}\mathrm{Re}(\hat{E}_0)
	= 2\pi T +g^2 T \frac{\CF}{2\pi}\left( \half + \mathrm{Re}(\hat{E}_0) \right),
\label{eq:fullmass}
\end{equation}
where in the last equality we have used the fact that for the lowest modes the real part of $M$ is independent of $\mu$.

The numerical solution is found by assuming that the ground state is cylindrically symmetric, solving the behavior of $\Psi_0(r)$ around the origin and then integrating out to larger $r$ and requiring square integrability. In zero density this is easy since both the wave function and $\hat{E}_0$ are real, and trying different values of $\hat{E}_0$ on the real axis with the condition that $\Psi_0(r)$ vanishes at large distances is very fast. We find, with somewhat superfluous precision,
\begin{equation}
 \setlength\arraycolsep{5pt}
 \hat{E}_0  =  \left\{ \begin{array}{lll} 0.16368014, & \rho = 2/3\, , & (N_F = 0) \\
	0.38237416, & \rho = 1/2 \,, & (N_F = 2) \\
	0.46939139, & \rho = 4/9 \,, & (N_F = 3)
 \end{array} \right.
\label{eq:muless_numbers}
\end{equation}
for different numbers of dynamical quarks $N_F$. The number of flavors only enters in $\mE$, so even the case $N_F=0$ makes sense if the creation and annihilation of quarks is suppressed, as in the so-called quenched simulations on lattice. When the chemical potentials are turned on, the numerical computation becomes more demanding, as all parameters and the wave function have complex values, and the solution has to be searched for in the complex plane instead of limiting to the real axis. Moreover, the dependence on $\mu/\piT$ cannot be solved analytically, but we have to repeat the process for each chemical potential separately.

\begin{figure}
\centering
\includegraphics*[width=0.92\textwidth]{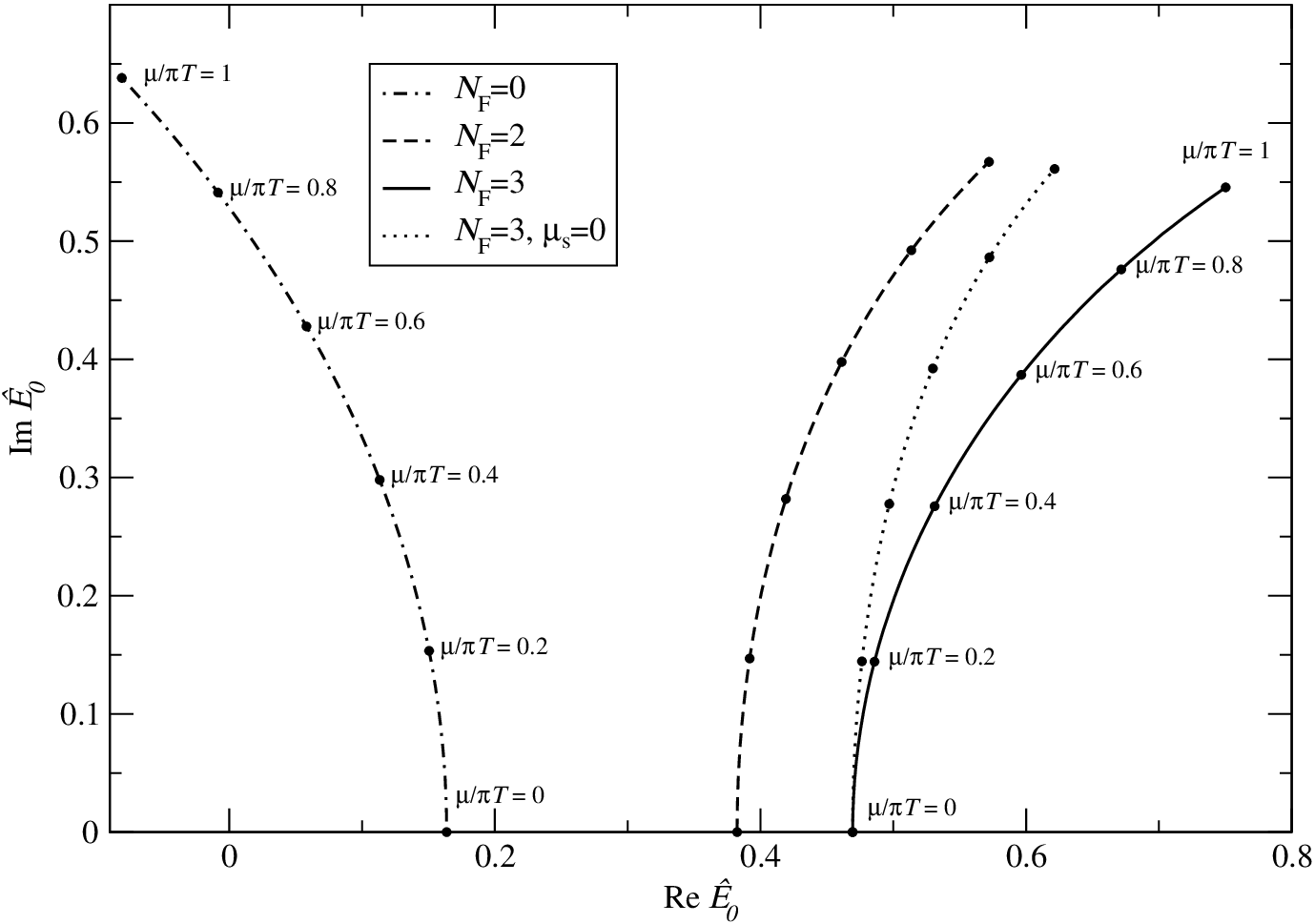}
\caption{The eigenvalue $\hat{E}_0$ with the lowest real part for isoscalar $\mu_S=0\ldots 1$.}
\label{fig:poles}
\end{figure}

In Fig.~\ref{fig:poles} we have plotted the eigenvalues $\hat{E}_0$ for isoscalar chemical potential with different numbers of dynamical quarks. We have also studied the case where $N_F=3$, but only the two flavors in the measured operator have nonzero chemical potentials. This should more closely correspond to the situation in heavy ion collisions, where the two colliding nuclei have finite up and down quark densities but vanishing net strangeness. The figure shows that the poles move along quadratic curves off the real axis, and the real part of $\hat{E}_0$, which is the contribution to the screening mass, grows with $\mu_S$ for dynamical quarks. In the quenched case the real part decreases and becomes negative at $\mu_S\approx 0.78 \piT$. At large values of the chemical potential the oscillations due to the imaginary part of $\hat{E}_0$ become strong and the numerical integration is unstable, so it is hard to go beyond $\mu\sim \piT$ numerically. On the other hand, in deriving the effective theory using dimensional reduction we assume that $T$ is larger than any other mass scale, so our results cannot be trusted for $\mu \gtrsim \piT$.

The chemical potential enters the dimensionless Schr\"odinger equation only through the parameter $\rho$. For the specific cases of isoscalar and isovector chemical potential that we study numerically, the dependence is
\begin{equation}
 \rho \propto \frac{1-i\hat{\mu}_S}{6+N_F+3N_F\hat{\mu}_S^2} \quad \textrm{(isoscalar)}, \quad
 \rho \propto \frac{1+\hat{\mu}_V^2}{6+N_F+3N_F\hat{\mu}_V^2} \quad \textrm{(isovector)}, \quad \hat{\mu}\equiv \mu/\piT\, .
\label{eq:rho_behavior}
\end{equation}
From Fig.~\ref{fig:poles} we can see the twofold influence of the chemical potential. When $N_F=0$, increasing the chemical potential just shifts $\rho$ into more imaginary values, while its real part stays constant. The following increase in the absolute value of $\rho$ decreases Re($\hat{E}_0$) like in the $\mu=0$ case, Eq.~(\ref{eq:muless_numbers}). For dynamical fermions this effect is more than compensated by the increase in $\mE$, which raises the potential and accordingly increases the energy eigenvalues.

\begin{figure}
\includegraphics[width=0.92\textwidth]{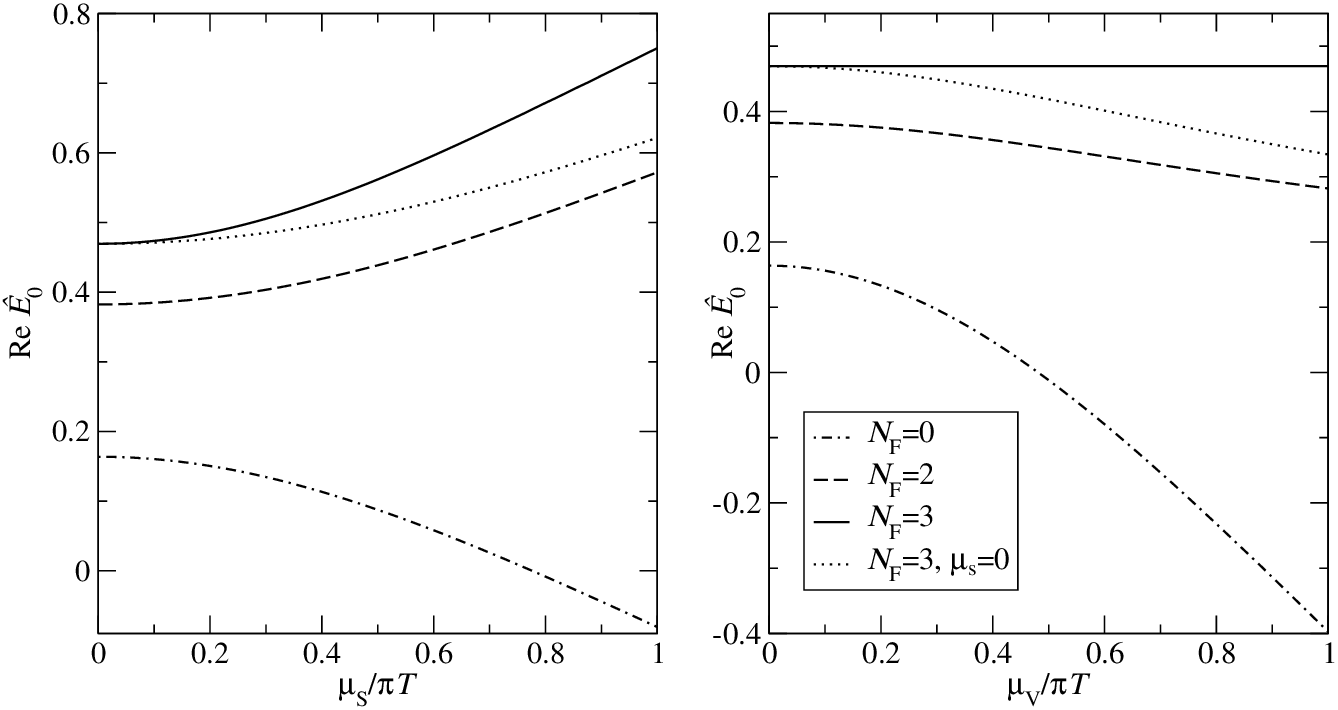}
\caption{The real part of the eigenvalue $\hat{E}_0$ for isoscalar (left) and isovector (right) chemical potentials.}
\label{fig:real_vs_mu}
\end{figure}

In Fig.~\ref{fig:real_vs_mu} we plot the real part of $\hat{E}_0$, which apart from scaling and an additive $\mu$-independent constant is the same as the screening mass, see Eq.~(\ref{eq:fullmass}). The isoscalar data is the same as in Fig.~\ref{fig:poles} discussed above, while for isovector chemical potential we see that the mass decreases with $\mu_V$ for $N_F<3$. This is in agreement with the previous discussion on the relation between $|\rho|$ and $\hat{E}_0$, and Eq.~(\ref{eq:rho_behavior}), where $\rho$ for isovector chemical potential increases with $\mu$ for small number of dynamical fermions.

Numerically the correction we have computed is small, even near the phase transition where the coupling constant itself is large. For $N_F=2$ the range of screening masses for different values of chemical potentials fits in the interval
\begin{equation}
 \mathrm{Re}(m_\mathrm{full}) \approx 2\pi T +\gE^2 \times \left\{ 
	\begin{array}{lr}
	 0.227, & \mu_S/\pi T = 1.0 \\
	 0.187, & \mu/\pi T = 0.0 \\
	 0.166, & \mu_V/\pi T = 1.0 \\
	\end{array} \right.
\label{eq:numbers_with_mu}
\end{equation}
The effective coupling $\gE$ is estimated to be $\gE^2/T \approx 2.2$ when $T\sim 2T_c$ \cite{Laine:2005ai}, giving the next-to-leading order corrections to the screening mass of about 6--8\%. Nevertheless, as long as the coupling is large there is no reason to expect that the next correction would be smaller by a factor of the same magnitude.

\begin{figure}
\centering
\includegraphics[width=0.92\textwidth]{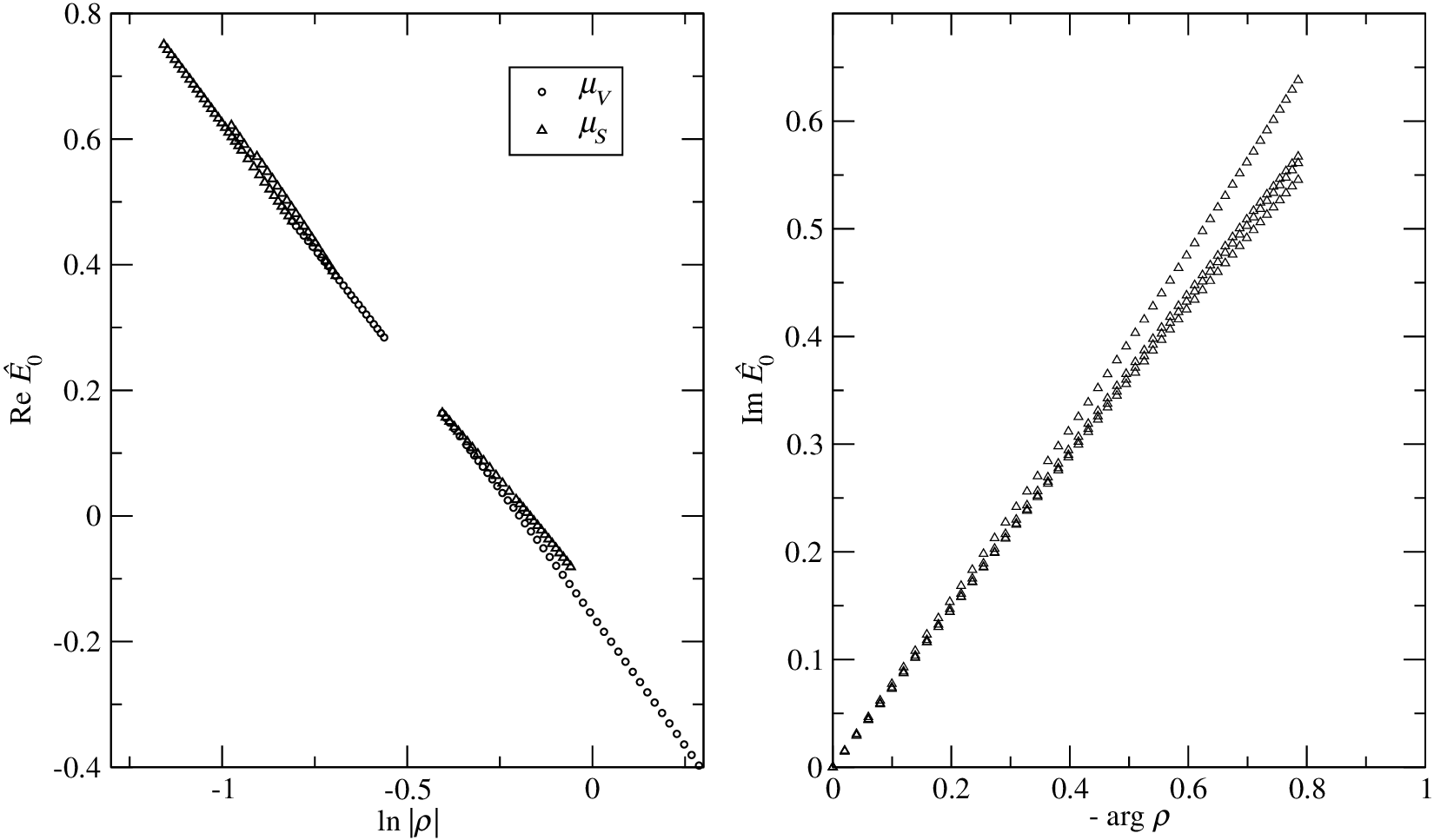}
\caption{The numerical data in Figs.\ \ref{fig:poles} and \ref{fig:real_vs_mu} parametrized by the logarithm of the dimensionless parameter $\rho$.}
\label{fig:logfit}
\end{figure}

While the Schr\"odinger equation Eq.~(\ref{eq:schrodinger}) cannot be solved exactly, we were able to find a simple approximate dependence on the parameters while writing this introductory part. Realizing that the modified Bessel function $K_0(\mE r)$ in the potential interpolates between $-\ln(\mE r/2)$ and 0, we can try to estimate the potential by
\begin{equation}
 V(\mathbf{r}) \approx \frac{\gE^2 \CF}{2\pi}\left( C_1\ln\frac{\mE r}{2} +C_2 \right),
\label{eq:appr_potential}
\end{equation}
where we expect $1\leq C_1\leq 2$. For real values of the dimensionless parameter $\rho$ defined in Eq.~(\ref{eq:dimless_parameters}) its effects can be scaled into the dimensionless variables, giving for the potential in Eq.~(\ref{eq:appr_potential})
\begin{equation}
 \hat{E}_0 (\rho) = \hat{E}_0 (1) -\frac{C_1}{2} \ln\rho\, .
\label{eq:E_as_lnrho}
\end{equation}
In Fig.~\ref{fig:logfit} we have plotted $\hat{E}_0$ vs.\ $\ln \rho$ for the numerical data we have computed, assuming that the expression in Eq.~(\ref{eq:E_as_lnrho}) can be extended for complex values of $\rho$ as well, if the branch cut is introduced on the negative real axis. As the figure shows, for isovector $\mu$ with real $\rho$ the behavior is extremely well described by Eq.~(\ref{eq:E_as_lnrho}), and the agreement is also good for the complex values of $\rho$. Fitting a line to both real and imaginary parts separately, we get slopes -0.78 and -0.74, respectively. To desired accuracy, the screening masses from our computations can then be summarized by the two-parameter fit from the left plot in Fig.~\ref{fig:logfit},
\begin{equation}
 \mathrm{Re}(m_\mathrm{full}) = 2\pi T +g^2 T \frac{\CF}{2\pi}( 0.39 -0.78\ln |\rho| ),
\end{equation}
where $\rho$ can be read from Eq.~(\ref{eq:dimless_parameters}).

\section{Comparison with other results}

The static mesonic correlators considered here have also been measured in the lattice simulations. In zero density there is a long tradition of measuring the screening masses, for recent results see \cite{deForcrand:2000jx,Nomura:2001hv,Wissel:2005pb,Gavai:2006fs}. Early measurements gave large differences between the masses of different spin structure operators, but in recent works the general picture is that already at $\sim 2T_c$ the masses come close to the ideal gas result, with $\rho$ (vector) meson slightly larger than $\pi$ (pseudoscalar) and both lying 5--10\% below the ideal gas result. The correction we have computed above is of the same magnitude, but with different sign. For all $N_F$ the $\mu=0$ results we have computed lie above the free theory result $2\piT$, approaching the ideal gas slowly as $\alpha_s(T)$ gets smaller due to asymptotic freedom.

The systematic errors in lattice simulations are related to dynamical quarks, light quark masses, and the difficulties in going to the infinite volume limit. In \cite{Wissel:2005pb} the analysis of the infinite volume extrapolation is carried out, resulting in slightly higher masses than those measured earlier, but even then the screening masses are clearly below the ideal gas result. It should be noted that our perturbative result is above $2\piT$ also for $N_F=0$, so the potential difficulties with dynamical quarks cannot completely explain the disagreement with lattice measurements. On the other hand, because the strong coupling is large near $T_c$, the higher order perturbative corrections can be as large as the $\order{g^2}$ term we have computed, at least at temperatures within the reach of lattice measurements. At asymptotically high temperatures, however, the perturbative calculation should be valid, so we expect the screening masses to cross above the free theory result at high enough temperature.

Recently the screening masses have also been computed by evaluating the meson spectral function in the HTL resummation scheme and determining the fall-off of the correlator through the spectral function \cite{Alberico:2007yj}. This method is very orthogonal to our computations, but the weak coupling limit of those results is very similar, and can be summarized in our terms by setting $\hat{E}_0=0$ in Eq.~(\ref{eq:fullmass}). The difference can be explained by the soft gluon contributions that were knowingly left out in \cite{Alberico:2007yj}. At temperatures close to the phase transition the meson properties have been studied analytically in the Nambu--Jona-Lasinio model, which gives screening masses well below the free theory result \cite{Florkowski:1997pi}. It should be noted that this method is valid at very different temperature region, and does not necessarily contradict our high-temperature results.

At finite density the lattice measurements of hadronic screening masses have been performed only recently \cite{Pushkina:2004wa}. Simulations at nonzero chemical potentials are difficult because of the complex fermion determinants, so these computations were carried out by expanding the masses as Taylor series in $\mu$ around $\mu=0$ and measuring the derivatives up to second order response. All measurements in \cite{Pushkina:2004wa} were carried out using $N_F=2$ flavors of staggered fermions. The leading order term is just the $\mu=0$ mass which behaves as described above, and the first derivative vanishes for both isovector and isoscalar chemical potentials. For the latter this follows from symmetry properties, whereas the response with isovector $\mu$ is explicitly measured to be zero. In our perturbative calculation the first derivative at $\mu=0$ vanishes as well for both chemical potentials, as required by the symmetry $\hat{E}_0(-\mu)=\hat{E}_0^*(\mu)$.

The second derivatives give the $\mu$-dependence of the masses as measured on lattice. For isoscalar $\mu$ the measured second order response rises steeply at $T_c$ and settles somewhat below $2/T$ at higher temperatures, for both $\pi$ and $\rho$. In the isovector case the response is small and negative, and approaches zero as temperature is raised. At small $\mu$ these results are in qualitative agreement with our perturbative calculation, but for isoscalar the actual numbers differ by orders of magnitude. For comparison, we have fitted quadratic curves to our data to extract the second order response, and the result is
\begin{equation}
 \frac{\dd^2 \mathrm{Re}(\hat{E}_0)}{\dd \hat{\mu}_S^2} = \left\{
	\begin{array}{rl}
	 -0.62, & (N_F=0) \\	0.45, & (N_F=2) \\	0.75, & (N_F=3) \\ 0.34, & (N_F=3, \mu_\mathrm{s}=0)
	\end{array} \right.
 \frac{\dd^2 \mathrm{Re}(\hat{E}_0)}{\dd \hat{\mu}_V^2} = \left\{
	\begin{array}{rl}
	 -1.42, & (N_F=0) \\	-0.31, & (N_F=2) \\	0.00, & (N_F=3) \\ -0.41, & (N_F=3, \mu_\mathrm{s}=0)
	\end{array} \right.
\end{equation}
for isoscalar and isovector, respectively. In terms of physical parameters
\begin{equation}
 T\frac{\dd^2 m}{\dd \mu^2} = \frac{\gE^2\CF}{2\pi^3 T}\frac{\dd^2\mathrm{Re}(\hat{E}_0)}{\dd \hat{\mu}^2},
\end{equation}
which gives second derivatives of order $T\dd^2 m/\dd \mu^2 \sim \pm 0.02$, two orders of magnitude smaller than the value measured on lattice. While the masses in both perturbative and lattice computations are consistent with the free theory result when all possible error sources are taken into account, the difference in the second derivatives with respect to isoscalar chemical potential is striking. One has to be careful, however, to define the mass the same way in both computations before making any comparison.

\subsection{Definition of the screening mass at nonzero density}

The correlator $C_z$ in Eq.~(\ref{eq:pointplanedef}) is a complicated function of $z$ already at leading order, as the exact free theory result in Eq.~(\ref{eq:iso_vapaa_skalaari}) shows. The screening masses appear as poles in the momentum space Green's functions, but in practice the correlator is measured in configuration space, and only exhibits simple exponential behavior at asymptotically large distances, the coefficient of this exponential decay being the screening mass. An effective $z$-dependent screening mass can be defined as
\begin{equation}
 m(z) \equiv -\frac{1}{C_z}\frac{\partial C_z}{\partial z},
\label{eq:effective_mass_def}
\end{equation}
which in free theory approaches the screening mass roughly as $1/z$. In lattice simulations this complication is often removed by measuring the one-dimensional correlations between planar sources. On a finite lattice with periodic boundary conditions the correlator then behaves as $C_\mathrm{1d}(z) \sim \cosh(-mz)$, which is used in fitting to extract $m$ from data.

When chemical potentials are turned on, the situation becomes more complicated. The definition of the effective mass in Eq.~(\ref{eq:effective_mass_def}) is useless because of the oscillations in the correlator which cause $C_z$ to periodically go through zero and to negative values, preventing us from taking the $z\to\infty$ limit. At short distances the fall-off is faster than at $\mu=0$ because of the cosine term $\cos(2\mu z) \approx 1-2\mu^2 z^2$, but this term does not contribute to the decay at longer distances. The asymptotic behavior of the correlator is dominated by the coefficient of the exponential fall-off, which corresponds to the real part of the momentum space pole location. At large distances this coefficient can be extracted from data, but usually one cannot do measurements at arbitrarily large separations because of the finite lattice size and the exponentially small value of the correlator.

When determining the mass from short-distance data the potential oscillatory terms have to be included in the fit in order to get reliable estimates. For example, the free plane-plane correlator at finite chemical potential behaves as $\sim \cos(\bar{\mu}z) \exp(-M z)$ with $\bar{\mu}\equiv \mu_\mathrm{i}+\mu_\mathrm{j}$ and $M=2\piT$. Simply fitting an exponential of the form $C\exp(-mz)$ overestimates the fall-off because of the oscillations, giving
\begin{equation}
 (m-M)(m+M)^2 = \bar{\mu}^2(3m+M) \quad \Rightarrow \quad
	m \approx M\left(1 + \frac{\bar{\mu}^2}{M^2} -\frac{\bar{\mu}^4}{4M^4}\right).
\label{eq:m_fit}
\end{equation}
The derivatives with respect to $\mu$ are also affected by the unfortunate choice of the fitting function. For isoscalar chemical potential $\bar{\mu}=2\mu_S$, and even for the $\mu_S$-independent screening mass $2\piT$ of the free theory Eq.~(\ref{eq:m_fit}) would give
\begin{equation}
 T\frac{\dd^2 m}{\dd \mu_S^2}\Big|_{\mu=0} = \frac{8T}{M} = \frac{4}{\pi},
\end{equation}
which incidentally is of the same magnitude as the derivatives measured in \cite{Pushkina:2004wa}. In the case of isovector chemical potentials $\bar{\mu}=0$ and the oscillations vanish at leading order in $1/z$, so this effect disappears for $\mu_V$.

As discussed above, the actual lattice simulations are carried out by measuring the derivatives of correlators at $\mu=0$ instead of the full correlator at $\mu \neq 0$. The expression for the second derivative of $C(z)=A\exp(-Mz)$, given that the first derivative vanishes at $\mu=0$, is \cite{Choe:2002mt}
\begin{equation}
 \frac{1}{C(z)}\frac{\dd^2 C(z)}{\dd \mu^2} = \frac{1}{A}\frac{\dd^2 A}{\dd \mu^2} -z \frac{\dd^2 M}{\dd \mu^2}
	= -4z^2  -z \frac{\dd^2 M}{\dd \mu^2},
\end{equation}
where in the last equality we have substituted $A=\cos(2\mu z)$ and the whole expression is written in the limit of infinitely long lattice. The oscillations should then show as quadratic $z$-dependence of the second derivative, but at least the data in  \cite{Choe:2002mt} does not support the existence of this kind of term.

Given that the second order response does not show signs of oscillation in lattice simulations, it is not clear if the cosine terms clearly visible at the leading order and in one-loop terms survive at the non-perturbative level. However, understanding the potentially complicated behavior of the correlator at short distances is necessary to extract reliable information from data measured at finite distances. In particular, the oscillatory terms suggested by the perturbative calculations should probably be taken into account when fitting the correlator at finite chemical potentials.

\chapter{Review and outlook}
\label{chap:lopetus}

Dimensional reduction is an approximate method which is perfectly suited for computing static observables in weakly coupled high temperature field theory. It is based on the observation that when the temperature is large and the coupling small, there is a clear scale hierarchy between the thermal, electric and magnetic scales $\pi T$, $gT$ and $g^2T$, respectively, and makes full use of that hierarchy by applying effective field theory techniques. In perturbation theory reorganizing the perturbative expansion by resummation of classes of diagrams is necessary in order to regulate the infrared divergences, and this can be carried out systematically using dimensional reduction, without worrying about double counting diagrams. For non-perturbative lattice simulations the dimensionally reduced effective theory is appealing because of the analytic treatment of fermions, lower space-time dimension and the integration out of scales $\sim \pi T$, allowing for larger lattice spacing. For these reasons, dimensional reduction has been extensively used to compute properties of both QCD and electroweak theory at high temperatures. However, it is not of use near the QCD phase transition, where $\alpha_s$ is large and the weak-coupling hierarchy of scales disappears. It should be also noted that dimensional reduction explicitly breaks the $Z(3)$ symmetry of degenerate QCD vacua above $T_c$, and the fluctuations between these states become important close to the phase transition.

In this thesis, we have studied two applications of dimensional reduction to the standard model physics. The pressure of the electroweak theory, which was previously undetermined beyond the first terms, was computed up to three-loop level, or $\order{g^5}$, in \cite{Gynther:2005dj,Gynther:2005av}. Combined with the previously known QCD pressure, this enables us to study the pressure of the full standard model above the electroweak phase transition. We have shown that the pressure of the symmetric phase goes smoothly through the transition point once the small value of the Higgs field mass is taken into account in the effective theory. The pressure of the whole standard model to this order lies 10--15\% below the ideal gas value, but the perturbative expansion does not converge well because of the large values of QCD and top quark Yukawa couplings. As a special case we have studied the weakly coupled SU(2) + fundamental scalar theory, where the convergence is good as expected.

The correlation lengths of mesonic operators at high temperatures have been computed in \cite{Laine:2003bd} and extended to finite chemical potentials in \cite{Vepsalainen:2007ke}. We have computed the leading order (free) correlators in full QCD, and then formulated a dimensionally reduced effective theory which includes the lowest fermionic modes in order to compute the first perturbative correction of order $g^2$ to masses. The fermionic theory is seen to correspond to a nonrelativistic theory for heavy quarks in 2+1 dimensions, and draws inspiration from the heavy quark effective theories used to compute the properties of quarkonia. The screening mass correction we find is small and positive, and it depends in a complicated way on the chemical potentials and the number of dynamical quarks. The finite density result is only in qualitative agreement with lattice results, and we discuss the potential differences in measuring the mass at finite distances.

Dimensional reduction has proved to be an efficient tool for computing bosonic observables at high temperatures both in perturbation theory and combined with non-perturbative methods. As a prominent example, the perturbative expansion of the QCD pressure has been computed as far as possible by purely perturbative means, and it shows good agreement with 4d lattice simulations if the unknown $g^6$ term related to magnetic scales turns out to have the right value. This is in accordance with the conjecture that the truncated perturbative series is close to the actual result once all dynamical scales have entered the computation \cite{Laine:2003ay}. Recently, first steps toward computing the missing $g^6$ correction have been taken by evaluating the four-loop pressure of massless O($N$) scalar theory, requiring for the first time the computation of four-loop sum-integrals \cite{Gynther:2007bw}. In the weak coupling regime dimensional reduction also extends to large values of $\mu/T$, for the actual requirement is simply that $\pi T$ should be larger than other mass scales, which for $\mu \gtrsim \pi T$ translates into $\pi T \gg \mE\sim g\mu$ as discussed in \cite{Ipp:2006ij}. An often overlooked part of the perturbative computations are finite quark masses \cite{Laine:2006cp}, which give sizeable contributions near the thresholds where they enter.

New directions for effective theory computations are provided by the possibility of studying also fermionic operators, as discussed in chapter \ref{chap:correlators}. To estimate the reliability of this method it would be central to understand the origin of differences between the perturbative and lattice results for the masses, possibly also implementing \NRQCD{} on lattice.

\end{fmffile}

\bibliographystyle{omatyyli}
\bibliography{/home/mtvepsal/texfiles/bibliography/articles.bib,/home/mtvepsal/texfiles/bibliography/books.bib}

\begin{thebibliography}{10}

\bibitem{Laine:2003bd}
M.~Laine and M.~Veps{\"{a}}l{\"{a}}inen,
\newblock JHEP {\bf 02}, 004 (2004) [hep-ph/0311268].

\bibitem{Gynther:2005dj}
A.~Gynther and M.~Veps{\"{a}}l{\"{a}}inen,
\newblock JHEP {\bf 01}, 060 (2006) [hep-ph/0510375].

\bibitem{Gynther:2005av}
A.~Gynther and M.~Veps{\"{a}}l{\"{a}}inen,
\newblock JHEP {\bf 03}, 011 (2006) [hep-ph/0512177].

\bibitem{Vepsalainen:2007ke}
M.~Veps{\"{a}}l{\"{a}}inen,
\newblock JHEP {\bf 03}, 022 (2007) [hep-ph/0701250].

\bibitem{Ginsparg:1980ef}
P.~Ginsparg,
\newblock Nucl. Phys. {\bf B170}, 388 (1980).

\bibitem{Appelquist:1981vg}
T.~Appelquist and R.~D. Pisarski,
\newblock Phys. Rev. {\bf D23}, 2305 (1981).

\bibitem{Shuryak:1977ut}
E.~V. Shuryak,
\newblock Sov. Phys. {JETP} {\bf 47}, 212 (1978).

\bibitem{Chin:1978gj}
S.~A. Chin,
\newblock Phys. Lett. {\bf B78}, 552 (1978).

\bibitem{Kapusta:1979fh}
J.~I. Kapusta,
\newblock Nucl. Phys. {\bf B148}, 461 (1979).

\bibitem{Toimela:1982hv}
T.~Toimela,
\newblock Phys. Lett. {\bf B124}, 407 (1983).

\bibitem{Arnold:1994ps}
P.~Arnold and C.~Zhai,
\newblock Phys. Rev. {\bf D50}, 7603 (1994) [hep-ph/9408276].

\bibitem{Arnold:1995eb}
P.~Arnold and C.~Zhai,
\newblock Phys. Rev. {\bf D51}, 1906 (1995) [hep-ph/9410360].

\bibitem{Zhai:1995ac}
C.~Zhai and B.~Kastening,
\newblock Phys. Rev. {\bf D52}, 7232 (1995) [hep-ph/9507380].

\bibitem{Braaten:1996jr}
E.~Braaten and A.~Nieto,
\newblock Phys. Rev. {\bf D53}, 3421 (1996) [hep-ph/9510408].

\bibitem{Kajantie:2002wa}
K.~Kajantie, M.~Laine, K.~Rummukainen and Y.~Schr{\"{o}}der,
\newblock Phys. Rev. {\bf D67}, 105008 (2003) [hep-ph/0211321].

\bibitem{Kajantie:2003ax}
K.~Kajantie, M.~Laine, K.~Rummukainen and Y.~Schr{\"{o}}der,
\newblock JHEP {\bf 04}, 036 (2003) [hep-ph/0304048].

\bibitem{Vuorinen:2003fs}
A.~Vuorinen,
\newblock Phys. Rev. {\bf D68}, 054017 (2003) [hep-ph/0305183].

\bibitem{Reisz:1992er}
T.~Reisz,
\newblock Z. Phys. {\bf C53}, 169 (1992).

\bibitem{Karkkainen:1992jh}
L.~K{\"{a}}rkk{\"{a}}inen, P.~Lacock, D.~E. Miller, B.~Petersson and T.~Reisz,
\newblock Phys. Lett. {\bf B282}, 121 (1992).

\bibitem{Karkkainen:1993wu}
L.~K{\"{a}}rkk{\"{a}}inen, P.~Lacock, B.~Petersson and T.~Reisz,
\newblock Nucl. Phys. {\bf B395}, 733 (1993).

\bibitem{Kajantie:1997tt}
K.~Kajantie, M.~Laine, K.~Rummukainen and M.~E. Shaposhnikov,
\newblock Nucl. Phys. {\bf B503}, 357 (1997) [hep-ph/9704416].

\bibitem{Laine:1998nq}
M.~Laine and O.~Philipsen,
\newblock Nucl. Phys. {\bf B523}, 267 (1998) [hep-lat/9711022].

\bibitem{Laine:1999hh}
M.~Laine and O.~Philipsen,
\newblock Phys. Lett. {\bf B459}, 259 (1999) [hep-lat/9905004].

\bibitem{Hart:1999dj}
A.~Hart and O.~Philipsen,
\newblock Nucl. Phys. {\bf B572}, 243 (2000) [hep-lat/9908041].

\bibitem{Hart:2000ha}
A.~Hart, M.~Laine and O.~Philipsen,
\newblock Nucl. Phys. {\bf B586}, 443 (2000) [hep-ph/0004060].

\bibitem{Cucchieri:2001tw}
A.~Cucchieri, F.~Karsch and P.~Petreczky,
\newblock Phys. Rev. {\bf D64}, 036001 (2001) [hep-lat/0103009].

\bibitem{Vuorinen:2006nz}
A.~Vuorinen and L.~G. Yaffe,
\newblock Phys. Rev. {\bf D74}, 025011 (2006) [hep-ph/0604100].

\bibitem{Kajantie:1998yc}
K.~Kajantie, M.~Laine, A.~Rajantie, K.~Rummukainen and M.~Tsypin,
\newblock JHEP {\bf 11}, 011 (1998) [hep-lat/9811004].

\bibitem{Kajantie:1995dw}
K.~Kajantie, M.~Laine, K.~Rummukainen and M.~E. Shaposhnikov,
\newblock Nucl. Phys. {\bf B458}, 90 (1996) [hep-ph/9508379].

\bibitem{Kajantie:1995kf}
K.~Kajantie, M.~Laine, K.~Rummukainen and M.~E. Shaposhnikov,
\newblock Nucl. Phys. {\bf B466}, 189 (1996) [hep-lat/9510020].

\bibitem{Kajantie:1996mn}
K.~Kajantie, M.~Laine, K.~Rummukainen and M.~E. Shaposhnikov,
\newblock Phys. Rev. Lett. {\bf 77}, 2887 (1996) [hep-ph/9605288].

\bibitem{Kajantie:1996qd}
K.~Kajantie, M.~Laine, K.~Rummukainen and M.~E. Shaposhnikov,
\newblock Nucl. Phys. {\bf B493}, 413 (1997) [hep-lat/9612006].

\bibitem{Karsch:1996yh}
F.~Karsch, T.~Neuhaus, A.~Patkos and J.~Rank,
\newblock Nucl. Phys. Proc. Suppl. {\bf 53}, 623 (1997) [hep-lat/9608087].

\bibitem{Gurtler:1997hr}
M.~Gurtler, E.-M. Ilgenfritz and A.~Schiller,
\newblock Phys. Rev. {\bf D56}, 3888 (1997) [hep-lat/9704013].

\bibitem{Vuorinen:2002ue}
A.~Vuorinen,
\newblock Phys. Rev. {\bf D67}, 074032 (2003) [hep-ph/0212283].

\bibitem{Huang:1996tz}
S.-z. Huang and M.~Lissia,
\newblock Nucl. Phys. {\bf B480}, 623 (1996) [hep-ph/9511383].

\bibitem{Braaten:1989mz}
E.~Braaten and R.~D. Pisarski,
\newblock Nucl. Phys. {\bf B337}, 569 (1990).

\bibitem{Braaten:1990az}
E.~Braaten and R.~D. Pisarski,
\newblock Nucl. Phys. {\bf B339}, 310 (1990).

\bibitem{Frenkel:1989br}
J.~Frenkel and J.~C. Taylor,
\newblock Nucl. Phys. {\bf B334}, 199 (1990).

\bibitem{Weinberg:qft2}
S.~Weinberg,
\newblock {\em The Quantum Theory of Fields, Vol.~2} (Cambridge University
  Press, 1996).

\bibitem{Kapusta:finite}
J.~I. Kapusta,
\newblock {\em Finite-temperature field theory} (Cambridge University Press,
  1989).

\bibitem{Weinberg:qft1}
S.~Weinberg,
\newblock {\em The Quantum Theory of Fields, Vol.~1} (Cambridge University
  Press, 1995).

\bibitem{Lebellac:thermal}
M.~Le~Bellac,
\newblock {\em Thermal field theory} (Cambridge University Press, 1996).

\bibitem{Cuniberti:2001hm}
G.~Cuniberti, E.~De~Micheli and G.~A. Viano,
\newblock Commun. Math. Phys. {\bf 216}, 59 (2001).

\bibitem{Matsubara:1955ws}
T.~Matsubara,
\newblock Prog. Theor. Phys. {\bf 14}, 351 (1955).

\bibitem{Collins:renormalization}
J.~Collins,
\newblock {\em Renormalization} (Cambridge University Press, 1984).

\bibitem{Appelquist:1974tg}
T.~Appelquist and J.~Carazzone,
\newblock Phys. Rev. {\bf D11}, 2856 (1975).

\bibitem{Lepage:1989hf}
G.~P. Lepage,
\newblock hep-ph/0506330.

\bibitem{Dobado:effective}
A.~Dobado, A.~G{\'o}mez-Nicola, A.~L. Maroto and J.~R. Pel{\'a}ez,
\newblock {\em Effective Lagrangians for the Standard Model} (Springer, 1997).

\bibitem{Ramond:1989yd}
P.~Ramond,
\newblock Front. Phys. {\bf 74}, 1 (1989).

\bibitem{Weinberg:1980wa}
S.~Weinberg,
\newblock Phys. Lett. {\bf B91}, 51 (1980).

\bibitem{Gross:1981br}
D.~J. Gross, R.~D. Pisarski and L.~G. Yaffe,
\newblock Rev. Mod. Phys. {\bf 53}, 43 (1981).

\bibitem{Linde:1980ts}
A.~D. Linde,
\newblock Phys. Lett. {\bf B96}, 289 (1980).

\bibitem{Arnold:1995bh}
P.~Arnold and L.~G. Yaffe,
\newblock Phys. Rev. {\bf D52}, 7208 (1995) [hep-ph/9508280].

\bibitem{Jakovac:1994xg}
A.~Jakovac, K.~Kajantie and A.~Patkos,
\newblock Phys. Rev. {\bf D49}, 6810 (1994) [hep-ph/9312355].

\bibitem{Braaten:1994na}
E.~Braaten,
\newblock Phys. Rev. Lett. {\bf 74}, 2164 (1995) [hep-ph/9409434].

\bibitem{Laine:2005ai}
M.~Laine and Y.~Schr{\"{o}}der,
\newblock JHEP {\bf 03}, 067 (2005) [hep-ph/0503061].

\bibitem{Anderson:1991zb}
G.~W. Anderson and L.~J. Hall,
\newblock Phys. Rev. {\bf D45}, 2685 (1992).

\bibitem{Carrington:1991hz}
M.~E. Carrington,
\newblock Phys. Rev. {\bf D45}, 2933 (1992).

\bibitem{Dine:1992wr}
M.~Dine, R.~G. Leigh, P.~Y. Huet, A.~D. Linde and D.~A. Linde,
\newblock Phys. Rev. {\bf D46}, 550 (1992) [hep-ph/9203203].

\bibitem{Arnold:1992rz}
P.~Arnold and O.~Espinosa,
\newblock Phys. Rev. {\bf D47}, 3546 (1993) [hep-ph/9212235].

\bibitem{Farakos:1994kx}
K.~Farakos, K.~Kajantie, K.~Rummukainen and M.~E. Shaposhnikov,
\newblock Nucl. Phys. {\bf B425}, 67 (1994) [hep-ph/9404201].

\bibitem{Fodor:1994bs}
Z.~Fodor and A.~Hebecker,
\newblock Nucl. Phys. {\bf B432}, 127 (1994) [hep-ph/9403219].

\bibitem{Bertone:2004pz}
G.~Bertone, D.~Hooper and J.~Silk,
\newblock Phys. Rept. {\bf 405}, 279 (2005) [hep-ph/0404175].

\bibitem{Hindmarsh:2005ix}
M.~Hindmarsh and O.~Philipsen,
\newblock Phys. Rev. {\bf D71}, 087302 (2005) [hep-ph/0501232].

\bibitem{Arnold:1993bq}
P.~Arnold and L.~G. Yaffe,
\newblock Phys. Rev. {\bf D49}, 3003 (1994) [hep-ph/9312221].

\bibitem{Eidelman:2004wy}
Particle Data Group, S.~Eidelman {\em et~al.},
\newblock Phys. Lett. {\bf B592}, 1 (2004).

\bibitem{Yao:2006px}
Particle Data Group, W.~M. Yao {\em et~al.},
\newblock J. Phys. {\bf G33}, 1 (2006).

\bibitem{Gynther:thesis}
A.~Gynther,
\newblock {\em Thermodynamics of electroweak matter},
\newblock PhD thesis, University of Helsinki, 2006, [hep-ph/0609226].

\bibitem{Csikor:1996sp}
F.~Csikor, Z.~Fodor, J.~Hein, A.~Jaster and I.~Montvay,
\newblock Nucl. Phys. {\bf B474}, 421 (1996) [hep-lat/9601016].

\bibitem{Weinberg:1987vp}
E.~J. Weinberg and A.-q. Wu,
\newblock Phys. Rev. {\bf D36}, 2474 (1987).

\bibitem{DeTar:1985kx}
C.~DeTar,
\newblock Phys. Rev. {\bf D32}, 276 (1985).

\bibitem{DeTar:1987ga}
C.~DeTar,
\newblock Phys. Rev. {\bf D37}, 2328 (1988).

\bibitem{DeTar:1987ar}
C.~DeTar and J.~B. Kogut,
\newblock Phys. Rev. Lett. {\bf 59}, 399 (1987).

\bibitem{DeTar:1987xb}
C.~DeTar and J.~B. Kogut,
\newblock Phys. Rev. {\bf D36}, 2828 (1987).

\bibitem{Gottlieb:1987gz}
S.~A. Gottlieb, W.~Liu, D.~Toussaint, R.~L. Renken and R.~L. Sugar,
\newblock Phys. Rev. Lett. {\bf 59}, 1881 (1987).

\bibitem{Hansson:1992kb}
T.~H. Hansson and I.~Zahed,
\newblock Nucl. Phys. {\bf B374}, 277 (1992).

\bibitem{Hansson:1994nb}
T.~H. Hansson, M.~Sporre and I.~Zahed,
\newblock Nucl. Phys. {\bf B427}, 545 (1994) [hep-ph/9401281].

\bibitem{Koch:1992nx}
V.~Koch, E.~V. Shuryak, G.~E. Brown and A.~D. Jackson,
\newblock Phys. Rev. {\bf D46}, 3169 (1992) [hep-ph/9204236].

\bibitem{Koch:1994zt}
V.~Koch,
\newblock Phys. Rev. {\bf D49}, 6063 (1994) [hep-ph/9401284].

\bibitem{Huang:1995yt}
S.-z. Huang and M.~Lissia,
\newblock Phys. Rev. {\bf D53}, 7270 (1996) [hep-ph/9509360].

\bibitem{Laine:2006ns}
M.~Laine, O.~Philipsen, P.~Romatschke and M.~Tassler,
\newblock JHEP {\bf 03}, 054 (2007) [hep-ph/0611300].

\bibitem{Florkowski:1993bq}
W.~Florkowski and B.~L. Friman,
\newblock Z. Phys. {\bf A347}, 271 (1994).

\bibitem{Luke:1999kz}
M.~E. Luke, A.~V. Manohar and I.~Z. Rothstein,
\newblock Phys. Rev. {\bf D61}, 074025 (2000) [hep-ph/9910209].

\bibitem{deForcrand:2000jx}
QCD-TARO, P.~de~Forcrand {\em et~al.},
\newblock Phys. Rev. {\bf D63}, 054501 (2001) [hep-lat/0008005].

\bibitem{Nomura:2001hv}
K.~Nomura, O.~Miyamura, T.~Umeda and H.~Matsufuru,
\newblock Nucl. Phys. Proc. Suppl. {\bf 106}, 507 (2002) [hep-lat/0110204].

\bibitem{Wissel:2005pb}
S.~Wissel, E.~Laermann, S.~Shcheredin, S.~Datta and F.~Karsch,
\newblock PoS {\bf LAT2005}, 164 (2006) [hep-lat/0510031].

\bibitem{Gavai:2006fs}
R.~V. Gavai, S.~Gupta and R.~Lacaze,
\newblock PoS {\bf LAT2006}, 135 (2006) [hep-lat/0609074].

\bibitem{Alberico:2007yj}
W.~M. Alberico, A.~Beraudo, A.~Czerska, P.~Czerski and A.~Molinari,
\newblock Nucl. Phys. {\bf A792}, 152 (2007) [hep-ph/0703298].

\bibitem{Florkowski:1997pi}
W.~Florkowski,
\newblock Acta Phys. Polon. {\bf B28}, 2079 (1997) [hep-ph/9701223].

\bibitem{Pushkina:2004wa}
QCD-TARO, I.~Pushkina {\em et~al.},
\newblock Phys. Lett. {\bf B609}, 265 (2005) [hep-lat/0410017].

\bibitem{Choe:2002mt}
S.~Choe {\em et~al.},
\newblock Phys. Rev. {\bf D65}, 054501 (2002).

\bibitem{Laine:2003ay}
M.~Laine,
\newblock hep-ph/0301011.

\bibitem{Gynther:2007bw}
A.~Gynther, M.~Laine, Y.~Schr{\"{o}}der, C.~Torrero and A.~Vuorinen,
\newblock JHEP {\bf 04}, 094 (2007) [hep-ph/0703307].

\bibitem{Ipp:2006ij}
A.~Ipp, K.~Kajantie, A.~Rebhan and A.~Vuorinen,
\newblock Phys. Rev. {\bf D74}, 045016 (2006) [hep-ph/0604060].

\bibitem{Laine:2006cp}
M.~Laine and Y.~Schr{\"{o}}der,
\newblock Phys. Rev. {\bf D73}, 085009 (2006) [hep-ph/0603048].

\end{thebibliography}

\end{document}